\documentclass[a4paper]{article}

\usepackage[english]{babel}
\usepackage[utf8]{inputenc}
\usepackage{braket} 
\usepackage{physics} 
\usepackage{amsmath}
\usepackage{graphicx}
\usepackage[colorinlistoftodos]{todonotes}
\usepackage{blochsphere}
\usepackage{tikz}
\usepackage{tikz-3dplot}
\usepackage{caption}
\usepackage{subcaption}
\usepackage{float}
\usepackage{xfrac}
\usepackage{rotating}
\usepackage{placeins}
\usepackage{soul}
\soulregister\ref7
\soulregister\cite7

\definecolor{darkpink}{rgb}{139, 0, 139}
\definecolor{darkbrown}{rgb}{101, 67, 33}

\title{Towards Two Bloch Sphere Representation of Pure Two Qubit States and Unitaries.}

\author{Stanislav Filatov and Marcis Auzinsh\\
\textit{University of Latvia, Department of Physics}\\
\textit{Raina boulevard 19, LV-1586, Riga, Latvia}\\
              E-mail: sutasu@tutanota.com
             }

\date{\today}

\begin{document}
\maketitle

\begin{abstract}

We extend Bloch Sphere formalism to pure two qubit systems. Combining insights from Geometric Algebra and analysis of entanglement in different conjugate bases we identify Two Bloch Sphere geometry that is suitable for representing maximally entangled states. It turns out that relative direction of coordinate axes of the two Bloch Spheres may be used to describe the states. Moreover, coordinate axes of one Bloch sphere should be rignt-handed and of the other one - left-handed. We describe and depict separable and maximally entangled states as well as entangling and non-entangling rotations. We also offer graphical representation of workings of a CNOT gate for different inputs. Finally we provide a way to also represent partially entangled states and describe entanglement measure related to the surface area of the sphere enclosing the state representation.

\end{abstract}

\maketitle 

\section{Introduction}

The Bloch sphere's historical roots extend to the early 20th century, driven by the contributions of physicist Felix Bloch. His work in nuclear magnetic resonance (NMR), which delves into the behavior of atomic nuclei in magnetic fields, led to the development of the Bloch sphere as a visualization tool for comprehending quantum properties, particularly those associated with quantum spins in the presence of magnetic fields. In 1946, Bloch published a seminal paper titled ``Nuclear Induction" \cite{Bloch1946}, which played a crucial role in advancing our knowledge of how quantum systems interact with magnetic fields. This work laid the groundwork for fundamental concepts in quantum physics, including nuclear magnetic resonance and spin coherence.

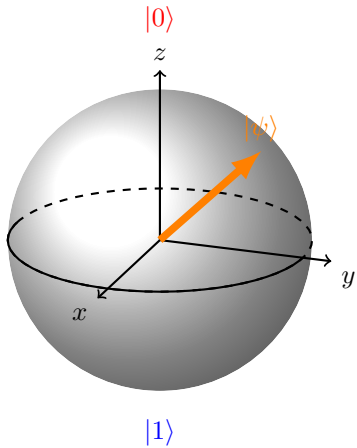
\begin{figure}
\centering
\tdplotsetmaincoords{70}{110}
\begin{tikzpicture}[tdplot_main_coords, scale=2]
    \shade[ball color=white!50, opacity=0.6] (0,0,0) circle (1cm);
    
    \draw[thick,->] (0,0,0) -- (1.2,0,0) node[anchor=north east]{$x$};
    \draw[thick,->] (0,0,0) -- (0,1.2,0) node[anchor=north west]{$y$};
    \draw[thick,->] (0,0,0) -- (0,0,1.2) node[anchor=south]{$z$};
    
    \node at (0,0,1.4) [above,red] {$\left|0\right>$};
    \node at (0,0,-1.2) [below,blue] {$\left|1\right>$};
    
    \tdplotdrawarc[thick]{(0,0,0)}{1}{-90}{90}{}{}
    
    \tdplotdrawarc[thick,dashed]{(0,0,0)}{1}{270}{-90}{}{}
    
    \coordinate (statevector) at (-0.3, 0.6, 0.6);
    
    \draw[orange, line width=1mm, -latex] (0,0,0) -- (statevector);
    
    \node at (statevector) [above,orange] {$\left|\psi\right>$};

\end{tikzpicture}
\caption{Arbitrary State Vector on Bloch Sphere}
\label{fig:arb-single}
\end{figure}

Mathematically, the Bloch sphere representation relies on the use of complex vector spaces to visually depict quantum states. This approach provides a geometric framework for understanding quantum systems while preserving essential information related to the amplitude and phase of these states \cite{NielsenChuang2000}. In this representation, every vector within the Bloch sphere corresponds to a specific quantum state (Figure \ref{fig:arb-single}). Pure, coherent states extend precisely to the surface of the Bloch sphere with a length of 1, while mixed, incoherent states fall short of reaching the sphere's surface with a length less than 1. The length of each representation's vector on the Bloch sphere provides insight into the degree of coherence within the quantum state. At the center of the sphere resides the maximally mixed state with length 0, characterized by a density matrix proportional to the identity matrix.

In addition to serving as a mathematical representation, the Bloch sphere finds practical applications in various physical fields. In the field of quantum optics, it facilitates the understanding of light polarization states, enabling insights into phenomena such as the polarization of single photons\cite{Loudon2000, GerryKnight2005}. In nuclear and atomic physics, the Bloch sphere serves as a crucial tool for describing atomic and nuclear spins. For instance, in nuclear magnetic resonance (NMR) experiments, it is employed to visualize the behavior of nuclear spins under the influence of magnetic fields, aiding in the analysis of spin dynamics\cite{Abragam1961, Slichter1996, CohenTannoudji2012}. In condensed matter physics, the Bloch sphere provides valuable insights into electron spin states in materials. It can be used to illustrate the behavior of electron spins in magnetic materials, enhancing our understanding of their magnetic properties\cite{Kittel2004,Sachdev2011} 

In the realm of quantum information, the Bloch sphere serves as a valuable tool for simplifying the visualization of quantum states, particularly in the context of qubits. Here, the concept of a qubit is abstracted away from its physical substrate, allowing us to focus solely on its logical and information-theoretical quantum mechanical properties\cite{NielsenChuang2000, Devitt2013}. Various implementations of qubits exist, ranging from superconducting circuits to trapped ions and quantum dots\cite{Devitt2013, Preskill1998}, but the Bloch sphere transcends the differences, emphasizing the fundamental quantum nature of these entities. Within the domain of quantum information processing, the Bloch sphere aids in understanding the workings of quantum gates and operations, which are fundamental components of quantum circuits\cite{Lidar2013}. This visualization simplifies the design and implementation of quantum algorithms and protocols, and it also highlights the challenges posed by decoherence, as it becomes evident how quantum states evolve and deviate from their ideal representations over time\cite{Preskill1998}.

This geometric representation simplifies the visualization of complex quantum states, allowing for an intuitive understanding of concepts related to quantum coherence and mixed states\cite{Preskill2018}. It provides a bridge between the abstract mathematics of quantum mechanics and the geometric visualization, enhancing our grasp of quantum phenomena without imposing judgments on their nature or significance.

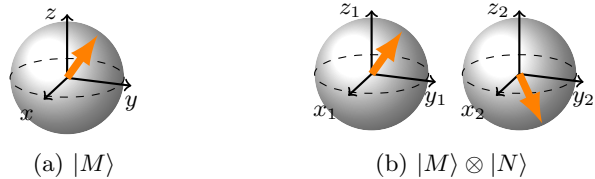
\begin{figure}
\begin{subfigure}{.4\textwidth}
  \centering
\tdplotsetmaincoords{70}{110}
\begin{tikzpicture}[tdplot_main_coords, scale=0.75]
    \shade[ball color=white!50, opacity=0.6] (0,0,0) circle (1cm);
    
    \draw[thick,->, scale=1] (0,0,0) -- (1.2,0,0) node[anchor=north east, font=\small]{$x$};
    \draw[thick,->, scale=1] (0,0,0) -- (0,1.2,0) node[anchor=north, font=\small]{$y$};
    \draw[thick,->, scale=1] (0,0,0) -- (0,0,1.2) node[anchor=east, font=\small]{$z$};
    
    
    \tdplotdrawarc[thin, dashed, scale=1]{(0,0,0)}{1}{-4}{356}{}{}

    \coordinate (statevector) at (-0.2, 0.5, 0.8);
    
    \draw[orange, line width=1mm, -latex] (0,0,0) -- (statevector);

\end{tikzpicture}
\caption{$\ket{M}$}
\end{subfigure}
\begin{subfigure}{.4\textwidth}
  \centering
\begin{minipage}[b][1.85cm][t]{0.4\textwidth}
\centering
\tdplotsetmaincoords{70}{110}
\begin{tikzpicture}[tdplot_main_coords, scale=0.75]
    \shade[ball color=white!50, opacity=0.6] (0,0,0) circle (1cm);
    
    \draw[thick,->, scale=1] (0,0,0) -- (1.2,0,0) node[anchor=north east, font=\small]{$x_1$};
    \draw[thick,->, scale=1] (0,0,0) -- (0,1.2,0) node[anchor=north, font=\small]{$y_1$};
    \draw[thick,->, scale=1] (0,0,0) -- (0,0,1.2) node[anchor=east, font=\small]{$z_1$};
    
    
    \tdplotdrawarc[thin, dashed, scale=1]{(0,0,0)}{1}{-4}{356}{}{}

    \coordinate (statevector) at (-0.2, 0.5, 0.8);
    
    \draw[orange, line width=1mm, -latex] (0,0,0) -- (statevector);

\end{tikzpicture}
\end{minipage}%
\begin{minipage}[b][1.85cm][t]{0.4\textwidth}
\centering
\tdplotsetmaincoords{70}{110}
\begin{tikzpicture}[tdplot_main_coords, scale=0.75]
    \shade[ball color=white!50, opacity=0.6] (0,0,0) circle (1cm);
    
    \draw[thick,->, scale=1] (0,0,0) -- (1.2,0,0) node[anchor=north east, font=\small]{$x_2$};
    \draw[thick,->, scale=1] (0,0,0) -- (0,1.2,0) node[anchor=north, font=\small]{$y_2$};
    \draw[thick,->, scale=1] (0,0,0) -- (0,0,1.2) node[anchor=east, font=\small]{$z_2$};
    
    
    \tdplotdrawarc[thin, dashed, scale=1]{(0,0,0)}{1}{-4}{356}{}{}

    \coordinate (statevector) at (0.2, 0.5, -0.8);
    
    \draw[orange, line width=1mm, -latex] (0,0,0) -- (statevector);

\end{tikzpicture}
\end{minipage}

  \caption{$\ket{M} \otimes \ket{N}$}
\end{subfigure}
\caption{(a) Arbitrary Single Qubit state. (b) Atribtrary Separable Two Qubit state}
\label{fig:arb-sep}
\end{figure}

Extending the BS formalism to two qubits is not only logical but also highly desirable. This extension would significantly enhance our ability to analyze experiments in the Foundations of Quantum Physics and comprehend the workings of Quantum Circuits. Two-qubit systems represent the simplest scenarios where entanglement, a quintessentially quantum phenomenon, may emerge.

In the context of quantum information, entanglement is pivotal for quantum communication protocols, ensuring secure information transmission through quantum key distribution \cite{Ekert1991, Nadlinger2022}. Moving to quantum computation, entanglement becomes a crucial resource for quantum gates, enabling parallel information processing and the potential to solve specific problems exponentially faster than classical computers \cite{Shor1997}. 

In the foundations of quantum mechanics, entanglement challenges classical intuitions by highlighting the non-local nature of quantum correlations. This, in turn, sparks debates about the fundamental principles of the quantum world \cite{Einstein1935, Bell1964}. Moreover, advancements in the interplay between entanglement and gravity have opened up new avenues for understanding the profound connections between quantum mechanics and the nature of spacetime \cite{MaldacenaSusskind2013}.

Entanglement, however, is what makes the two BS representation difficult to obtain. While separable states are easily visualized (Figure \ref{fig:arb-sep}), entanglement introduces complexity without a corresponding structural representation. The single-qubit reduced density matrix of any maximally entangled state is proportional to the identity. Simply depicting Bloch sphere with a dot in the middle for every qubit of entangled pair results in the same representation for all maximally entangled states — two spheres with dots in the middle. Which is not informative at all. While one can visualize clouds of different color around the two Bloch spheres to signify entanglement and its different kinds, this approach still fails to reveal the distinct geometries of the underlying states. 

There have been attempts at creating thorus geometry visualizations \cite{Thorus2022} or even sphere geometries \cite{Wie2020}, an important milestone in creating the two BS representation. We haven't seen any, though, that would use only two Bloch Spheres and connect straightforward representation of separable states with representation of entanglement. Moreover, there has been an extensive research into the geometry and symmetries of the SU(N) and specifically SU(4) groups \cite{Uskov2008, Rau2021, Bengtsson2006, Khaneja2000}. In particular, \cite{Uskov2008} present an extension of the Bloch sphere to two qubits involving a four-sphere at each point of which is a fiber consisting of two two-spheres of SU(2).

At the core of our work lies the uncovering of the structure that describes maximally entangled states within two Bloch Sphere formalism aided by the insights from Geometric Algebra. The structure is the relative direction of the Bloch Sphere's coordinate axes, one being right-handed, the other one being left-handed. Insights from Geometric Algebra are association of each of the 15 two-qubit pauli matrices with the 15 planes of rotation formed by the coordinate axes of two Bloch Spheres. Combining everything together we get representations of separable and maximally entangled states as well as dynamics of all the two-qubit unitaries. We navigate the space of two qubit states with a step of $\pi/2$. Not only these rotations are the most widespread in quantum computation but, importantly, they allow to obtain a set of visually orthogonal representations (Stabilizer set of two-qubit states) that can further be merged to create representation of an arbitrary pure state and rotation. 

In this work, we specifically focus on pure two-qubit states, so unless otherwise stated, the term "state" refers to pure states. If not specified otherwise, "entanglement" denotes maximal entanglement, and an "entangled state" refers to a Maximally Entangled State (MES), while Partially Entangled States (PES) will be explicitly identified or referred to as such to make a clear distinction from MES.

The subsequent sections are organized as follows: Chapter 2 provides a brief overview of the relevant concepts in Geometric Algebra; Chapter 3 focuses on representing two-qubit entanglement; Chapter 4 elucidates the visual dynamics of a two-qubit system, detailing the process of inferring eigenmatrices from graphical representations and performing various Unitary Operations without relying on matrix formalism; Chapter 5 demonstrates usefulness of this approach for analysis of logical operations and as an example presents a step-by-step breakdown of the workings of the CNOT gate for three different inputs; Chapter 6 addresses the representation of arbitrary two-qubit pure states and rotations and provides a way to measure entanglement within the state graphically; the graphical representation of the entanglement and disentanglement dynamics can be found in Figure \ref{fig:ent-full-rot} of that chapter; In Chapter 7 we discuss possible questions regarding the model and Chapter 8 serves as the conclusion of this work. 
\FloatBarrier

\section{Geometric Algebra of two qubits}

Geometric algebra offers tools for visually interpreting the matrix formalism of Quantum Mechanics. It provides the structure to represent four-dimensional complex Hilbert space of two qubit system in six-dimensional real Euclidean space. Visualizing what is going on in three dimensions is simple, but it gets more complex when trying to capture even a facet of what's unfolding in six dimensions. GA serves as a bridge between abstract mathematics and embodied intuition by employing concepts such as ``bivector," ``plane of rotation," and ``handedness". At the same time, GA remains an exceptionally versatile system, offering a unified language for diverse areas of mathematics, physics, and engineering that traditionally employ field-specific formalisms \cite{Hestenes1984, Lasenby2017}. Therefore, we will take some time to introduce the essential concepts within this system.

Several sources \cite{Gull1993, Hestenes2003, MacDonald2011, Hestenes1999, Doran1993} offer valuable insights into GA, but our primary reference is the work by Havel and Doran \cite{HavelDoran2004}. The introduction of their work explains the fundamental principles of GA, while its core employs GA to create a Bloch-sphere-like GA model for a two-qubit system in a six-dimensional Real Euclidean space. This work has been the main inspiration and guide in our explorations. Although the authors do not explicitly address a representation involving two Bloch Spheres, their graph model and the clear connection between GA concepts and standard quantum mechanical formalism were crucial in shaping the representations that will be developed later in this work. For detailed mathematical information and formal proofs, we refer the reader to this specific work.

They note, among other things, that just as in the case of two-dimensional complex Hilbert space of a single qubit where Lie algebra isomorphism so(3) $\approx$ su(2) allows to represent states of a qubit in a three-dimensional real Euclidean vector space, in the case of four-dimensional complex Hilbert space of two qubits, the Lie algebra isomorphisms spin(6) $\approx$ so(6) $\approx$ su(4) allow to represent states of two-qubit system in a six-dimensional real Euclidean vector space. More precisely, the respective Geometric algebras of real Euclidean three- and six- dimensional vector spaces are suitable to represent single- and two- qubit states and operators. 

The fundamental component of geometric algebra (GA) is the bivector concept. For a rigorous account of GA concepts we refer the reader to the literature on GA \cite{Gull1993, Hestenes2003, MacDonald2011, Hestenes1999, Doran1993, HavelDoran2004}. We will give an intuitive description of concepts relevant to our work. Bivectors result from the wedge product (notated as $a\land b$), involving two vectors $a$ and $b$. Similar to how vectors define an oriented line, bivectors define an oriented plane. They also specify a rotation within that plane, with the rotation direction determined by the order of the vectors in the wedge product. For example,  $a\land b$ signifies a rotation from vector $a$ to vector $b$. In three-dimensional space, each rotation plane possesses a single perpendicular vector, which can be termed an axis of rotation. In higher-dimensional spaces, multiple normals exist for each rotation plane, rendering the concept of an ``axis of rotation" ill-defined for rotations in spaces exceeding three dimensions. Conversely, the notion of a ``plane of rotation" remains well-defined in spaces of any dimensionality. Bivectors are anticommutative $a \land b = - (b \land a)$ and $-(b \land a) = -b \land a = b \land -a$. One could interpret the anticommutativity relations as follows: To reverse a rotation from $a$ to $b$ we need to make a rotation from $b$ to $a$.

We employ the concept of plane of rotation to visualize rotations of the statevector on a Bloch sphere. Moreover, we use passive rotations instead of active ones. Namely, the arrow representing the statevector remains fixed and the coordinate axes are rotated instead. While such an approach arguably doesn't add more clarity when representing the separable states, it will be seen later that it is essential in representing the entangled states. 

Following \cite{HavelDoran2004} we associate each generator of rotations, which has the form of a tensor product of two elements of the set $\{ I; \sigma_x ; \sigma_y ; \sigma_z \}$\ with a plane of rotation formed by two out of six axes of the two Bloch spheres $\{ x_1 ; y_1 ; z_1; x_2; y_2; z_2 \}$. See equations below that list all the correspondences. The correspondences for local rotations (i.e. rotations in plane formed by the axes of the same Bloch Sphere) are seen in eq. \ref{eq:1} and correspondences for double-Pauli rotations (rotations in plane formed by the axes of different Bloch Spheres) in eq. \ref{eq:2}. 

We refrain from the term ``non-local" when speaking about the plane or rotation. Instead to describe those planes we use term ``double-Pauli" to emphasize the nature of the rotation when expressed in matrix formalism. However, if we interpret wedge products of the form $x_1 \land y_2$ as planes and forget the matrix description of the rotation these planes may be characterized as ``non-local" to reflect the fact that they are formed by coordinate axes of different Bloch Spheres. Furthermore because those planes are formed by axes of different BS they won't be visible in the two-Bloch-sphere visualization unlike the planes formed by two axes of the same Bloch sphere. 
\newline

\begin{samepage}
\begin{center}
local rotations
\end{center}
\begin{equation} \label{eq:1}
\begin{split}
I \otimes \sigma_x & \leftrightarrow y_2 \land z_2 \\
I \otimes \sigma_y & \leftrightarrow z_2 \land x_2 \\
I \otimes \sigma_z & \leftrightarrow x_2 \land y_2 \\ 
\end{split}
\quad
\begin{split}
\sigma_x \otimes I & \leftrightarrow y_1 \land z_1 \\
\sigma_y \otimes I & \leftrightarrow z_1 \land x_1 \\
\sigma_z \otimes I & \leftrightarrow x_1 \land y_1 \\
\end{split}
\end{equation}
\end{samepage}
\newline

\begin{figure}
\begin{subfigure}{.4\textwidth}
  \centering
\begin{minipage}[b][2cm][t]{0.4\textwidth}
\centering
\tdplotsetmaincoords{70}{110}
\begin{tikzpicture}[tdplot_main_coords, scale=0.75]
    \shade[ball color=white!50, opacity=0.6] (0,0,0) circle (1cm);
    
    \draw[thick,->, scale=1] (0,0,0) -- (1.2,0,0) node[anchor=north east, font=\small]{$x_1$};
    \draw[thick,->, scale=1] (0,0,0) -- (0,1.2,0) node[anchor=north, font=\small]{$y_1$};
    \draw[thick,->, scale=1] (0,0,0) -- (0,0,1.2) node[anchor=east, font=\small]{$z_1$};
    
    
    \tdplotdrawarc[thin, dashed, scale=1]{(0,0,0)}{1}{-4}{356}{}{}

    \coordinate (statevector) at (0, 0, 1);
    
    \draw[orange, line width=1mm, -latex] (0,0,0) -- (statevector);

\end{tikzpicture}
\end{minipage}%
\begin{minipage}[b][2cm][t]{0.4\textwidth}
\centering
\tdplotsetmaincoords{70}{110}
\begin{tikzpicture}[tdplot_main_coords, scale=0.75]
    \shade[ball color=white!50, opacity=0.6] (0,0,0) circle (1cm);
    
    \draw[thick,->, scale=1] (0,0,0) -- (1.2,0,0) node[anchor=north east, font=\small]{$x_2$};
    \draw[thick,->, scale=1] (0,0,0) -- (0,1.2,0) node[anchor=north, font=\small]{$y_2$};
    \draw[thick,->, scale=1] (0,0,0) -- (0,0,1.2) node[anchor=east, font=\small]{$z_2$};
    
    
    \tdplotdrawarc[thin, dashed, scale=1]{(0,0,0)}{1}{-4}{356}{}{}

    \coordinate (statevector) at (0, 0, 1);
    
    \draw[orange, line width=1mm, -latex] (0,0,0) -- (statevector);

\end{tikzpicture}
\end{minipage}

  \caption{$\ket{\uparrow\uparrow}$}
\end{subfigure}
\raisebox{2\height}{\LARGE$\xrightarrow[e^{i\frac{\pi}{4}(I \otimes \sigma_x)}]{y_2 \land z_2}$}%
\begin{subfigure}{.4\textwidth}
  \centering
\begin{minipage}[b][2cm][t]{0.4\textwidth}
\centering
\tdplotsetmaincoords{70}{110}
\begin{tikzpicture}[tdplot_main_coords, scale=0.75]
    \shade[ball color=white!50, opacity=0.6] (0,0,0) circle (1cm);
    
    \draw[thick,->, scale=1] (0,0,0) -- (1.2,0,0) node[anchor=north east, font=\small]{$x_1$};
    \draw[thick,->, scale=1] (0,0,0) -- (0,1.2,0) node[anchor=north, font=\small]{$y_1$};
    \draw[thick,->, scale=1] (0,0,0) -- (0,0,1.2) node[anchor=east, font=\small]{$z_1$};
    
    
    \tdplotdrawarc[thin, dashed, scale=1]{(0,0,0)}{1}{-4}{356}{}{}

    \coordinate (statevector) at (0, 0, 1);
    
    \draw[orange, line width=1mm, -latex] (0,0,0) -- (statevector);

\end{tikzpicture}
\end{minipage}%
\begin{minipage}[b][2cm][t]{0.4\textwidth}
\centering
\tdplotsetmaincoords{70}{110}
\begin{tikzpicture}[tdplot_main_coords, scale=0.75]
    \shade[ball color=white!50, opacity=0.6] (0,0,0) circle (1cm);
    
    \draw[thick,->, scale=1] (0,0,0) -- (1.2,0,0) node[anchor=north east, font=\small]{$x_2$};
    \draw[thick,->, scale=1] (0,0,0) -- (0,-1.2,0) node[anchor=south, font=\small]{$z_2$};
    \draw[thick,->, scale=1] (0,0,0) -- (0,0,1.2) node[anchor=west, font=\small]{$y_2$};
    
    
    \tdplotdrawarc[thin, dashed, scale=1]{(0,0,0)}{1}{-4}{356}{}{}

    \coordinate (statevector) at (0, 0, 1);
    
    \draw[orange, line width=1mm, -latex] (0,0,0) -- (statevector);

\end{tikzpicture}
\end{minipage}

  \caption{$\ket{\uparrow} \otimes \ket{\uparrow + i \downarrow}$}
\end{subfigure}

\vspace{0.5cm}

\begin{subfigure}{.4\textwidth}
  \centering
\begin{minipage}[b][2cm][t]{0.4\textwidth}
\centering
\tdplotsetmaincoords{70}{110}
\begin{tikzpicture}[tdplot_main_coords, scale=0.75]
    \shade[ball color=white!50, opacity=0.6] (0,0,0) circle (1cm);
    
    \draw[thick,->, scale=1] (0,0,0) -- (1.2,0,0) node[anchor=north east, font=\small]{$x_1$};
    \draw[thick,->, scale=1] (0,0,0) -- (0,1.2,0) node[anchor=north, font=\small]{$y_1$};
    \draw[thick,->, scale=1] (0,0,0) -- (0,0,1.2) node[anchor=east, font=\small]{$z_1$};
    
    
    \tdplotdrawarc[thin, dashed, scale=1]{(0,0,0)}{1}{-4}{356}{}{}

    \coordinate (statevector) at (0, 0, 1);
    
    \draw[orange, line width=1mm, -latex] (0,0,0) -- (statevector);

\end{tikzpicture}
\end{minipage}%
\begin{minipage}[b][2cm][t]{0.4\textwidth}
\centering
\tdplotsetmaincoords{70}{110}
\begin{tikzpicture}[tdplot_main_coords, scale=0.75]
    \shade[ball color=white!50, opacity=0.6] (0,0,0) circle (1cm);
    
    \draw[thick,->, scale=1] (0,0,0) -- (1.2,0,0) node[anchor=north east, font=\small]{$x_2$};
    \draw[thick,->, scale=1] (0,0,0) -- (0,1.2,0) node[anchor=north, font=\small]{$y_2$};
    \draw[thick,->, scale=1] (0,0,0) -- (0,0,1.2) node[anchor=east, font=\small]{$z_2$};
    
    
    \tdplotdrawarc[thin, dashed, scale=1]{(0,0,0)}{1}{-4}{356}{}{}

    \coordinate (statevector) at (0, 0, 1);
    
    \draw[orange, line width=1mm, -latex] (0,0,0) -- (statevector);

\end{tikzpicture}
\end{minipage}

  \caption{$\ket{\uparrow\uparrow}$}
\end{subfigure}
\raisebox{2\height}{\LARGE$\xrightarrow[e^{i\frac{\pi}{4}(\sigma_x \otimes I)}]{y_1 \land z_1}$}%
\begin{subfigure}{.4\textwidth}
  \centering
\begin{minipage}[b][2cm][t]{0.4\textwidth}
\centering
\tdplotsetmaincoords{70}{110}
\begin{tikzpicture}[tdplot_main_coords, scale=0.75]
    \shade[ball color=white!50, opacity=0.6] (0,0,0) circle (1cm);

    \draw[thick,->, scale=1] (0,0,0) -- (1.2,0,0) node[anchor=north east, font=\small]{$x_1$};
    \draw[thick,->, scale=1] (0,0,0) -- (0,-1.2,0) node[anchor=south, font=\small]{$z_1$};
    \draw[thick,->, scale=1] (0,0,0) -- (0,0,1.2) node[anchor=west, font=\small]{$y_1$};

    
    \tdplotdrawarc[thin, dashed, scale=1]{(0,0,0)}{1}{-4}{356}{}{}

    \coordinate (statevector) at (0, 0, 1);
    
    \draw[orange, line width=1mm, -latex] (0,0,0) -- (statevector);

\end{tikzpicture}
\end{minipage}%
\begin{minipage}[b][2cm][t]{0.4\textwidth}
\centering
\tdplotsetmaincoords{70}{110}
\begin{tikzpicture}[tdplot_main_coords, scale=0.75]
    \shade[ball color=white!50, opacity=0.6] (0,0,0) circle (1cm);
    
    \draw[thick,->, scale=1] (0,0,0) -- (1.2,0,0) node[anchor=north east, font=\small]{$x_2$};
    \draw[thick,->, scale=1] (0,0,0) -- (0,1.2,0) node[anchor=north, font=\small]{$y_2$};
    \draw[thick,->, scale=1] (0,0,0) -- (0,0,1.2) node[anchor=east, font=\small]{$z_2$};
    
    
    \tdplotdrawarc[thin, dashed, scale=1]{(0,0,0)}{1}{-4}{356}{}{}

    \coordinate (statevector) at (0, 0, 1);
    
    \draw[orange, line width=1mm, -latex] (0,0,0) -- (statevector);

\end{tikzpicture}
\end{minipage}

  \caption{$\ket{\uparrow + i \downarrow} \otimes \ket{\uparrow}$}
\end{subfigure}

\vspace{0.5cm}

\begin{subfigure}{.4\textwidth}
  \centering
\begin{minipage}[b][2cm][t]{0.4\textwidth}
\centering
\tdplotsetmaincoords{70}{110}
\begin{tikzpicture}[tdplot_main_coords, scale=0.75]
    \shade[ball color=white!50, opacity=0.6] (0,0,0) circle (1cm);
    
    \draw[thick,->, scale=1] (0,0,0) -- (1.2,0,0) node[anchor=north east, font=\small]{$x_1$};
    \draw[thick,->, scale=1] (0,0,0) -- (0,1.2,0) node[anchor=north, font=\small]{$y_1$};
    \draw[thick,->, scale=1] (0,0,0) -- (0,0,1.2) node[anchor=east, font=\small]{$z_1$};
    
    
    \tdplotdrawarc[thin, dashed, scale=1]{(0,0,0)}{1}{-4}{356}{}{}

    \coordinate (statevector) at (0, 0, 1);
    
    \draw[orange, line width=1mm, -latex] (0,0,0) -- (statevector);

\end{tikzpicture}
\end{minipage}%
\begin{minipage}[b][2cm][t]{0.4\textwidth}
\centering
\tdplotsetmaincoords{70}{110}
\begin{tikzpicture}[tdplot_main_coords, scale=0.75]
    \shade[ball color=white!50, opacity=0.6] (0,0,0) circle (1cm);
    
    \draw[thick,->, scale=1] (0,0,0) -- (1.2,0,0) node[anchor=north east, font=\small]{$x_2$};
    \draw[thick,->, scale=1] (0,0,0) -- (0,1.2,0) node[anchor=north, font=\small]{$y_2$};
    \draw[thick,->, scale=1] (0,0,0) -- (0,0,1.2) node[anchor=east, font=\small]{$z_2$};
    
    
    \tdplotdrawarc[thin, dashed, scale=1]{(0,0,0)}{1}{-4}{356}{}{}

    \coordinate (statevector) at (0, 0, 1);
    
    \draw[orange, line width=1mm, -latex] (0,0,0) -- (statevector);

\end{tikzpicture}
\end{minipage}

  \caption{$\ket{\uparrow\uparrow}$}
\end{subfigure}
\raisebox{2\height}{\LARGE$\xrightarrow[e^{i\frac{\pi}{4}(\sigma_x \otimes I)}]{y_1 \land z_1}$}%
\begin{subfigure}{.4\textwidth}
  \centering
\begin{minipage}[b][2cm][t]{0.4\textwidth}
\centering
\tdplotsetmaincoords{70}{110}
\begin{tikzpicture}[tdplot_main_coords, scale=0.75]
    \shade[ball color=white!50, opacity=0.6] (0,0,0) circle (1cm);

    \draw[thick,->, scale=1] (0,0,0) -- (1.2,0,0) node[anchor=north east, font=\small]{$x_1$};
    \draw[thick,->, scale=1] (0,0,0) -- (0,1.2,0) node[anchor=south, font=\small]{$y_1$};
    \draw[thick,->, scale=1] (0,0,0) -- (0,0,1.2) node[anchor=west, font=\small]{$z_1$};

    
    \tdplotdrawarc[thin, dashed, scale=1]{(0,0,0)}{1}{-4}{356}{}{}

    \coordinate (statevector) at (0, 1, 0);
    
    \draw[orange, line width=1mm, -latex] (0,0,0) -- (statevector);

\end{tikzpicture}
\end{minipage}%
\begin{minipage}[b][1.75cm][t]{0.4\textwidth}
\centering
\tdplotsetmaincoords{70}{110}
\begin{tikzpicture}[tdplot_main_coords, scale=0.75]
    \shade[ball color=white!50, opacity=0.6] (0,0,0) circle (1cm);
    
    \draw[thick,->, scale=1] (0,0,0) -- (1.2,0,0) node[anchor=north east, font=\small]{$x_2$};
    \draw[thick,->, scale=1] (0,0,0) -- (0,1.2,0) node[anchor=north, font=\small]{$z_2$};
    \draw[thick,->, scale=1] (0,0,0) -- (0,0,-1.2) node[anchor=east, font=\small]{$y_2$};
    
    
    \tdplotdrawarc[thin, dashed, scale=1]{(0,0,0)}{1}{-4}{356}{}{}

    \coordinate (statevector) at (0, 1, 0);
    
    \draw[orange, line width=1mm, -latex] (0,0,0) -- (statevector);

\end{tikzpicture}
\end{minipage}

  \caption{$\ket{\uparrow + i \downarrow} \otimes \ket{\uparrow}$}
\end{subfigure}

\vspace{0.5cm}

\begin{subfigure}{.4\textwidth}
  \centering
\begin{minipage}[b][2cm][t]{0.4\textwidth}
\centering
\tdplotsetmaincoords{70}{110}
\begin{tikzpicture}[tdplot_main_coords, scale=0.75]
    \shade[ball color=white!50, opacity=0.6] (0,0,0) circle (1cm);
    
    \draw[thick,->, scale=1] (0,0,0) -- (1.2,0,0) node[anchor=north east, font=\small]{$x_1$};
    \draw[thick,->, scale=1] (0,0,0) -- (0,1.2,0) node[anchor=north, font=\small]{$y_1$};
    \draw[thick,->, scale=1] (0,0,0) -- (0,0,1.2) node[anchor=east, font=\small]{$z_1$};
    
    
    \tdplotdrawarc[thin, dashed, scale=1]{(0,0,0)}{1}{-4}{356}{}{}

    \coordinate (statevector) at (0, 0, 1);
    
    \draw[orange, line width=1mm, -latex] (0,0,0) -- (statevector);

\end{tikzpicture}
\end{minipage}%
\begin{minipage}[b][2cm][t]{0.4\textwidth}
\centering
\tdplotsetmaincoords{70}{110}
\begin{tikzpicture}[tdplot_main_coords, scale=0.75]
    \shade[ball color=white!50, opacity=0.6] (0,0,0) circle (1cm);
    
    \draw[thick,->, scale=1] (0,0,0) -- (1.2,0,0) node[anchor=north east, font=\small]{$x_2$};
    \draw[thick,->, scale=1] (0,0,0) -- (0,1.2,0) node[anchor=north, font=\small]{$y_2$};
    \draw[thick,->, scale=1] (0,0,0) -- (0,0,1.2) node[anchor=east, font=\small]{$z_2$};
    
    
    \tdplotdrawarc[thin, dashed, scale=1]{(0,0,0)}{1}{-4}{356}{}{}

    \coordinate (statevector) at (0, 0, 1);
    
    \draw[orange, line width=1mm, -latex] (0,0,0) -- (statevector);

\end{tikzpicture}
\end{minipage}

  \caption{$\ket{\uparrow\uparrow}$}
\end{subfigure}
\raisebox{2\height}{\LARGE$\xrightarrow[e^{i\frac{\pi}{4}(I \otimes \sigma_z)}]{x_2 \land y_2}$}%
\begin{subfigure}{.4\textwidth}
  \centering
\begin{minipage}[b][2cm][t]{0.4\textwidth}
\centering
\tdplotsetmaincoords{70}{110}
\begin{tikzpicture}[tdplot_main_coords, scale=0.75]
    \shade[ball color=white!50, opacity=0.6] (0,0,0) circle (1cm);
    
    \draw[thick,->, scale=1] (0,0,0) -- (1.2,0,0) node[anchor=north east, font=\small]{$x_1$};
    \draw[thick,->, scale=1] (0,0,0) -- (0,1.2,0) node[anchor=north, font=\small]{$y_1$};
    \draw[thick,->, scale=1] (0,0,0) -- (0,0,1.2) node[anchor=east, font=\small]{$z_1$};
    
    
    \tdplotdrawarc[thin, dashed, scale=1]{(0,0,0)}{1}{-4}{356}{}{}

    \coordinate (statevector) at (0, 0, 1);
    
    \draw[orange, line width=1mm, -latex] (0,0,0) -- (statevector);

\end{tikzpicture}
\end{minipage}%
\begin{minipage}[b][2cm][t]{0.4\textwidth}
\centering
\tdplotsetmaincoords{70}{110}
\begin{tikzpicture}[tdplot_main_coords, scale=0.75]
    \shade[ball color=white!50, opacity=0.6] (0,0,0) circle (1cm);
    
    \draw[thick,->, scale=1] (0,0,0) -- (-1.2,0,0) node[anchor=south west, font=\small]{$y_2$};
    \draw[thick,->, scale=1] (0,0,0) -- (0,1.2,0) node[anchor=north, font=\small]{$x_2$};
    \draw[thick,->, scale=1] (0,0,0) -- (0,0,1.2) node[anchor=east, font=\small]{$z_2$};
    
    
    \tdplotdrawarc[thin, dashed, scale=1]{(0,0,0)}{1}{-4}{356}{}{}

    \coordinate (statevector) at (0, 0, 1);
    
    \draw[orange, line width=1mm, -latex] (0,0,0) -- (statevector);

\end{tikzpicture}
\end{minipage}

  \caption{$\ket{\uparrow\uparrow}$}
\end{subfigure}
\caption{Local rotations of separable states. (a) to (b) rotation of plane defined by $y \land z$ of the second BS. (c) to (d) rotation of plane defined by $y \land z$ of the first BS. (e) to (f) same rotation as in (c) to (d) depicted in the notation we will be using. (g) to (h) eigenrotaion that doesn't change the state.}
\label{fig:RotSep}
\end{figure}
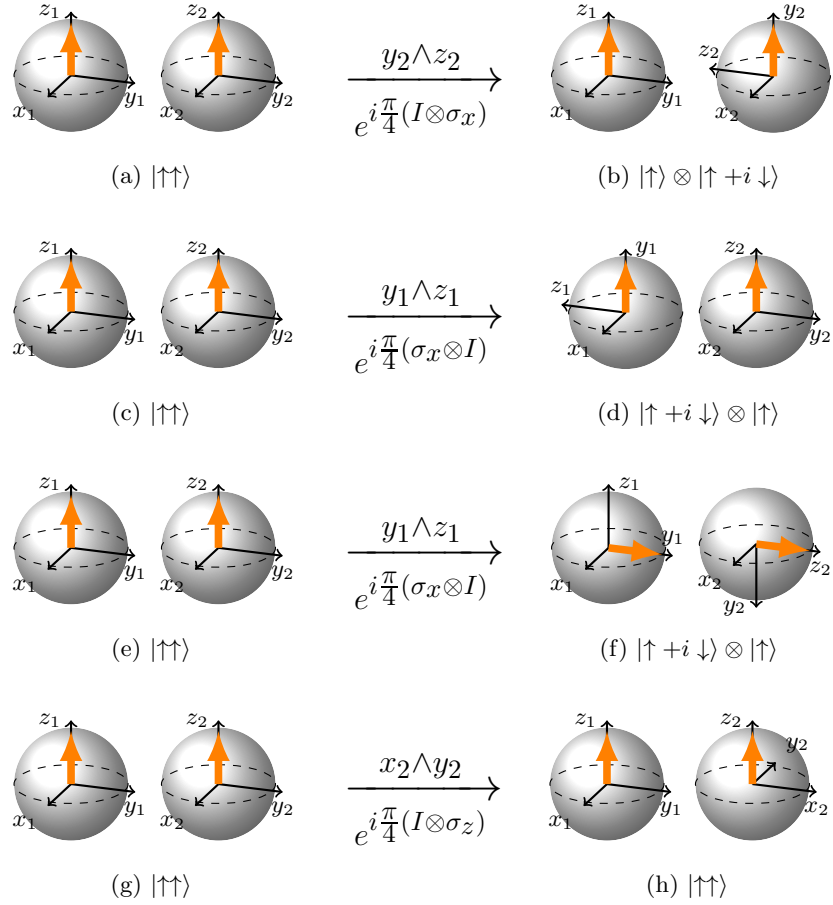

The local rotations of separable states are seen in Figure \ref{fig:RotSep}. The order of axes in the wedge product determines the directionality of the plane they define and the direction of rotation of that plane on the picture. Once the internal coordinates of each BS are explicitly added to the picture there is considerable freedom in how to represent the states because rigid rotations of the whole BS (coordinates and the statevector) don't change the state represented. Rather they amount to a change of point of view from which we are looking at the BS. 

Convention we have chosen is following: we always represent two BS in such a way that the statevector arrows are pointing in the same absolute direction; the coordinates of the first BS are always fixed. These rules make it easier to obtain and visually recognize the entangled states later. Due to such notational choice local rotations in planes that belong to the first BS require not only rotation of the plane relative to the statevector, but also a rigid rotation of both BS so that the two aforementioned rules are satisfied (see Figure \ref{fig:RotSep} (c) through (f)). Rotations depicted in (c) to (d) are physically the same rotations as (e) to (f). (f) is representing the final state in a convention we have chosen: the coordinates of the first BS should be always drawn in the same way and both statevector arrows must face in the same absolute direction.. A move from (d) to (f) is essentially a change of point of view on both BS. 

Eigentorations, rotations that don't change the state of the system are rotations that happen in the plane that is perpendicular to the statevector arrow. In other words, any rotation of the plane perpendicular to the statevector arrow may be considered a change of point of view on the BS. Strictly speaking, the direction of the axes perpendicular to the statevector arrow cannot be well defined, only the bivector formed by them can be - without language of GA it would be difficult to formulate such an observation. We will see later that such ambiguity in where those coordinate axes should be drawn is an important ingredient in the mechanism of rotation between separable and entangled states.

\begin{samepage}
\begin{center}
double-Pauli rotations
\end{center}
\begin{equation} \label{eq:2}
\begin{split}
\sigma_x \otimes \sigma_x & \leftrightarrow x_1 \land x_2 \\
\sigma_x \otimes \sigma_y & \leftrightarrow x_1 \land y_2 \\
\sigma_x \otimes \sigma_z & \leftrightarrow x_1 \land z_2 \\
\end{split}
\quad
\begin{split}
\sigma_y \otimes \sigma_x & \leftrightarrow y_1 \land x_2 \\
\sigma_y \otimes \sigma_y & \leftrightarrow y_1 \land y_2 \\
\sigma_y \otimes \sigma_z & \leftrightarrow y_1 \land z_2 \\
\end{split}
\quad
\begin{split}
\sigma_z \otimes \sigma_x & \leftrightarrow z_1 \land x_2 \\
\sigma_z \otimes \sigma_y & \leftrightarrow z_1 \land y_2 \\
\sigma_z \otimes \sigma_z & \leftrightarrow z_1 \land z_2 \\
\end{split}
\end{equation}
\end{samepage}
\newline

Furthermore, it can be shown using Cartan decomposition \cite{HavelDoran2004, FuchsSchweigert1997} that any rotation in our six-dimensional space (in other words, any unitary transformation) may be written as a combination of rotations in six local planes and rotations in only three double-Pauli planes. Which leaves us with 9 relevant planes of rotation. The most symmetric choice of three double-Pauli planes of rotation would be $ \{ x_1 \land x_2; y_1 \land y_2; z_1 \land z_2 \} $. In fact, such decomposition may be done using the formalism of Pauli Matrices (see Chapter 4 in \cite{Khaneja2000}).

\FloatBarrier

\section{Representing Two-Qubit Entanglement}

Creating a representation for a two-qubit system requires careful consideration, especially when dealing with entanglement. The structure needs to reflect that an entangled state doesn't provide information about individual qubits but rather conveys information about both qubits as a unified system. Moreover, since entangled states are non-separable, there aren't specific directions in which the qubits face. While we can make statements about the relative direction of the qubits, like ``qubits always face in opposite directions for the state $\ket{\Psi^-}$," this holds true for some entangled states and not for others. Despite adhering to certain symmetries, translating these symmetries into a clear representation in a two Bloch Sphere framework is not straightforward. Thus, graphically illustrating two-partite entanglement becomes a challenge.

Surprisingly, the key element for constructing the representation of entangled states did not originate from GA. Instead, it emerged through the process of rewriting the four Bell basis states in all three conjugate bases. Disregarding normalization constant and having $\uparrow_x = (\uparrow + \downarrow)_z$; $\downarrow_x = (\uparrow - \downarrow)_z$; $\uparrow_y = (\uparrow +\ i\downarrow)_z$; $\uparrow_y = (\uparrow -\ i\downarrow)_z$ one obtains:

\begin{equation}
{
\renewcommand{\arraystretch}{5} 
\setlength{\arraycolsep}{15pt} 
\begin{array}{cc}
\begin{aligned}
\Psi^- &= (\uparrow\downarrow - \downarrow\uparrow)_z \\
    &= (\uparrow\downarrow - \downarrow\uparrow)_x \\
    &= (\uparrow\downarrow - \downarrow\uparrow)_y
\end{aligned}
&
\begin{aligned}
\Psi^+ &= (\uparrow\downarrow + \downarrow\uparrow)_z \\
    &= (\uparrow\uparrow - \downarrow\downarrow)_x \\
    &= (\uparrow\uparrow - \downarrow\downarrow)_y
\end{aligned} \\
\begin{aligned}
\Phi^- &= (\uparrow\uparrow - \downarrow\downarrow)_z \\
    &= (\uparrow\downarrow + \downarrow\uparrow)_x \\
    &= (\uparrow\uparrow + \downarrow\downarrow)_y
\end{aligned}
&
\begin{aligned}
\Phi^+ &= (\uparrow\uparrow + \downarrow\downarrow)_z \\
    &= (\uparrow\uparrow + \downarrow\downarrow)_x \\
    &= (\uparrow\downarrow + \downarrow\uparrow)_y
\end{aligned}
\end{array}
}
\end{equation}

\begin{figure}[H]
\begin{subfigure}{.5\textwidth}
  \centering
\begin{minipage}[b][2cm][t]{0.4\textwidth}
\centering
\tdplotsetmaincoords{70}{110}
\begin{tikzpicture}[tdplot_main_coords, scale=0.75]
    \shade[ball color=white!50, opacity=0.6] (0,0,0) circle (1cm);
    
    \draw[thick,->, scale=1] (0,0,0) -- (1.2,0,0) node[anchor=north east, font=\small]{$x_1$};
    \draw[thick,->, scale=1] (0,0,0) -- (0,1.2,0) node[anchor=north, font=\small]{$y_1$};
    \draw[thick,->, scale=1] (0,0,0) -- (0,0,1.2) node[anchor=east, font=\small]{$z_1$};
    
    
    \tdplotdrawarc[thin, dashed, scale=1]{(0,0,0)}{1}{-4}{356}{}{}
    
    \shade[ball color=orange, opacity=1, scale=2] (0,0,0) circle (0.05cm);
\end{tikzpicture}
\end{minipage}%
\begin{minipage}[b][1.75cm][t]{0.4\textwidth}
\centering
\tdplotsetmaincoords{70}{110}
\begin{tikzpicture}[tdplot_main_coords, scale=0.75]
    \shade[ball color=white!50, opacity=0.6] (0,0,0) circle (1cm);
    
    \draw[thick,->, scale=1] (0,0,0) -- (-1.2,0,0) node[anchor=south west, font=\small]{$x_2$};
    \draw[thick,->, scale=1] (0,0,0) -- (0,-1.2,0) node[anchor=south, font=\small]{$y_2$};
    \draw[thick,->, scale=1] (0,0,0) -- (0,0,-1.2) node[anchor=west, font=\small]{$z_2$};
    
    
    \tdplotdrawarc[thin, dashed, scale=1]{(0,0,0)}{1}{-4}{356}{}{}
    
    \shade[ball color=orange, opacity=1, scale=2] (0,0,0) circle (0.05cm);
\end{tikzpicture}
\end{minipage}

  \caption{$\Psi^-$}
  \label{fig:sub-first}
\end{subfigure}
\begin{subfigure}{.5\textwidth}
  \centering
\begin{minipage}[b][2cm][t]{0.4\textwidth}
\centering
\tdplotsetmaincoords{70}{110}
\begin{tikzpicture}[tdplot_main_coords, scale=0.75]
    \shade[ball color=white!50, opacity=0.6] (0,0,0) circle (1cm);
    
    \draw[thick,->, scale=1] (0,0,0) -- (1.2,0,0) node[anchor=north east, font=\small]{$x_1$};
    \draw[thick,->, scale=1] (0,0,0) -- (0,1.2,0) node[anchor=north, font=\small]{$y_1$};
    \draw[thick,->, scale=1] (0,0,0) -- (0,0,1.2) node[anchor=east, font=\small]{$z_1$};
    
    
    \tdplotdrawarc[thin, dashed, scale=1]{(0,0,0)}{1}{-4}{356}{}{}
    
    \shade[ball color=orange, opacity=1, scale=2] (0,0,0) circle (0.05cm);
\end{tikzpicture}
\end{minipage}
\begin{minipage}[b][1.75cm][t]{0.4\textwidth}
\centering
\tdplotsetmaincoords{70}{110}
\begin{tikzpicture}[tdplot_main_coords, scale=0.75]
    \shade[ball color=white!50, opacity=0.6] (0,0,0) circle (1cm);
    
    \draw[thick,->, scale=1] (0,0,0) -- (1.2,0,0) node[anchor=north east, font=\small]{$x_2$};
    \draw[thick,->, scale=1] (0,0,0) -- (0,1.2,0) node[anchor=north, font=\small]{$y_2$};
    \draw[thick,->, scale=1] (0,0,0) -- (0,0,-1.2) node[anchor=west, font=\small]{$z_2$};
    
    
    \tdplotdrawarc[thin, dashed, scale=1]{(0,0,0)}{1}{-4}{356}{}{}
    
    \shade[ball color=orange, opacity=1, scale=2] (0,0,0) circle (0.05cm);
\end{tikzpicture}
\end{minipage}

  \caption{$\Psi^+$}
  \label{fig:sub-second}
\end{subfigure}

\vspace{0.5cm}

\begin{subfigure}{.5\textwidth}
  \centering
\begin{minipage}[b][2cm][t]{0.4\textwidth}
\centering
\tdplotsetmaincoords{70}{110}
\begin{tikzpicture}[tdplot_main_coords, scale=0.75]
    \shade[ball color=white!50, opacity=0.6] (0,0,0) circle (1cm);
    
    \draw[thick,->, scale=1] (0,0,0) -- (1.2,0,0) node[anchor=north east, font=\small]{$x_1$};
    \draw[thick,->, scale=1] (0,0,0) -- (0,1.2,0) node[anchor=north, font=\small]{$y_1$};
    \draw[thick,->, scale=1] (0,0,0) -- (0,0,1.2) node[anchor=east, font=\small]{$z_1$};
    
    
    \tdplotdrawarc[thin, dashed, scale=1]{(0,0,0)}{1}{-4}{356}{}{}
    
    \shade[ball color=orange, opacity=1, scale=2] (0,0,0) circle (0.05cm);
\end{tikzpicture}
\end{minipage}%
\begin{minipage}[b][2cm][t]{0.4\textwidth}
\centering
\tdplotsetmaincoords{70}{110}
\begin{tikzpicture}[tdplot_main_coords, scale=0.75]
    \shade[ball color=white!50, opacity=0.6] (0,0,0) circle (1cm);
    
    \draw[thick,->, scale=1] (0,0,0) -- (-1.2,0,0) node[anchor=south west, font=\small]{$x_2$};
    \draw[thick,->, scale=1] (0,0,0) -- (0,1.2,0) node[anchor=north, font=\small]{$y_2$};
    \draw[thick,->, scale=1] (0,0,0) -- (0,0,1.2) node[anchor=east, font=\small]{$z_2$};
    
    
    \tdplotdrawarc[thin, dashed, scale=1]{(0,0,0)}{1}{-4}{356}{}{}
    
    \shade[ball color=orange, opacity=1, scale=2] (0,0,0) circle (0.05cm);
\end{tikzpicture}
\end{minipage}
  \caption{$\Phi^-$}
  \label{fig:sub-third}
\end{subfigure}
\begin{subfigure}{.5\textwidth}
  \centering
\begin{minipage}[b][2cm][t]{0.4\textwidth}
\centering
\tdplotsetmaincoords{70}{110}
\begin{tikzpicture}[tdplot_main_coords, scale=0.75]
    \shade[ball color=white!50, opacity=0.6] (0,0,0) circle (1cm);
    
    \draw[thick,->, scale=1] (0,0,0) -- (1.2,0,0) node[anchor=north east, font=\small]{$x_1$};
    \draw[thick,->, scale=1] (0,0,0) -- (0,1.2,0) node[anchor=north, font=\small]{$y_1$};
    \draw[thick,->, scale=1] (0,0,0) -- (0,0,1.2) node[anchor=east, font=\small]{$z_1$};
    
    
    \tdplotdrawarc[thin, dashed, scale=1]{(0,0,0)}{1}{-4}{356}{}{}
    
    \shade[ball color=orange, opacity=1, scale=2] (0,0,0) circle (0.05cm);
\end{tikzpicture}
\end{minipage}%
\begin{minipage}[b][2cm][t]{0.4\textwidth}
\centering
\tdplotsetmaincoords{70}{110}
\begin{tikzpicture}[tdplot_main_coords, scale=0.75]
    \shade[ball color=white!50, opacity=0.6] (0,0,0) circle (1cm);
    
    \draw[thick,->, scale=1] (0,0,0) -- (1.2,0,0) node[anchor=north east, font=\small]{$x_2$};
    \draw[thick,->, scale=1] (0,0,0) -- (0,-1.2,0) node[anchor=south, font=\small]{$y_2$};
    \draw[thick,->, scale=1] (0,0,0) -- (0,0,1.2) node[anchor=east, font=\small]{$z_2$};
    
    
    \tdplotdrawarc[thin, dashed, scale=1]{(0,0,0)}{1}{-4}{356}{}{}
    
    \shade[ball color=orange, opacity=1, scale=2] (0,0,0) circle (0.05cm);
\end{tikzpicture}
\end{minipage}
  \caption{$\Phi^+$}
  \label{fig:sub-fourth}
\end{subfigure}
\caption{Bell states depicted using two Bloch spheres. A pair of Bloch Spheres, with differing handedness in their coordinate axes, represents each state. Odd number of coordinate axes is always inverted on the second BS. For example, $\Phi^-$ has the following representations (up to a global phase and dropping the normalization constant): $(z) \ket{\uparrow\uparrow} - \ket{\downarrow\downarrow}$; $(x) \ket{\uparrow\downarrow} + \ket{\downarrow\uparrow}$; $(y) \ket{\uparrow\uparrow} + \ket{\downarrow\downarrow}$. Therefore the Bloch spheres will have $z_1$ and $z_2$; $y_1$ and $y_2$ aligned and $x_1$ and $x_2$ anti-aligned. Similarly for the rest of Bell states.}
\label{fig:Bell}
\end{figure}
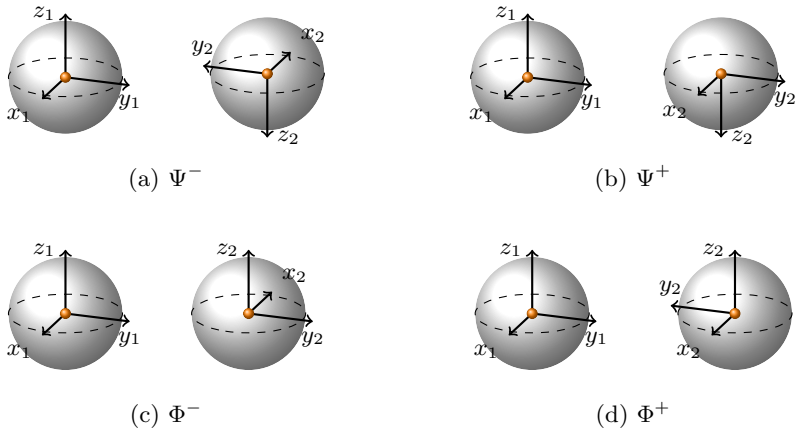

Now, recalling that the reduced density matrix for each of the qubits in a maximally entangled state is Identity and therefore corresponds to the dot in the middle of the Bloch sphere, let us try to use the information provided by different conjugate basis expansions to create a representation of the four Bell basis states. Let us use the following rule: whenever we see the arrows in the description of a state being aligned, e.g. $(\uparrow\uparrow + \downarrow\downarrow)$, we look at the label and draw the corresponding coordinate axes of the two Bloch spheres as facing in the same direction. And whenever they are anti-aligned in the description - we draw the corresponding axes to be anti-aligned. Going through all the labels ($x,y,z$) we draw all the axes on both Bloch spheres. We put a dot in the middle of the Bloch sphere instead of an arrow of the statevector. If we follow these simple rules we get the following visual representations (Figure \ref{fig:Bell}).

Let us note that the four representations reflect exactly the four ways in which the right-handed coordinates may be transformed into left-handed using only inversion of axes: there are three ways of inverting a single axis ($x, y$ or $z$) and there is one way to invert all three axes simultaneously. If one inverts two axes the coordinates will stay right-handed and such an inversion would be equivalent to some rotation. Also, it is impossible to go from right- to left-handed coordinates by rotation of axes only. 

The next important thing to note about this representation is that the choice of the orientation of axes on the first Bloch sphere is arbitrary, and the rules of drawing the axes specify only the aligned/anti-aligned relationships between the axes, not their absolute orientation. Hence is the next difference from the separable state representation: in the representation of entangled states only the relative directions of coordinate axes matter. This property together with a dot in the centre of the Bloch spheres reflects precisely the fact that looking at a single qubit of an entangled pair doesn't reveal anything about the entangled state.

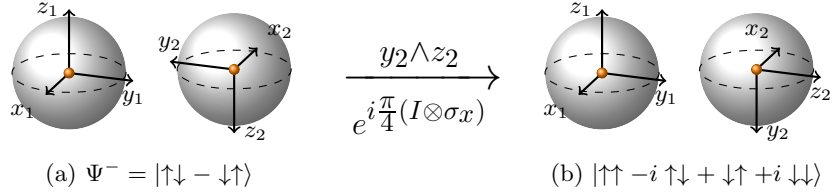
\begin{figure}
\begin{subfigure}{.4\textwidth}
  \centering
\begin{minipage}[b][2cm][t]{0.4\textwidth}
\centering
\tdplotsetmaincoords{70}{110}
\begin{tikzpicture}[tdplot_main_coords, scale=0.75]
    \shade[ball color=white!50, opacity=0.6] (0,0,0) circle (1cm);
    
    \draw[thick,->, scale=1] (0,0,0) -- (1.2,0,0) node[anchor=north east, font=\small]{$x_1$};
    \draw[thick,->, scale=1] (0,0,0) -- (0,1.2,0) node[anchor=north, font=\small]{$y_1$};
    \draw[thick,->, scale=1] (0,0,0) -- (0,0,1.2) node[anchor=east, font=\small]{$z_1$};
    
    
    \tdplotdrawarc[thin, dashed, scale=1]{(0,0,0)}{1}{-4}{356}{}{}
    
    \shade[ball color=orange, opacity=1, scale=2] (0,0,0) circle (0.05cm);
\end{tikzpicture}
\end{minipage}%
\begin{minipage}[b][1.75cm][t]{0.4\textwidth}
\centering
\tdplotsetmaincoords{70}{110}
\begin{tikzpicture}[tdplot_main_coords, scale=0.75]
    \shade[ball color=white!50, opacity=0.6] (0,0,0) circle (1cm);
    
    \draw[thick,->, scale=1] (0,0,0) -- (-1.2,0,0) node[anchor=south west, font=\small]{$x_2$};
    \draw[thick,->, scale=1] (0,0,0) -- (0,-1.2,0) node[anchor=south, font=\small]{$y_2$};
    \draw[thick,->, scale=1] (0,0,0) -- (0,0,-1.2) node[anchor=west, font=\small]{$z_2$};
    
    
    \tdplotdrawarc[thin, dashed, scale=1]{(0,0,0)}{1}{-4}{356}{}{}
    
    \shade[ball color=orange, opacity=1, scale=2] (0,0,0) circle (0.05cm);
\end{tikzpicture}
\end{minipage}

  \caption{$\Psi^-= \ket{\uparrow\downarrow - \downarrow\uparrow} $}
\end{subfigure}
\raisebox{2\height}{\LARGE$\xrightarrow[e^{i\frac{\pi}{4}(I \otimes \sigma_x)}]{y_2 \land z_2}$}%
\begin{subfigure}{.4\textwidth}
  \centering
\begin{minipage}[b][2cm][t]{0.4\textwidth}
\centering
\tdplotsetmaincoords{70}{110}
\begin{tikzpicture}[tdplot_main_coords, scale=0.75]
    \shade[ball color=white!50, opacity=0.6] (0,0,0) circle (1cm);
    
    \draw[thick,->, scale=1] (0,0,0) -- (1.2,0,0) node[anchor=north east, font=\small]{$x_1$};
    \draw[thick,->, scale=1] (0,0,0) -- (0,1.2,0) node[anchor=north, font=\small]{$y_1$};
    \draw[thick,->, scale=1] (0,0,0) -- (0,0,1.2) node[anchor=east, font=\small]{$z_1$};
    
    
    \tdplotdrawarc[thin, dashed, scale=1]{(0,0,0)}{1}{-4}{356}{}{}
    
    \shade[ball color=orange, opacity=1, scale=2] (0,0,0) circle (0.05cm);
\end{tikzpicture}
\end{minipage}
\begin{minipage}[b][1.75cm][t]{0.4\textwidth}
\centering
\tdplotsetmaincoords{70}{110}
\begin{tikzpicture}[tdplot_main_coords, scale=0.75]
    \shade[ball color=white!50, opacity=0.6] (0,0,0) circle (1cm);
    
    \draw[thick,->, scale=1] (0,0,0) -- (0,1.2,0) node[anchor=north, font=\small]{$z_2$};
    \draw[thick,->, scale=1] (0,0,0) -- (-1.2,0,0) node[anchor=south east, font=\small]{$x_2$};
    \draw[thick,->, scale=1] (0,0,0) -- (0,0,-1.2) node[anchor=west, font=\small]{$y_2$};
    
    
    \tdplotdrawarc[thin, dashed, scale=1]{(0,0,0)}{1}{-4}{356}{}{}
    
    \shade[ball color=orange, opacity=1, scale=2] (0,0,0) circle (0.05cm);
\end{tikzpicture}
\end{minipage}

\caption{$\ket{\uparrow\uparrow - i\uparrow\downarrow + \downarrow\uparrow + i\downarrow\downarrow}$}
\end{subfigure}
\caption{In order to graphically apply $\pi/2$ rotation using $I \otimes \sigma_x$ to the $\Psi^-$ state we take the plane that corresponds to such a rotation (see eq. \ref{eq:1}), in this case it's $y_2 \land z_2$, and rotate this plane by $\pi/2$ in the direction indicated by the letter order in the wedge product. In this case the direction may be described as ``from $y_2$ to $z_2$".}
\label{fig:PsiXRot}
\end{figure}

The next question one may ask is ``How to depict other maximally entangled states?" This is the moment when the concept of plane of rotation and all the relations between GA and standard Matrix representation of quantum mechanics come into play. Every maximally entangled state may be reached by single-qubit local rotations from any other entangled state \cite{Preskill-notes}. From our representation it becomes visually obvious why - because any relative configuration of coordinate axes may be arrived at through rotations on a single sphere. That is why we can use the correspondences from eq. \ref{eq:1} and just rotate the relevant planes of relevant Bell state to obtain all the maximally entangled state representations. 
A great benefit of GA isomorphism is that even in left-handed coordinates and even without a statevector arrow to rotate (!) rotations are very simple. One just needs to rotate the plane of rotation formed by the relevant axes in the direction specified by the order of axes in the wedge product. Further as an example we use $\Psi^-$ state to obtain other maximally entangled state. For clarity of representation we drop the normalization constants. In eq. \ref{eq:PsiRotX} we apply $\pi/2$ rotation to the state $\Psi^-$ using the $I \otimes \sigma_x$ generator of rotation and obtain the state $\ket{\uparrow\uparrow - i\uparrow\downarrow + i\downarrow\uparrow - \downarrow\downarrow}$. In fig. \ref{fig:PsiXRot} we show a graphical representation of this rotation.

\begin{equation}
\begin{aligned}
Rot(\theta) = e^{i(I \otimes \sigma_x)\frac{\theta}{2}} &= \cos{\frac{\theta}{2}} \ (I \otimes I) + i\sin{\frac{\theta}{2}} (I \otimes \sigma_x) \Rightarrow \\
Rot(\frac{\pi}{2}) = e^{i(I \otimes \sigma_x)\frac{\pi}{4}} &= \cos{\frac{\pi}{4}} \ (I \otimes I) + i\sin{\frac{\pi}{4}} (I \otimes \sigma_x) \Rightarrow\\
Rot(\frac{\pi}{2})\ket{\Psi^-} &=\\
\frac{(I \otimes I + i I \otimes \sigma_x)}{\sqrt{2}}\ket{\Psi^-} &= \frac{1}{2}\left[\begin{pmatrix}
1 & 0 & 0 & 0 \\
0 & 1 & 0 & 0 \\
0 & 0 & 1 & 0 \\
0 & 0 & 0 & 1
\end{pmatrix} 
+ i \begin{pmatrix}
0 & 1 & 0 & 0 \\
1 & 0 & 0 & 0 \\
0 & 0 & 0 & 1 \\
0 & 0 & 1 & 0
\end{pmatrix} \right] \cdot \begin{pmatrix}
0 \\
1 \\
-1 \\
0
\end{pmatrix} \\
&= \frac{1}{2}\begin{pmatrix}
1 & i & 0 & 0 \\
i & 1 & 0 & 0 \\
0 & 0 & 1 & i \\
0 & 0 & i & 1
\end{pmatrix} \cdot \begin{pmatrix}
0 \\
1 \\
-1 \\
0
\end{pmatrix} 
= \frac{i}{2}\begin{pmatrix}
1 \\
-i \\
i \\
-1
\end{pmatrix}
\end{aligned}
\label{eq:PsiRotX}
\end{equation}

Following this procedure we can generate representations of all maximally entangled states. See appendices for representations of 24 maximally entangled states generated by $\pi/2$ rotations. Incidentally, one could examine stabilizer group for a given maximally entangled state and use it to obtain our graphical representation. We focus on $\pi/2$ rotations because they produce a set of visually orthogonal representations on the Bloch Spheres (Stabilizer set) that can later be combined with each other to obtain representation of any state. As opposed to $\pi$ rotations which produce mathematically orthogonal vectors in Hilbert Space, but create too few representations for unambiguous interpolation. The set of separable states may be generated in a similar way. It consists of 36 states. The simplest way to visualize it is to note that there are 6 visually orthogonal directions which an arrow of the BS might be facing (along or opposite each of the three coordinate axes). Squaring this number to account for two BS and therefore two qubits - gives us 36. 

The differences in the number of states are reflected in the different symmetries of representations. For the separable states absolute direction of the single arrow matters while for entangled states relative direction of coordinate axes is important. In the language of GA this observation may be expressed as follows. For separable states absolute direction of two vectors are important while for entangled states relative directions of two trivectors are important. 

For every separable state there are 3 planes - two of them local - whose rotation doesn't result in the change of state (eigenplanes); 8 planes - 4 local, 4 double-Pauli - that result in separable states only and 4 planes - all double-Pauli - that connect the state to entangled states. For every entangled state the 15 planes of rotation are distributed, respectively as 3 eigenplanes - all double-Pauli; 6 planes that result in entangled states only - all local; 6 planes that connect to separable states - all double-Pauli. 

One could say that it is easier to stay inside the set of separable states or that it is easier to get out of the set of entangled states, hence - the difference in the size of sets (36 vs 24). The total number 60 confirms that we have obtained the whole set of Stabilizer states for two qubits \cite{Gross2006}. 

\FloatBarrier

\section{Exploring Two Qubit Stabilizer Set Dynamics Graphically}

After examining the structure of bipartite entanglement, our focus now shifts to exploring the capabilities within our framework. Within this framework, we can visually perform rotations without resorting to matrix formalism, and we can  deduce the eigenmatrices of a given state just by looking at its representataion. To systematize this information, we've established a set of visual rules. While these rules may appear somewhat extensive, one should keep in mind that they accommodate the intricacies of rotations in six-dimensional space. Indeed, after a bit of practice one realizes that behind the words lay straightforward intuitive moves. Importantly, the articulation of these rules not only aids in manual application but also lays the groundwork for their computational implementation. This coding aspect enables the creation of visualizations that are even more efficient than manual rule application. These rules unveil interesting symmetries within two-qubit dynamics, providing an intuitive understanding of such systems.

We have already mentioned rule for performing visual rotations in local planes. We just take the coordinate axes corresponding to a local rotation (eq. \ref{eq:1}) and rotate the plane defined by the coordinate axes pair in the wedge product. This rule works both in entangled mode and separable mode and accounts for 6 out of 15 planes planes of rotation (Figures \ref{fig:RotSep}, \ref{fig:PsiXRot}).

\subsection{Eigenrotations}
Before speaking about double-Pauli rotations, let's speak about visual rules for guessing the eigenplanes (planes of rotation that correspond to the matrix that doesn't change the state up to an overall phase). There are always three eigenmatrices for every separable and entangled state. Two local and one double-Pauli for separable states and three double-Pauli for entangled.

For separable states eigenmatrices will be a) tensor product of eigenmatrices of each individual qubit; b,c) tensor product of each individual qubit's eigenmatrix with Identity. In other words, planes that are perpendicular to the direction of statevector on the Bloch spheres will be eigenplanes. For example, for the state $\ket{\uparrow \leftarrow} = \ket{\uparrow \uparrow - \uparrow \downarrow}$ the arrow of the first qubit is facing in $z$ direction and the arrow of the second qubit is facing in $-x$ direction. Therefore the eigenmatrices will be $\{ \sigma_z \otimes \sigma_x$; $\sigma_z \otimes I$; $I \otimes \sigma_x \}$, which translates into GA language as the eigenplanes are $\{ z_1 \land x_2$; $x_1 \land y_1$; $y_2 \land z_2 \}$. The alignment or anti-alignment along the axis doesn't play a role in determining the eigenplane, only the axis itself which is logical because if it is an eigenrotation it doesn't matter whether it's performed clockwise or counterclockwise - it won't change the state.

For maximally entangled states everything is about the relative direction of coordinate axes, which also determines the eigenplanes. Here aligned axes determine the eigenplanes. For example, the Bell states (Figure \ref{fig:Bell}) all have $x_1$ aligned with $x_2$, $y_1$ with $y_2$, $z_1$ with $z_2$, therefore all of them will have the same sets of eigenplanes and eigenmatrices: $\{ x_1 \land x_2$; $y_1 \land y_2$; $z_1 \land z_2 \}$ and $\{ \sigma_x \otimes \sigma_x$; $\sigma_y \otimes \sigma_y$; $\sigma_z \otimes \sigma_z \}$. State $\ket{\uparrow\uparrow - i\uparrow\downarrow + \downarrow\uparrow + i\downarrow\downarrow}$ which we've been describing in Figure \ref{fig:PsiXRot} will have the following sets of eigneplanes and eigenmatrices: $\{ z_1 \land y_2$; $x_1 \land x_2$; $y_1 \land z_2 \}$ and $\{ \sigma_z \otimes \sigma_y$; $\sigma_x \otimes \sigma_x$; $\sigma_y \otimes \sigma_z \}$.

\subsection{Rules for Double-Pauli Rotations.}

Rules for performing rotations in double-Pauli planes are somewhat more intricate, but most of the times they are dealing with axis inversion. Essentially there are three kinds of dynamics that are obtainable with rotations generated by tensor product of two Pauli matrices: movement from Entanglement to Separability; movement from Separability to Entanglement and movement from Separability to Separability. 

\subsubsection{Entanglement - Separability}

\textbf{Rule ``E-S":} \textit{Take the matrix for rotation and the wedge product corresponding to that rotation. The axes of the wedge product belong to different BS, and are orthogonal to each other. If translated to originate from the same point they are forming an imaginary plane. Second axis in the wedge product inverts and state arrows appear on both BS in direction opposite to the directionality of imaginary plane.}

We call it an imaginary plane because this plane is formed by axes of different BS. Directionality of this plane can be settled by applying right-hand-rule or taking a cross product of the two vectors in the order specified by the wedge product. The vector resulting from the cross product is the directionality of the plane. In case rotation matrix and therefore the resulting wedge product has a minus sign in it, use negative direction of one of the axes in the wedge product and apply the rules as usual.

\begin{figure}
\begin{subfigure}{.4\textwidth}
  \centering
\begin{minipage}[b][2cm][t]{0.4\textwidth}
\centering
\tdplotsetmaincoords{70}{110}
\begin{tikzpicture}[tdplot_main_coords, scale=0.75]

    \shade[ball color=white!50, opacity=0.6] (0,0,0) circle (1cm);
    
    \draw[line width=0.6mm,->, scale=1] (0,0,0) -- (1.2,0,0) node[anchor=north east, font=\small]{$x_1$};
    \draw[thick,->, scale=1] (0,0,0) -- (0,1.2,0) node[anchor=north, font=\small]{$y_1$};
    \draw[thick,->, scale=1] (0,0,0) -- (0,0,1.2) node[anchor=east, font=\small]{$z_1$};
    
    
    \tdplotdrawarc[thin, dashed, scale=1]{(0,0,0)}{1}{-4}{356}{}{}
    
    \shade[ball color=orange, opacity=1, scale=2] (0,0,0) circle (0.05cm);

\end{tikzpicture}
\end{minipage}%
\begin{minipage}[b][1.75cm][t]{0.4\textwidth}
\centering
\tdplotsetmaincoords{70}{110}
\begin{tikzpicture}[tdplot_main_coords, scale=0.75]

    \shade[ball color=white!50, opacity=0.6] (0,0,0) circle (1cm);
    
    \draw[thick,->, scale=1] (0,0,0) -- (-1.2,0,0) node[anchor=south west, font=\small]{$x_2$};
    \draw[line width=0.6mm,->, scale=1] (0,0,0) -- (0,-1.2,0) node[anchor=south, font=\small]{$y_2$};
    \draw[thick,->, scale=1] (0,0,0) -- (0,0,-1.2) node[anchor=west, font=\small]{$z_2$};
    
    
    \tdplotdrawarc[thin, dashed, scale=1]{(0,0,0)}{1}{-4}{356}{}{}
    
    \shade[ball color=orange, opacity=1, scale=2] (0,0,0) circle (0.05cm);
   
\end{tikzpicture}
\end{minipage}

  \caption{$\Psi^-= \ket{\uparrow\downarrow - \downarrow\uparrow} $}
\end{subfigure}
\raisebox{2\height}{\LARGE$\xrightarrow[e^{i\frac{\pi}{4}(\sigma_x \otimes \sigma_y)}]{x_1 \land y_2}$}%
\begin{subfigure}{.4\textwidth}
  \centering
\begin{minipage}[b][2cm][t]{0.4\textwidth}
\centering
\tdplotsetmaincoords{70}{110}
\begin{tikzpicture}[tdplot_main_coords, scale=0.75]

    \shade[ball color=white!50, opacity=0.6] (0,0,0) circle (1cm);
    
    \draw[line width=0.6mm,->, scale=1] (0,0,0) -- (1.2,0,0) node[anchor=north east, font=\small]{$x_1$};
    \draw[thick,->, scale=1] (0,0,0) -- (0,1.2,0) node[anchor=north, font=\small]{$y_1$};
    \draw[thick,->, scale=1] (0,0,0) -- (0,0,1.2) node[anchor=east, font=\small]{$z_1$};
    
    
    \tdplotdrawarc[thin, dashed, scale=1]{(0,0,0)}{1}{-4}{356}{}{}
    
    \coordinate (statevector) at (0, 0, 1);
    
    \draw[orange, line width=1mm, -latex] (0,0,0) -- (statevector);

\end{tikzpicture}
\end{minipage}
\begin{minipage}[b][1.75cm][t]{0.4\textwidth}
\centering
\tdplotsetmaincoords{70}{110}
\begin{tikzpicture}[tdplot_main_coords, scale=0.75]

    \shade[ball color=white!50, opacity=0.6] (0,0,0) circle (1cm);
    
    \draw[line width=0.6mm,->, scale=1] (0,0,0) -- (0,1.2,0) node[anchor=north, font=\small]{$y_2$};
    \draw[thick,->, scale=1] (0,0,0) -- (-1.2,0,0) node[anchor=south west, font=\small]{$x_2$};
    \draw[thick,->, scale=1] (0,0,0) -- (0,0,-1.2) node[anchor=west, font=\small]{$z_2$};
    
    
    \tdplotdrawarc[thin, dashed, scale=1]{(0,0,0)}{1}{-4}{356}{}{}
    
    \coordinate (statevector) at (0, 0, 1);

    \draw[orange, line width=1mm, -latex] (0,0,0) -- (statevector);

\end{tikzpicture}
\end{minipage}

\caption{$\ket{\uparrow\downarrow}$}
\end{subfigure}
\vspace{0.5cm}

\begin{subfigure}{.4\textwidth}
  \centering
\begin{minipage}[b][2cm][t]{0.4\textwidth}
\centering
\tdplotsetmaincoords{70}{110}
\begin{tikzpicture}[tdplot_main_coords, scale=0.75]
    \shade[ball color=white!50, opacity=0.6] (0,0,0) circle (1cm);
    
    \draw[line width=0.6mm,->, scale=1] (0,0,0) -- (-1.2,0,0) ;
    \draw[thick,->, scale=1] (0,0,0) -- (1.2,0,0) node[anchor=north east, font=\small]{$x_1$};
    \draw[thick,->, scale=1] (0,0,0) -- (0,1.2,0) node[anchor=north, font=\small]{$y_1$};
    \draw[thick,->, scale=1] (0,0,0) -- (0,0,1.2) node[anchor=east, font=\small]{$z_1$};
    
    
    \tdplotdrawarc[thin, dashed, scale=1]{(0,0,0)}{1}{-4}{356}{}{}
    
    \shade[ball color=orange, opacity=1, scale=2] (0,0,0) circle (0.05cm);
\end{tikzpicture}
\end{minipage}%
\begin{minipage}[b][2cm][t]{0.4\textwidth}
\centering
\tdplotsetmaincoords{70}{110}
\begin{tikzpicture}[tdplot_main_coords, scale=0.75]
    \shade[ball color=white!50, opacity=0.6] (0,0,0) circle (1cm);
    
    \draw[thick,->, scale=1] (0,0,0) -- (1.2,0,0) node[anchor=north east, font=\small]{$y_2$};
    \draw[thick,->, scale=1] (0,0,0) -- (0,-1.2,0) node[anchor=south, font=\small]{$z_2$};
    \draw[line width=0.6mm,->, scale=1] (0,0,0) -- (0,0,1.2) node[anchor=west, font=\small]{$x_2$};
    
    
    \tdplotdrawarc[thin, dashed, scale=1]{(0,0,0)}{1}{-4}{356}{}{}
    
    \shade[ball color=orange, opacity=1, scale=2] (0,0,0) circle (0.05cm);
\end{tikzpicture}
\end{minipage}

  \caption{$\ket{\uparrow\uparrow + \uparrow\downarrow - i\downarrow\uparrow + i\downarrow\downarrow}$}
\end{subfigure}
\raisebox{2\height}{\LARGE$\xrightarrow[e^{-i\frac{\pi}{4}(\sigma_x \otimes \sigma_x)}]{-x_1 \land x_2}$}%
\begin{subfigure}{.4\textwidth}
  \centering
\begin{minipage}[b][2cm][t]{0.4\textwidth}
\centering
\tdplotsetmaincoords{70}{110}
\begin{tikzpicture}[tdplot_main_coords, scale=0.75]
    \shade[ball color=white!50, opacity=0.6] (0,0,0) circle (1cm);
    
    \draw[line width=0.6mm,->, scale=1] (0,0,0) -- (-1.2,0,0) ;
    \draw[thick,->, scale=1] (0,0,0) -- (1.2,0,0) node[anchor=north east, font=\small]{$x_1$};
    \draw[thick,->, scale=1] (0,0,0) -- (0,1.2,0) node[anchor=north, font=\small]{$y_1$};
    \draw[thick,->, scale=1] (0,0,0) -- (0,0,1.2) node[anchor=east, font=\small]{$z_1$};
    
    
    \tdplotdrawarc[thin, dashed, scale=1]{(0,0,0)}{1}{-4}{356}{}{}
    
    \coordinate (statevector) at (0, -1, 0);
    
    \draw[orange, line width=1mm, -latex] (0,0,0) -- (statevector);
\end{tikzpicture}
\end{minipage}
\begin{minipage}[b][1.75cm][t]{0.4\textwidth}
\centering
\tdplotsetmaincoords{70}{110}
\begin{tikzpicture}[tdplot_main_coords, scale=0.75]
    \shade[ball color=white!50, opacity=0.6] (0,0,0) circle (1cm);
    
    \draw[thick,->, scale=1] (0,0,0) -- (1.2,0,0) node[anchor=north east, font=\small]{$y_2$};
    \draw[thick,->, scale=1] (0,0,0) -- (0,-1.2,0) node[anchor=south, font=\small]{$z_2$};
    \draw[line width=0.6mm,->, scale=1] (0,0,0) -- (0,0,-1.2) node[anchor=west, font=\small]{$x_2$};
    
    
    \tdplotdrawarc[thin, dashed, scale=1]{(0,0,0)}{1}{-4}{356}{}{}
    
    \coordinate (statevector) at (0, -1, 0);
    
    \draw[orange, line width=1mm, -latex] (0,0,0) -- (statevector);
\end{tikzpicture}
\end{minipage}

\caption{$\ket{\uparrow\uparrow - i\downarrow\uparrow} = \ket{\uparrow - i\downarrow} \otimes \ket{\uparrow}$}
\end{subfigure}
\caption{(a) to (b): $e^{i\frac{\pi}{4}(\sigma_x \otimes \sigma_y)} \ket{\uparrow\downarrow - \downarrow\uparrow} = \ket{\uparrow\downarrow}$ Imaginary plane formed by the vectors $x_1$ and $y_2$ through the wedge product $x_1 \land y_2$ is facing down on this picture. Therefore the orange statevectors appear in the opposite direction - up. The second vector in the wedge product ($y_2$) inverts. \\ (c) to (d): $e^{-i\frac{\pi}{4}(\sigma_x \otimes \sigma_x)} \ket{\uparrow\uparrow + \uparrow\downarrow - i\downarrow\uparrow + i\downarrow\downarrow} = \ket{\uparrow\uparrow - i\downarrow\uparrow} = \ket{\uparrow - i\downarrow} \otimes \ket{\uparrow}$ Imaginary plane formed by the vectors $-x_1$ and $x_2$ through the wedge product $-x_1 \land x_2$ is facing in the $y_1$ direction -right, so both statevector arrows will appear in the opposite direction - left. Second vector of the wedge product $-x_1 \land x_2$ inverts.}
\label{fig:Ent-Sep}
\end{figure}
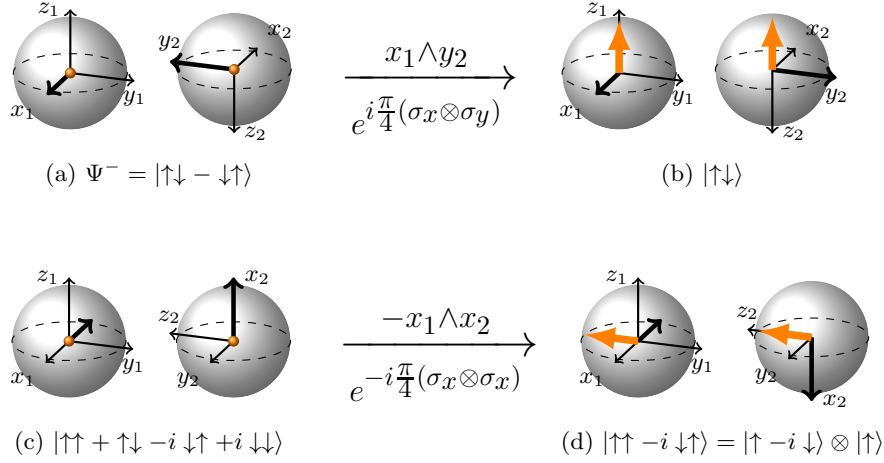

Now let us apply this rule to two rotations illustrated in Figure \ref{fig:Ent-Sep}. First of all (a to b) we will be rotating state $\Psi^-= \ket{\uparrow\downarrow - \downarrow\uparrow}$ in the plane $x_1 \land y_2$ by an angle of $\pi/2$ to obtain state $\ket{\uparrow\downarrow}$. The vectors $x_1$ and $y_2$ forming a wedge product are drawn as thicker arrows. If we imagine that they originate from the same point (say, origin of the first BS) they will be defining a plane that is perpendicular to $z_1$ axis. The directionality of this plane may be determined by taking cross product of vectors $x_1$ and $y_2$ or applying a right-hand rule. If we did so, we would see that the plane is facing down or in the direction opposite to $z_1$ axis. Let's call this direction $-z_1$. According to the rule the statevector arrows should appear in the direction opposite of the imaginary plane direction. Hence in this case the statevector arrows will appear in the $z_1$ direction, in other words they will be facing up (Figure \ref{fig:Ent-Sep} (b)). Also according to the rule the second vector of the wedge product $x_1 \land y_2$, will invert. That is why $y_2$ got inverted moving from (a) to (b). 

As a result we have the state $\ket{\uparrow\downarrow}$ (note that the second statevector is facing in the $-z_2$ direction). If we were to rotate the second BS rigidly with all the coordinates and statevectors around the $y_2$ axis we would obtain a more conventional representation of the same state.

Now let us look at a more intricate rotation in the same Figure \ref{fig:Ent-Sep} (c) to (d). Here we are rotating Maximally entangled state $\ket{\uparrow\uparrow + \uparrow\downarrow - i\downarrow\uparrow + i\downarrow\downarrow}$ to obtain a separable state $\ket{\uparrow - i\downarrow} \otimes \ket{\uparrow}$. Having a minus sign in the exponent of the Pauli Matrix description of rotation we are free to choose the bivector corresponding to it: $-x_1 \land x_2$ or $x_1 \land -x_2$. This choice does not affect the final state. We decided to choose the former. The plane of rotation is now $-x_1 \land x_2$. As in previous case we have used thicker arrows to explicitly show the vectors forming the wedge product: those are vector $-x_1$ on the first BS and vector $x_2$ on the second. If we imagine them originating from the same point (say, origin of the first BS) the wedge product $-x_1 \land x_2$ is forming a plane that is facing in the $y_1$ direction. Let's call this direction Right. According to the rule the statevector arrows will appear in the direction opposite. Therefore in (d) both statevector arrows are facing in the direction Left or the direction opposite to $y_1$. The second vector of the wedge product, in this case $x_2$, inverts. 

Note that if we were to interpret the wedge product as $x_1 \land -x_2$ nothing would change in the analysis because they define the same plane with same directionality and inversion of $-x_2$ is equivalent to inversion of $x_2$. Also one should remember that it is important to pay attention to the axes along which the statevectors are pointing, not their absolute direction to read out the final separable state. In (d) the first statevector is facing in $-y_1$ direction and the second statevector is facing in $z_2$ direction. Hence the final state: $\ket{\uparrow - i\downarrow} \otimes \ket{\uparrow}$

\FloatBarrier

\subsubsection{Separability - Entanglement.} 

\textbf{Rule ``S-E":} \textit{First draw the coordinate axes of the first BS in a usual way. Now draw the statevector arrows so that they are pointing in the same absolute direction. Next, draw the axes of the second BS so that they are right-handed, the axes of the wedge product corresponding to rotation are perpendicular to each other and form imaginary plane whose directionality is opposite to the directionality of the statevector arrows. Now invert the second axis of the wedge product and statevector arrows are pulled into the center of BS.}

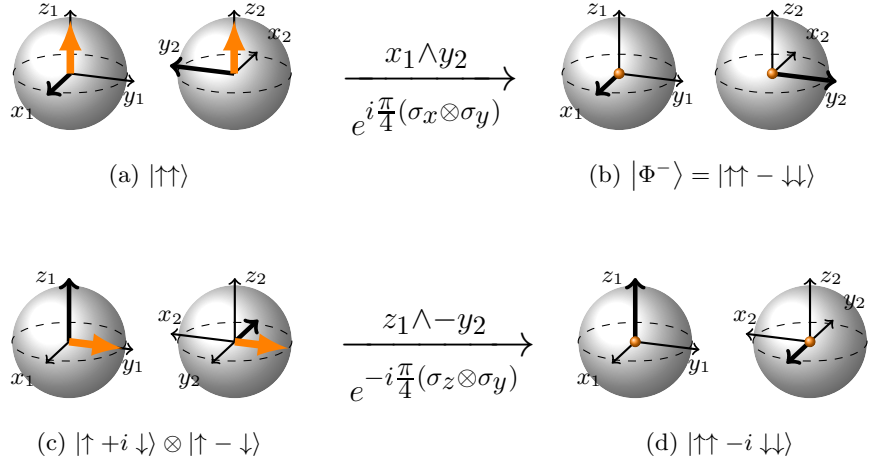
\begin{figure}
\begin{subfigure}{.4\textwidth}
  \centering
\begin{minipage}[b][2cm][t]{0.4\textwidth}
\centering
\tdplotsetmaincoords{70}{110}
\begin{tikzpicture}[tdplot_main_coords, scale=0.75]
    \shade[ball color=white!50, opacity=0.6] (0,0,0) circle (1cm);
    
    \draw[line width=0.6mm,->, scale=1] (0,0,0) -- (1.2,0,0) node[anchor=north east, font=\small]{$x_1$};
    \draw[thick,->, scale=1] (0,0,0) -- (0,1.2,0) node[anchor=north, font=\small]{$y_1$};
    \draw[thick,->, scale=1] (0,0,0) -- (0,0,1.2) node[anchor=east, font=\small]{$z_1$};
    
    
    \tdplotdrawarc[thin, dashed, scale=1]{(0,0,0)}{1}{-4}{356}{}{}

    \coordinate (statevector) at (0, 0, 1);
    
    \draw[orange, line width=1mm, -latex] (0,0,0) -- (statevector);

\end{tikzpicture}
\end{minipage}%
\begin{minipage}[b][2cm][t]{0.4\textwidth}
\centering
\tdplotsetmaincoords{70}{110}
\begin{tikzpicture}[tdplot_main_coords, scale=0.75]
    \shade[ball color=white!50, opacity=0.6] (0,0,0) circle (1cm);
    
    \draw[thick,->, scale=1] (0,0,0) -- (-1.2,0,0) node[anchor=south west, font=\small]{$x_2$};
    \draw[line width=0.6mm,->, scale=1] (0,0,0) -- (0,-1.2,0) node[anchor=south, font=\small]{$y_2$};
    \draw[thick,->, scale=1] (0,0,0) -- (0,0,1.2) node[anchor=west, font=\small]{$z_2$};
    
    
    \tdplotdrawarc[thin, dashed, scale=1]{(0,0,0)}{1}{-4}{356}{}{}

    \coordinate (statevector) at (0, 0, 1);
    
    \draw[orange, line width=1mm, -latex] (0,0,0) -- (statevector);

\end{tikzpicture}
\end{minipage}

  \caption{$\ket{\uparrow\uparrow}$}
\end{subfigure}
\raisebox{2\height}{\LARGE$\xrightarrow[e^{i\frac{\pi}{4}(\sigma_x \otimes \sigma_y)}]{x_1 \land y_2}$}%
\begin{subfigure}{.4\textwidth}
  \centering
\begin{minipage}[b][2cm][t]{0.4\textwidth}
\centering
\tdplotsetmaincoords{70}{110}
\begin{tikzpicture}[tdplot_main_coords, scale=0.75]
    \shade[ball color=white!50, opacity=0.6] (0,0,0) circle (1cm);
    
    \draw[line width=0.6mm,->, scale=1] (0,0,0) -- (1.2,0,0) node[anchor=north east, font=\small]{$x_1$};
    \draw[thick,->, scale=1] (0,0,0) -- (0,1.2,0) node[anchor=north, font=\small]{$y_1$};
    \draw[thick,->, scale=1] (0,0,0) -- (0,0,1.2) node[anchor=east, font=\small]{$z_1$};
    
    
    \tdplotdrawarc[thin, dashed, scale=1]{(0,0,0)}{1}{-4}{356}{}{}
    
    \shade[ball color=orange, opacity=1, scale=2] (0,0,0) circle (0.05cm);
\end{tikzpicture}
\end{minipage}
\begin{minipage}[b][2cm][t]{0.4\textwidth}
\centering
\tdplotsetmaincoords{70}{110}
\begin{tikzpicture}[tdplot_main_coords, scale=0.75]
    \shade[ball color=white!50, opacity=0.6] (0,0,0) circle (1cm);
    
    \draw[line width=0.6mm,->, scale=1] (0,0,0) -- (0,1.2,0) node[anchor=north, font=\small]{$y_2$};
    \draw[thick,->, scale=1] (0,0,0) -- (-1.2,0,0) node[anchor=south west, font=\small]{$x_2$};
    \draw[thick,->, scale=1] (0,0,0) -- (0,0,1.2) node[anchor=west, font=\small]{$z_2$};
    
    
    \tdplotdrawarc[thin, dashed, scale=1]{(0,0,0)}{1}{-4}{356}{}{}
    
    \shade[ball color=orange, opacity=1, scale=2] (0,0,0) circle (0.05cm);
\end{tikzpicture}
\end{minipage}

\caption{$\ket{\Phi^-} = \ket{\uparrow\uparrow - \downarrow\downarrow}$}
\end{subfigure}
\vspace{0.5cm}

\begin{subfigure}{.4\textwidth}
  \centering
\begin{minipage}[b][2cm][t]{0.4\textwidth}
\centering
\tdplotsetmaincoords{70}{110}
\begin{tikzpicture}[tdplot_main_coords, scale=0.75]
    \shade[ball color=white!50, opacity=0.6] (0,0,0) circle (1cm);
    
    \draw[thick,->, scale=1] (0,0,0) -- (1.2,0,0) node[anchor=north east, font=\small]{$x_1$};
    \draw[thick,->, scale=1] (0,0,0) -- (0,1.2,0) node[anchor=north, font=\small]{$y_1$};
    \draw[line width=0.6mm,->, scale=1] (0,0,0) -- (0,0,1.2) node[anchor=east, font=\small]{$z_1$};
    
    
    \tdplotdrawarc[thin, dashed, scale=1]{(0,0,0)}{1}{-4}{356}{}{}

    \coordinate (statevector) at (0, 1, 0);
    
    \draw[orange, line width=1mm, -latex] (0,0,0) -- (statevector);
\end{tikzpicture}
\end{minipage}%
\begin{minipage}[b][2cm][t]{0.4\textwidth}
\centering
\tdplotsetmaincoords{70}{110}
\begin{tikzpicture}[tdplot_main_coords, scale=0.75]
    \shade[ball color=white!50, opacity=0.6] (0,0,0) circle (1cm);
    
    \draw[line width=0.6mm,->, scale=1] (0,0,0) -- (-1.2,0,0);
    \draw[thick,->, scale=1] (0,0,0) -- (1.2,0,0) node[anchor=north east, font=\small]{$y_2$};
    \draw[thick,->, scale=1] (0,0,0) -- (0,-1.2,0) node[anchor=south, font=\small]{$x_2$};
    \draw[thick,->, scale=1] (0,0,0) -- (0,0,1.2) node[anchor=west, font=\small]{$z_2$};
    
    
    \tdplotdrawarc[thin, dashed, scale=1]{(0,0,0)}{1}{-4}{356}{}{}

    \coordinate (statevector) at (0, 1, 0);
    
    \draw[orange, line width=1mm, -latex] (0,0,0) -- (statevector);
\end{tikzpicture}
\end{minipage}

  \caption{$\ket{\uparrow + i\downarrow} \otimes \ket{\uparrow - \downarrow}$}
\end{subfigure}
\raisebox{2\height}{\LARGE$\xrightarrow[e^{-i\frac{\pi}{4}(\sigma_z \otimes \sigma_y)}]{z_1 \land -y_2}$}%
\begin{subfigure}{.4\textwidth}
  \centering
\begin{minipage}[b][2cm][t]{0.4\textwidth}
\centering
\tdplotsetmaincoords{70}{110}
\begin{tikzpicture}[tdplot_main_coords, scale=0.75]
    \shade[ball color=white!50, opacity=0.6] (0,0,0) circle (1cm);
    
    \draw[thick,->, scale=1] (0,0,0) -- (1.2,0,0) node[anchor=north east, font=\small]{$x_1$};
    \draw[thick,->, scale=1] (0,0,0) -- (0,1.2,0) node[anchor=north, font=\small]{$y_1$};
    \draw[line width=0.6mm,->, scale=1] (0,0,0) -- (0,0,1.2) node[anchor=east, font=\small]{$z_1$};
    
    
    \tdplotdrawarc[thin, dashed, scale=1]{(0,0,0)}{1}{-4}{356}{}{}
    
    \shade[ball color=orange, opacity=1, scale=2] (0,0,0) circle (0.05cm);
\end{tikzpicture}
\end{minipage}
\begin{minipage}[b][2cm][t]{0.4\textwidth}
\centering
\tdplotsetmaincoords{70}{110}
\begin{tikzpicture}[tdplot_main_coords, scale=0.75]
    \shade[ball color=white!50, opacity=0.6] (0,0,0) circle (1cm);
    
    \draw[line width=0.6mm,->, scale=1] (0,0,0) -- (1.2,0,0);
    \draw[thick,->, scale=1] (0,0,0) -- (-1.2,0,0) node[anchor=south west, font=\small]{$y_2$};
    \draw[thick,->, scale=1] (0,0,0) -- (0,-1.2,0) node[anchor=south, font=\small]{$x_2$};
    \draw[thick,->, scale=1] (0,0,0) -- (0,0,1.2) node[anchor=west, font=\small]{$z_2$};
    
    
    \tdplotdrawarc[thin, dashed, scale=1]{(0,0,0)}{1}{-4}{356}{}{}
    
    \shade[ball color=orange, opacity=1, scale=2] (0,0,0) circle (0.05cm);
\end{tikzpicture}
\end{minipage}

\caption{$\ket{\uparrow\uparrow - i\downarrow\downarrow}$}
\end{subfigure}
\caption{(a) to (b): $e^{i\frac{\pi}{4}(\sigma_x \otimes \sigma_y)} \ket{\uparrow\downarrow} = \ket{\uparrow\uparrow - \downarrow\downarrow}$ The coordinates are drawn in such a way that both orange statevectors are facing the same direction and imaginary plane formed by the wedge product mutually orthogonal vectors $x_1$ and $y_2$ is facing the opposite direction. The second vector of the wedge product ($y_2$) inverts and statevectors are pulled into the center of the BS. \\ (c) to (d): $e^{-i\frac{\pi}{4}(\sigma_z \otimes \sigma_y)} \ket{\uparrow + i\downarrow} \otimes \ket{\uparrow - \downarrow} = e^{-i\frac{\pi}{4}(\sigma_z \otimes \sigma_y)} \ket{\uparrow\uparrow -\uparrow\downarrow + i\downarrow\uparrow - i \downarrow\downarrow} = \ket{\uparrow\uparrow - i\downarrow\downarrow}$ This time we have chosen the second vector of the wedge product to bear the minus sign. Plane formed by the vectors of the wedge product is directed in the opposite direction from the orange statevector arrows; $z_1$ and $-y_2$ are orthogonal. $-y_2$ inverts, arrows are again pulled into the center.}
\label{fig:Sep-Ent}
\end{figure}

Let us now turn to Figure \ref{fig:Sep-Ent} for examples. In (a) to (b) we will be rotating the state $\ket{\uparrow\uparrow}$ in the plane $x_1 \land y_2$ to obtain the state $\ket{\Phi^-}$. Following the rule we draw the two BS with statevectors. Note how the $x_2$ and $y_2$ axes are rotated with respect to $x_1$ and $y_1$. This is done to comply with the \textit{...the axes of the wedge product corresponding to rotation are perpendicular to each other and form imaginary plane whose directionality is opposite to the directionality of the statevector arrows} part of the rule. The vectors forming the wedge product $x_1$ and $y_2$ are drawn as thicker arrows. If we were to translate them so that they originate from the same point, say the origin of the first BS, they would form an (imaginary) plane. It is facing in the direction $-z_1$ or Down, which is opposite to the direction statevector arrows are pointing. We have complied with all the requirements of the rule. Now we just pull the statevectors into the center of BS drawing a dot instead and invert the second vector of the wedge product, $y_2$. We have obtained a representation for the state $\ket{\uparrow\uparrow - \downarrow\downarrow}$.

Now let us turn to a more intricate (c) - (d) case in Figure \ref{fig:Sep-Ent}. Here we start with the state $\ket{\uparrow + i\downarrow} \otimes \ket{\uparrow - \downarrow}$. Using this information according to the Rule ``S-E" we can draw the coordinates of the first BS and both statevectors. The first statevector corresponds to the single-qubit state $\ket{\uparrow + i\downarrow}$ meaning it is facing the $y_1$ direction. We draw the first statevector in the appropriate direction and immediately draw the second statevector arrow in the empty second BS in such a way that it faces the same absolute direction. Now we can draw one coordinate axis of the second BS. We know that the second qubit is in the state $\ket{\uparrow - \downarrow}$ meaning it is facing in the $-x_2$ direction. Hence we draw $x_2$ axis as pointing opposite to statevector direction. 

To draw the $z_2$ and $y_2$ axes we should look at the rotation bivector. In this case it is $z_1 \land -y_2$.  Therefore we have to draw $-y_2$ axis in such a way that were it to originate from the same origin with $z_1$ the plane formed by the wedge product $z_1 \land -y_2$ would face in the direction opposite to the statevector arrows. This could be done in only one way: $-y_2$ (thicker arrow) should face into the page and therefore $y_2$ should face out of the page. In other words $y_2$ should face in the same absolute direction as $x_1$. Having $x_2$ and $y_2$ axes in place there is only one choice for the $z_2$ axis to keep the coordinates of the second BS right-handed.

Now, the most difficult part past us, we pull the statevectors inside the Bloch Spheres and invert the second axis of the wedge product, $-y_2$. That of course implies that $y_2$ is also inverted. If we check appendices where all 24 stabilizer maximally entangled states are depicted and find the one labelled (g), we will see that this state corresponds to $\ket{\uparrow\uparrow - i\downarrow\downarrow}$.

\FloatBarrier

\subsubsection{Separability - Separability.} 

\textbf{Rule ``S-S":} \textit{When the rotation matrix is a tensor product of two Pauli matrices, but the state is such that one of those Pauli matrices is an eigenmatrix, double-Pauli rotations act like local rotation. We can substitute the eigenmatrix by I multiplied by an eigenvalue.}

For example $(\sigma_x \otimes \sigma_z ) \ket{\uparrow \downarrow} = -(\sigma_x \otimes I ) \ket{\uparrow \downarrow}$ Thus the eigenvalue ($\pm 1$) determines the direction of rotation of the other qubit. We could also formulate this rule in terms of axes of the wedge product. \textit{If one axis of the rotation wedge product aligns with the state vector arrow, only rotate the other qubit around the remaining axis of the wedge product. Reverse the rotation direction if the state vector anti-aligns with the former axis of the wedge product.}

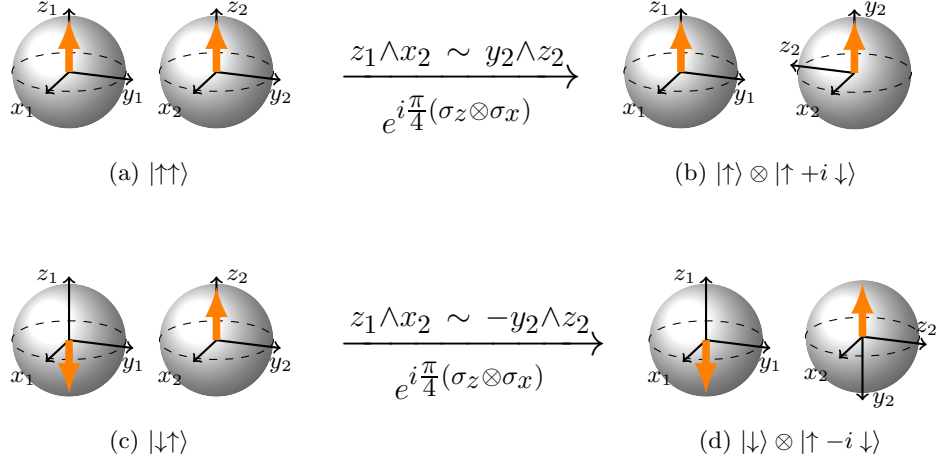
\begin{figure}
\begin{subfigure}{.4\textwidth}
  \centering
\begin{minipage}[b][2cm][t]{0.4\textwidth}
\centering
\tdplotsetmaincoords{70}{110}
\begin{tikzpicture}[tdplot_main_coords, scale=0.75]
    \shade[ball color=white!50, opacity=0.6] (0,0,0) circle (1cm);
    
    \draw[thick,->, scale=1] (0,0,0) -- (1.2,0,0) node[anchor=north east, font=\small]{$x_1$};
    \draw[thick,->, scale=1] (0,0,0) -- (0,1.2,0) node[anchor=north, font=\small]{$y_1$};
    \draw[thick,->, scale=1] (0,0,0) -- (0,0,1.2) node[anchor=east, font=\small]{$z_1$};
    
    
    \tdplotdrawarc[thin, dashed, scale=1]{(0,0,0)}{1}{-4}{356}{}{}

    \coordinate (statevector) at (0, 0, 1);
    
    \draw[orange, line width=1mm, -latex] (0,0,0) -- (statevector);

\end{tikzpicture}
\end{minipage}%
\begin{minipage}[b][2cm][t]{0.4\textwidth}
\centering
\tdplotsetmaincoords{70}{110}
\begin{tikzpicture}[tdplot_main_coords, scale=0.75]
    \shade[ball color=white!50, opacity=0.6] (0,0,0) circle (1cm);
    
    \draw[thick,->, scale=1] (0,0,0) -- (1.2,0,0) node[anchor=north east, font=\small]{$x_2$};
    \draw[thick,->, scale=1] (0,0,0) -- (0,1.2,0) node[anchor=north, font=\small]{$y_2$};
    \draw[thick,->, scale=1] (0,0,0) -- (0,0,1.2) node[anchor=west, font=\small]{$z_2$};
    
    
    \tdplotdrawarc[thin, dashed, scale=1]{(0,0,0)}{1}{-4}{356}{}{}

    \coordinate (statevector) at (0, 0, 1);
    
    \draw[orange, line width=1mm, -latex] (0,0,0) -- (statevector);

\end{tikzpicture}
\end{minipage}

  \caption{$\ket{\uparrow\uparrow}$}
\end{subfigure}
\raisebox{2\height}{\LARGE$\xrightarrow[e^{i\frac{\pi}{4}(\sigma_z \otimes \sigma_x)}]{z_1 \land x_2 \ \sim \ y_2 \land z_2}$}%
\begin{subfigure}{.4\textwidth}
  \centering
\begin{minipage}[b][2cm][t]{0.4\textwidth}
\centering
\tdplotsetmaincoords{70}{110}
\begin{tikzpicture}[tdplot_main_coords, scale=0.75]
    \shade[ball color=white!50, opacity=0.6] (0,0,0) circle (1cm);
    
    \draw[thick,->, scale=1] (0,0,0) -- (1.2,0,0) node[anchor=north east, font=\small]{$x_1$};
    \draw[thick,->, scale=1] (0,0,0) -- (0,1.2,0) node[anchor=north, font=\small]{$y_1$};
    \draw[thick,->, scale=1] (0,0,0) -- (0,0,1.2) node[anchor=east, font=\small]{$z_1$};
    
    
    \tdplotdrawarc[thin, dashed, scale=1]{(0,0,0)}{1}{-4}{356}{}{}
  
    \coordinate (statevector) at (0, 0, 1);
    
    \draw[orange, line width=1mm, -latex] (0,0,0) -- (statevector);
\end{tikzpicture}
\end{minipage}
\begin{minipage}[b][2cm][t]{0.4\textwidth}
\centering
\tdplotsetmaincoords{70}{110}
\begin{tikzpicture}[tdplot_main_coords, scale=0.75]
    \shade[ball color=white!50, opacity=0.6] (0,0,0) circle (1cm);
    
    \draw[thick,->, scale=1] (0,0,0) -- (0,-1.2,0) node[anchor=south, font=\small]{$z_2$};
    \draw[thick,->, scale=1] (0,0,0) -- (1.2,0,0) node[anchor=north east, font=\small]{$x_2$};
    \draw[thick,->, scale=1] (0,0,0) -- (0,0,1.2) node[anchor=west, font=\small]{$y_2$};
    
    
    \tdplotdrawarc[thin, dashed, scale=1]{(0,0,0)}{1}{-4}{356}{}{}
    
    \coordinate (statevector) at (0, 0, 1);
    
    \draw[orange, line width=1mm, -latex] (0,0,0) -- (statevector);
\end{tikzpicture}
\end{minipage}

\caption{$\ket{\uparrow} \otimes \ket{\uparrow + i\downarrow}$}
\end{subfigure}
\vspace{0.5cm}

\begin{subfigure}{.4\textwidth}
  \centering
\begin{minipage}[b][2cm][t]{0.4\textwidth}
\centering
\tdplotsetmaincoords{70}{110}
\begin{tikzpicture}[tdplot_main_coords, scale=0.75]
    \shade[ball color=white!50, opacity=0.6] (0,0,0) circle (1cm);
    
    \draw[thick,->, scale=1] (0,0,0) -- (1.2,0,0) node[anchor=north east, font=\small]{$x_1$};
    \draw[thick,->, scale=1] (0,0,0) -- (0,1.2,0) node[anchor=north, font=\small]{$y_1$};
    \draw[thick,->, scale=1] (0,0,0) -- (0,0,1.2) node[anchor=east, font=\small]{$z_1$};
    
    
    \tdplotdrawarc[thin, dashed, scale=1]{(0,0,0)}{1}{-4}{356}{}{}

    \coordinate (statevector) at (0, 0, -1);
    
    \draw[orange, line width=1mm, -latex] (0,0,0) -- (statevector);

\end{tikzpicture}
\end{minipage}%
\begin{minipage}[b][2cm][t]{0.4\textwidth}
\centering
\tdplotsetmaincoords{70}{110}
\begin{tikzpicture}[tdplot_main_coords, scale=0.75]
    \shade[ball color=white!50, opacity=0.6] (0,0,0) circle (1cm);
    
    \draw[thick,->, scale=1] (0,0,0) -- (1.2,0,0) node[anchor=north east, font=\small]{$x_2$};
    \draw[thick,->, scale=1] (0,0,0) -- (0,1.2,0) node[anchor=north, font=\small]{$y_2$};
    \draw[thick,->, scale=1] (0,0,0) -- (0,0,1.2) node[anchor=west, font=\small]{$z_2$};
    
    
    \tdplotdrawarc[thin, dashed, scale=1]{(0,0,0)}{1}{-4}{356}{}{}

    \coordinate (statevector) at (0, 0, 1);
    
    \draw[orange, line width=1mm, -latex] (0,0,0) -- (statevector);

\end{tikzpicture}
\end{minipage}

  \caption{$\ket{\downarrow\uparrow}$}
\end{subfigure}
\raisebox{2\height}{\LARGE$\xrightarrow[e^{i\frac{\pi}{4}(\sigma_z \otimes \sigma_x)}]{z_1 \land x_2 \ \sim \ -y_2 \land z_2}$}%
\begin{subfigure}{.4\textwidth}
  \centering
\begin{minipage}[b][2cm][t]{0.4\textwidth}
\centering
\tdplotsetmaincoords{70}{110}
\begin{tikzpicture}[tdplot_main_coords, scale=0.75]
    \shade[ball color=white!50, opacity=0.6] (0,0,0) circle (1cm);
    
    \draw[thick,->, scale=1] (0,0,0) -- (1.2,0,0) node[anchor=north east, font=\small]{$x_1$};
    \draw[thick,->, scale=1] (0,0,0) -- (0,1.2,0) node[anchor=north, font=\small]{$y_1$};
    \draw[thick,->, scale=1] (0,0,0) -- (0,0,1.2) node[anchor=east, font=\small]{$z_1$};
    
    
    \tdplotdrawarc[thin, dashed, scale=1]{(0,0,0)}{1}{-4}{356}{}{}
  
    \coordinate (statevector) at (0, 0, -1);
    
    \draw[orange, line width=1mm, -latex] (0,0,0) -- (statevector);
\end{tikzpicture}
\end{minipage}
\begin{minipage}[b][1.75cm][t]{0.4\textwidth}
\centering
\tdplotsetmaincoords{70}{110}
\begin{tikzpicture}[tdplot_main_coords, scale=0.75]
    \shade[ball color=white!50, opacity=0.6] (0,0,0) circle (1cm);
    
    \draw[thick,->, scale=1] (0,0,0) -- (0,1.2,0) node[anchor=south, font=\small]{$z_2$};
    \draw[thick,->, scale=1] (0,0,0) -- (1.2,0,0) node[anchor=north east, font=\small]{$x_2$};
    \draw[thick,->, scale=1] (0,0,0) -- (0,0,-1.2) node[anchor=west, font=\small]{$y_2$};
    
    
    \tdplotdrawarc[thin, dashed, scale=1]{(0,0,0)}{1}{-4}{356}{}{}
    
    \coordinate (statevector) at (0, 0, 1);
    
    \draw[orange, line width=1mm, -latex] (0,0,0) -- (statevector);
\end{tikzpicture}
\end{minipage}

\caption{$\ket{\downarrow} \otimes \ket{\uparrow - i\downarrow}$}
\end{subfigure}
\caption{(a) to (b): $e^{i\frac{\pi}{4}(\sigma_z \otimes \sigma_x)} \ket{\uparrow\uparrow}$ becomes equivalent to $e^{i\frac{\pi}{4}(I \otimes \sigma_x)} \ket{\uparrow\uparrow}$ and therefore to a rotation of the plane $y_2 \land z_2$. \\ (c) to (d): $e^{i\frac{\pi}{4}(\sigma_z \otimes \sigma_x)} \ket{\downarrow\uparrow}$ becomes equivalent to $e^{-i\frac{\pi}{4}(I \otimes \sigma_x)} \ket{\downarrow\uparrow}$ and therefore to a rotation of the plane $- y_2 \land z_2$. The direction of rotation has reversed. We do not stick to the convention of representation for more vivid difference between two initial state representations.}
\label{fig:Sep-Sep}
\end{figure}

Let us take a look at the Figure \ref{fig:Sep-Sep}. (a) to (b): We are trying to rotate the state $\ket{\uparrow\uparrow}$ in the plane defined by a wedge product $z_1 \land x_2$. The statevector of the first BS is aligned with $z_1$, in other words $\sigma_z$ from the tensor $\sigma_z \otimes \sigma_x$ is an eigenrotation for the state $\ket{\uparrow}$ with an eigenvalue $+1$. Therefore we are rotating the second qubit around the remaining axis of the wedge product, $x_2$. That is equivalent to a rotation using the wedge product $y_2 \land z_2$. As a result we obtain the state $\ket{\uparrow} \otimes \ket{\uparrow + i\downarrow}$.

In case the initial state is $\ket{\downarrow\uparrow}$ then rotation using the same Paulis and the same wedge product $z_1 \land x_2$ ((c) to (d)) results in the reversal of direction of rotation for the second qubit. Let us see how it happens. The statevector of the first BS is anti-aligned with the first vector in the wedge product. In other words the eigenvalue of the $\sigma_z$ acting on state $\ket{\downarrow}$ is $-1$. Therefore the rotation around the relaining axis of the wedge product will be reversed. That is equivalent to a rotation of the second BS via the wedge product $-y_2 \land z_2$ resulting in the state $\ket{\uparrow} \otimes \ket{\uparrow - i\downarrow}$.

\subsection{Summary}

The aforementioned rules combined with rules for local rotation and restricted to $\pi/2$ angles of rotation allow to navigate the whole stabilizer set of two-qubit states starting from a single state of the set. We have seen how double-pauli rotations can entangle, disentangle and be single-qubit rotators. In the latter case the direction of rotation of the qubit is conditioned on the state of the qubit that doesn't get rotated. Such behaviour reminds of the action of CNOT gate. In the next chapter we will see explicitly how such dynamics combined with local rotations produces CNOT gate.

\FloatBarrier

\section{Representation of CNOT Gate}

The CNOT gate holds significant importance in quantum computation due to its entangling nature. Understanding its role is essential for quantum computing practitioners, especially in quantum algorithms. This chapter delves into visualizing the CNOT gate's operations within our established framework. Observing the CNOT gate in action offers valuable insights, explicitly revealing why it entangles specific states and leaves others unaltered. The dynamics of this entanglement systematically unfold from the visual rules discussed in the preceding chapter.

We will show how CNOT gate acts on three different input states. First where the control qubit is such that the target qubit doesn't flip $CNOT \ket{\uparrow\uparrow} = \ket{\uparrow\uparrow}$; second when the control qubit is such that the target qubit gets flipped $CNOT \ket{\downarrow\uparrow} = \ket{\downarrow\downarrow}$; third - a superposition of the two previous cases which results in entanglement $CNOT (\ket{\uparrow\uparrow} + \ket{\downarrow\uparrow}) = \ket{\uparrow\uparrow} + \ket{\downarrow\downarrow}$. Being Unitary, CNOT may be expressed in terms of a combination of rotations of some of the 15 rotation planes mentioned earlier. 

\begin{equation}
    CNOT = e^{-i\tfrac{\pi}{4}} e^{i\tfrac{\pi}{4} \sigma_y \otimes I}  e^{i\tfrac{\pi}{4} \sigma_x \otimes I}  e^{i\tfrac{\pi}{4} I \otimes \sigma_x}  e^{-i\tfrac{\pi}{4} \sigma_x \otimes  \sigma_x}  e^{-i\tfrac{\pi}{4} \sigma_y \otimes I}  
\end{equation}

\begin{figure}
\begin{subfigure}{.4\textwidth}
  \centering
\begin{minipage}[b][2cm][t]{0.4\textwidth}
\centering
\tdplotsetmaincoords{70}{110}
\begin{tikzpicture}[tdplot_main_coords, scale=0.75]
    \shade[ball color=white!50, opacity=0.6] (0,0,0) circle (1cm);
    
    \draw[thick,->, scale=1] (0,0,0) -- (1.2,0,0) node[anchor=north east, font=\small]{$x_1$};
    \draw[thick,->, scale=1] (0,0,0) -- (0,1.2,0) node[anchor=north, font=\small]{$y_1$};
    \draw[thick,->, scale=1] (0,0,0) -- (0,0,1.2) node[anchor=east, font=\small]{$z_1$};
    
    
    \tdplotdrawarc[thin, dashed, scale=1]{(0,0,0)}{1}{-4}{356}{}{}

    \coordinate (statevector) at (0, 0, 1);
    
    \draw[orange, line width=1mm, -latex] (0,0,0) -- (statevector);

\end{tikzpicture}
\end{minipage}%
\begin{minipage}[b][2cm][t]{0.4\textwidth}
\centering
\tdplotsetmaincoords{70}{110}
\begin{tikzpicture}[tdplot_main_coords, scale=0.75]
    \shade[ball color=white!50, opacity=0.6] (0,0,0) circle (1cm);
    
    \draw[thick,->, scale=1] (0,0,0) -- (1.2,0,0) node[anchor=north east, font=\small]{$x_2$};
    \draw[thick,->, scale=1] (0,0,0) -- (0,1.2,0) node[anchor=north, font=\small]{$y_2$};
    \draw[thick,->, scale=1] (0,0,0) -- (0,0,1.2) node[anchor=west, font=\small]{$z_2$};
    
    
    \tdplotdrawarc[thin, dashed, scale=1]{(0,0,0)}{1}{-4}{356}{}{}

    \coordinate (statevector) at (0, 0, 1);
    
    \draw[orange, line width=1mm, -latex] (0,0,0) -- (statevector);

\end{tikzpicture}
\end{minipage}

  \caption{$\ket{\uparrow\uparrow}$}
\end{subfigure}
\raisebox{2\height}{\LARGE$\xrightarrow[e^{-i\frac{\pi}{4}(\sigma_y \otimes I)}]{-z_1 \land x_1}$}%
\begin{subfigure}{.4\textwidth}
  \centering
\begin{minipage}[b][2cm][t]{0.4\textwidth}
\centering
\tdplotsetmaincoords{70}{110}
\begin{tikzpicture}[tdplot_main_coords, scale=0.75]
    \shade[ball color=white!50, opacity=0.6] (0,0,0) circle (1cm);
    
    \draw[thick,->, scale=1] (0,0,0) -- (1.2,0,0) node[anchor=north east, font=\small]{$x_1$};
    \draw[thick,->, scale=1] (0,0,0) -- (0,1.2,0) node[anchor=north, font=\small]{$y_1$};
    \draw[thick,->, scale=1] (0,0,0) -- (0,0,1.2) node[anchor=east, font=\small]{$z_1$};
    
    
    \tdplotdrawarc[thin, dashed, scale=1]{(0,0,0)}{1}{-4}{356}{}{}

    \coordinate (statevector) at (1, 0, 0);
    
    \draw[orange, line width=1mm, -latex] (0,0,0) -- (statevector);

\end{tikzpicture}
\end{minipage}%
\begin{minipage}[b][1.75cm][t]{0.4\textwidth}
\centering
\tdplotsetmaincoords{70}{110}
\begin{tikzpicture}[tdplot_main_coords, scale=0.75]
    \shade[ball color=white!50, opacity=0.6] (0,0,0) circle (1cm);
    
    \draw[thick,->, scale=1] (0,0,0) -- (1.2,0,0) node[anchor=north east, font=\small]{$z_2$};
    \draw[thick,->, scale=1] (0,0,0) -- (0,1.2,0) node[anchor=north, font=\small]{$y_2$};
    \draw[thick,->, scale=1] (0,0,0) -- (0,0,-1.2) node[anchor=west, font=\small]{$x_2$};
    
    
    \tdplotdrawarc[thin, dashed, scale=1]{(0,0,0)}{1}{-4}{356}{}{}

    \coordinate (statevector) at (1, 0, 0);
    
    \draw[orange, line width=1mm, -latex] (0,0,0) -- (statevector);

\end{tikzpicture}
\end{minipage}

  \caption{$\ket{\uparrow + \downarrow} \otimes \ket{\uparrow}$}
\end{subfigure}
\vspace{0.5cm}
\begin{subfigure}{\textwidth}
\centering
\begin{turn}{0}
\LARGE$\xleftarrow[e^{-i\frac{\pi}{4}(\sigma_x \otimes \sigma_x )} \ \sim \ e^{-i\frac{\pi}{4}(I \otimes \sigma_x )}]{-x_1 \land x_2\ \sim \ -y_2 \land z_2}$%
\end{turn}
\end{subfigure}

\begin{subfigure}{.4\textwidth}
  \centering
\begin{minipage}[b][2cm][t]{0.4\textwidth}
\centering
\tdplotsetmaincoords{70}{110}
\begin{tikzpicture}[tdplot_main_coords, scale=0.75]
    \shade[ball color=white!50, opacity=0.6] (0,0,0) circle (1cm);
    
    \draw[thick,->, scale=1] (0,0,0) -- (1.2,0,0) node[anchor=north east, font=\small]{$x_1$};
    \draw[thick,->, scale=1] (0,0,0) -- (0,1.2,0) node[anchor=north, font=\small]{$y_1$};
    \draw[thick,->, scale=1] (0,0,0) -- (0,0,1.2) node[anchor=east, font=\small]{$z_1$};
    
    
    \tdplotdrawarc[thin, dashed, scale=1]{(0,0,0)}{1}{-4}{356}{}{}

    \coordinate (statevector) at (1, 0, 0);
    
    \draw[orange, line width=1mm, -latex] (0,0,0) -- (statevector);
\end{tikzpicture}
\end{minipage}
\begin{minipage}[b][1.75cm][t]{0.4\textwidth}
\centering
\tdplotsetmaincoords{70}{110}
\begin{tikzpicture}[tdplot_main_coords, scale=0.75]
    \shade[ball color=white!50, opacity=0.6] (0,0,0) circle (1cm);
    
    \draw[thick,->, scale=1] (0,0,0) -- (-1.2,0,0) node[anchor=south west, font=\small]{$y_2$};
    \draw[thick,->, scale=1] (0,0,0) -- (0,1.2,0) node[anchor=north, font=\small]{$z_2$};
    \draw[thick,->, scale=1] (0,0,0) -- (0,0,-1.2) node[anchor=west, font=\small]{$x_2$};
    
    
    \tdplotdrawarc[thin, dashed, scale=1]{(0,0,0)}{1}{-4}{356}{}{}

    \coordinate (statevector) at (1, 0, 0);
    
    \draw[orange, line width=1mm, -latex] (0,0,0) -- (statevector);
    
\end{tikzpicture}
\end{minipage}

\caption{$\ket{\uparrow + \downarrow} \otimes \ket{\uparrow - i\downarrow}$}
\end{subfigure}
\raisebox{2\height}{\LARGE$\xrightarrow[e^{i\frac{\pi}{4}(I \otimes \sigma_x)}]{y_2 \land z_2}$}%
\begin{subfigure}{.4\textwidth}
  \centering
\begin{minipage}[b][2cm][t]{0.4\textwidth}
\centering
\tdplotsetmaincoords{70}{110}
\begin{tikzpicture}[tdplot_main_coords, scale=0.75]
    \shade[ball color=white!50, opacity=0.6] (0,0,0) circle (1cm);
    
    \draw[thick,->, scale=1] (0,0,0) -- (1.2,0,0) node[anchor=north east, font=\small]{$x_1$};
    \draw[thick,->, scale=1] (0,0,0) -- (0,1.2,0) node[anchor=north, font=\small]{$y_1$};
    \draw[thick,->, scale=1] (0,0,0) -- (0,0,1.2) node[anchor=east, font=\small]{$z_1$};
    
    
    \tdplotdrawarc[thin, dashed, scale=1]{(0,0,0)}{1}{-4}{356}{}{}

    \coordinate (statevector) at (1, 0, 0);
    
    \draw[orange, line width=1mm, -latex] (0,0,0) -- (statevector);

\end{tikzpicture}
\end{minipage}%
\begin{minipage}[b][1.75cm][t]{0.4\textwidth}
\centering
\tdplotsetmaincoords{70}{110}
\begin{tikzpicture}[tdplot_main_coords, scale=0.75]
    \shade[ball color=white!50, opacity=0.6] (0,0,0) circle (1cm);
    
    \draw[thick,->, scale=1] (0,0,0) -- (1.2,0,0) node[anchor=north east, font=\small]{$z_2$};
    \draw[thick,->, scale=1] (0,0,0) -- (0,1.2,0) node[anchor=north, font=\small]{$y_2$};
    \draw[thick,->, scale=1] (0,0,0) -- (0,0,-1.2) node[anchor=west, font=\small]{$x_2$};
    
    
    \tdplotdrawarc[thin, dashed, scale=1]{(0,0,0)}{1}{-4}{356}{}{}

    \coordinate (statevector) at (1, 0, 0);
    
    \draw[orange, line width=1mm, -latex] (0,0,0) -- (statevector);

\end{tikzpicture}
\end{minipage}

  \caption{$\ket{\uparrow + \downarrow} \otimes \ket{\uparrow}$}
\end{subfigure}

\begin{subfigure}{\textwidth}
\centering
\begin{turn}{0}
\LARGE$\xleftarrow[e^{i\frac{\pi}{4}(\sigma_x \otimes I)}]{y_1 \land z_1}$%
\end{turn}
\end{subfigure}

\begin{subfigure}{.4\textwidth}
  \centering
\begin{minipage}[b][2cm][t]{0.4\textwidth}
\centering
\tdplotsetmaincoords{70}{110}
\begin{tikzpicture}[tdplot_main_coords, scale=0.75]
    \shade[ball color=white!50, opacity=0.6] (0,0,0) circle (1cm);
    
    \draw[thick,->, scale=1] (0,0,0) -- (1.2,0,0) node[anchor=north east, font=\small]{$x_1$};
    \draw[thick,->, scale=1] (0,0,0) -- (0,1.2,0) node[anchor=north, font=\small]{$y_1$};
    \draw[thick,->, scale=1] (0,0,0) -- (0,0,1.2) node[anchor=east, font=\small]{$z_1$};
    
    
    \tdplotdrawarc[thin, dashed, scale=1]{(0,0,0)}{1}{-4}{356}{}{}

    \coordinate (statevector) at (1, 0, 0);
    
    \draw[orange, line width=1mm, -latex] (0,0,0) -- (statevector);

\end{tikzpicture}
\end{minipage}%
\begin{minipage}[b][1.75cm][t]{0.4\textwidth}
\centering
\tdplotsetmaincoords{70}{110}
\begin{tikzpicture}[tdplot_main_coords, scale=0.75]
    \shade[ball color=white!50, opacity=0.6] (0,0,0) circle (1cm);
    
    \draw[thick,->, scale=1] (0,0,0) -- (1.2,0,0) node[anchor=north east, font=\small]{$z_2$};
    \draw[thick,->, scale=1] (0,0,0) -- (0,1.2,0) node[anchor=north, font=\small]{$y_2$};
    \draw[thick,->, scale=1] (0,0,0) -- (0,0,-1.2) node[anchor=west, font=\small]{$x_2$};
    
    
    \tdplotdrawarc[thin, dashed, scale=1]{(0,0,0)}{1}{-4}{356}{}{}

    \coordinate (statevector) at (1, 0, 0);
    
    \draw[orange, line width=1mm, -latex] (0,0,0) -- (statevector);

\end{tikzpicture}
\end{minipage}

  \caption{$\ket{\uparrow + \downarrow} \otimes \ket{\uparrow}$}
\end{subfigure}
\raisebox{2\height}{\LARGE$\xrightarrow[e^{i\frac{\pi}{4}(\sigma_y \otimes I)}]{z_1 \land x_1}$}%
\begin{subfigure}{.4\textwidth}
  \centering
\begin{minipage}[b][2cm][t]{0.4\textwidth}
\centering
\tdplotsetmaincoords{70}{110}
\begin{tikzpicture}[tdplot_main_coords, scale=0.75]
    \shade[ball color=white!50, opacity=0.6] (0,0,0) circle (1cm);
    
    \draw[thick,->, scale=1] (0,0,0) -- (1.2,0,0) node[anchor=north east, font=\small]{$x_1$};
    \draw[thick,->, scale=1] (0,0,0) -- (0,1.2,0) node[anchor=north, font=\small]{$y_1$};
    \draw[thick,->, scale=1] (0,0,0) -- (0,0,1.2) node[anchor=east, font=\small]{$z_1$};
    
    
    \tdplotdrawarc[thin, dashed, scale=1]{(0,0,0)}{1}{-4}{356}{}{}

    \coordinate (statevector) at (0, 0, 1);
    
    \draw[orange, line width=1mm, -latex] (0,0,0) -- (statevector);

\end{tikzpicture}
\end{minipage}%
\begin{minipage}[b][2cm][t]{0.4\textwidth}
\centering
\tdplotsetmaincoords{70}{110}
\begin{tikzpicture}[tdplot_main_coords, scale=0.75]
    \shade[ball color=white!50, opacity=0.6] (0,0,0) circle (1cm);
    
    \draw[thick,->, scale=1] (0,0,0) -- (1.2,0,0) node[anchor=north east, font=\small]{$x_2$};
    \draw[thick,->, scale=1] (0,0,0) -- (0,1.2,0) node[anchor=north, font=\small]{$y_2$};
    \draw[thick,->, scale=1] (0,0,0) -- (0,0,1.2) node[anchor=west, font=\small]{$z_2$};
    
    
    \tdplotdrawarc[thin, dashed, scale=1]{(0,0,0)}{1}{-4}{356}{}{}

    \coordinate (statevector) at (0, 0, 1);
    
    \draw[orange, line width=1mm, -latex] (0,0,0) -- (statevector);

\end{tikzpicture}
\end{minipage}

  \caption{$\ket{\uparrow\uparrow}$}
\end{subfigure}

\caption{Step-by-step representation of CNOT acting on the state $\ket{\uparrow\uparrow}$. (a) to (b): this rotation changes the state of the control qubit to $\ket{\rightarrow}$. (b) to (c) is a rotation by $e^{-i\frac{\pi}{4}(\sigma_x \otimes \sigma_x)}$ which in this case (see rules) is the same as rotation by $e^{-i\frac{\pi}{4}(I \otimes \sigma_x)}$ or the plane $ -y_2 \land z_2 = z_2 \land y_2$ so only the second BS rotates. State of the target qubit has changed. (c) to (d) $e^{i\frac{\pi}{4}(I \otimes \sigma_x)}$ or the plane $y_2 \land z_2$ which undoes the rotation of the previous step so the state of the target qubit is back to initial. (d) to (e)  is an eigenrotation. (e) to (f) rotates the state of the control qubit to its initial state.}
\label{fig:CNOTuu}
\end{figure}
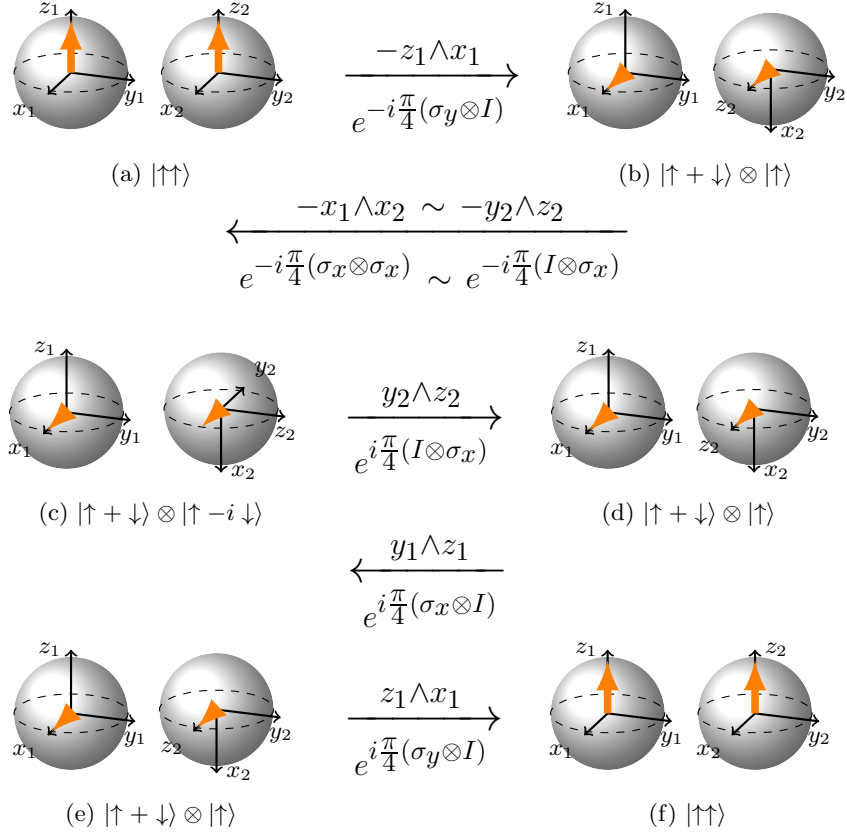

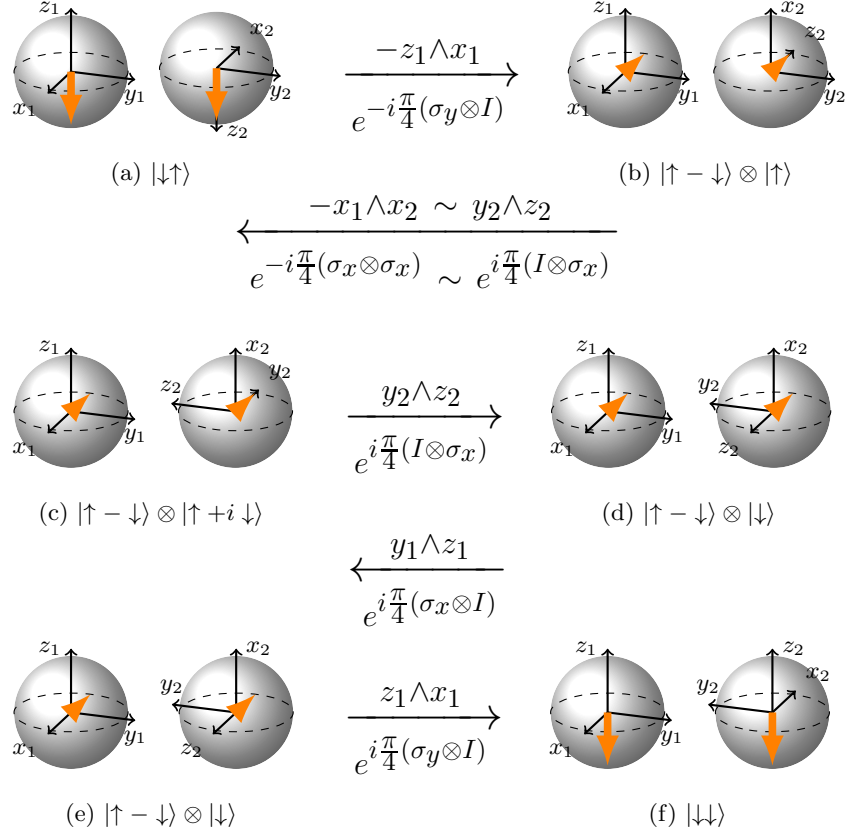
\begin{figure}
\begin{subfigure}{.4\textwidth}
  \centering
\begin{minipage}[b][2cm][t]{0.4\textwidth}
\centering
\tdplotsetmaincoords{70}{110}
\begin{tikzpicture}[tdplot_main_coords, scale=0.75]
    \shade[ball color=white!50, opacity=0.6] (0,0,0) circle (1cm);
    
    \draw[thick,->, scale=1] (0,0,0) -- (1.2,0,0) node[anchor=north east, font=\small]{$x_1$};
    \draw[thick,->, scale=1] (0,0,0) -- (0,1.2,0) node[anchor=north, font=\small]{$y_1$};
    \draw[thick,->, scale=1] (0,0,0) -- (0,0,1.2) node[anchor=east, font=\small]{$z_1$};
    
    
    \tdplotdrawarc[thin, dashed, scale=1]{(0,0,0)}{1}{-4}{356}{}{}

    \coordinate (statevector) at (0, 0, -1);
    
    \draw[orange, line width=1mm, -latex] (0,0,0) -- (statevector);

\end{tikzpicture}
\end{minipage}%
\begin{minipage}[b][1.75cm][t]{0.4\textwidth}
\centering
\tdplotsetmaincoords{70}{110}
\begin{tikzpicture}[tdplot_main_coords, scale=0.75]
    \shade[ball color=white!50, opacity=0.6] (0,0,0) circle (1cm);
    
    \draw[thick,->, scale=1] (0,0,0) -- (-1.2,0,0) node[anchor=south west, font=\small]{$x_2$};
    \draw[thick,->, scale=1] (0,0,0) -- (0,1.2,0) node[anchor=north, font=\small]{$y_2$};
    \draw[thick,->, scale=1] (0,0,0) -- (0,0,-1.2) node[anchor=west, font=\small]{$z_2$};
    
    
    \tdplotdrawarc[thin, dashed, scale=1]{(0,0,0)}{1}{-4}{356}{}{}

    \coordinate (statevector) at (0, 0, -1);
    
    \draw[orange, line width=1mm, -latex] (0,0,0) -- (statevector);

\end{tikzpicture}
\end{minipage}

  \caption{$\ket{\downarrow\uparrow}$}
\end{subfigure}
\raisebox{2\height}{\LARGE$\xrightarrow[e^{-i\frac{\pi}{4}(\sigma_y \otimes I)}]{-z_1 \land x_1}$}%
\begin{subfigure}{.4\textwidth}
  \centering
\begin{minipage}[b][2cm][t]{0.4\textwidth}
\centering
\tdplotsetmaincoords{70}{110}
\begin{tikzpicture}[tdplot_main_coords, scale=0.75]
    \shade[ball color=white!50, opacity=0.6] (0,0,0) circle (1cm);
    
    \draw[thick,->, scale=1] (0,0,0) -- (1.2,0,0) node[anchor=north east, font=\small]{$x_1$};
    \draw[thick,->, scale=1] (0,0,0) -- (0,1.2,0) node[anchor=north, font=\small]{$y_1$};
    \draw[thick,->, scale=1] (0,0,0) -- (0,0,1.2) node[anchor=east, font=\small]{$z_1$};
    
    
    \tdplotdrawarc[thin, dashed, scale=1]{(0,0,0)}{1}{-4}{356}{}{}

    \coordinate (statevector) at (-1, 0, 0);
    
    \draw[orange, line width=1mm, -latex] (0,0,0) -- (statevector);

\end{tikzpicture}
\end{minipage}%
\begin{minipage}[b][2cm][t]{0.4\textwidth}
\centering
\tdplotsetmaincoords{70}{110}
\begin{tikzpicture}[tdplot_main_coords, scale=0.75]
    \shade[ball color=white!50, opacity=0.6] (0,0,0) circle (1cm);
    
    \draw[thick,->, scale=1] (0,0,0) -- (-1.2,0,0) node[anchor=south west, font=\small]{$z_2$};
    \draw[thick,->, scale=1] (0,0,0) -- (0,1.2,0) node[anchor=north, font=\small]{$y_2$};
    \draw[thick,->, scale=1] (0,0,0) -- (0,0,1.2) node[anchor=west, font=\small]{$x_2$};
    
    
    \tdplotdrawarc[thin, dashed, scale=1]{(0,0,0)}{1}{-4}{356}{}{}

    \coordinate (statevector) at (-1, 0, 0);
    
    \draw[orange, line width=1mm, -latex] (0,0,0) -- (statevector);

\end{tikzpicture}
\end{minipage}

  \caption{$\ket{\uparrow - \downarrow} \otimes \ket{\uparrow}$}
\end{subfigure}
\vspace{0.5cm}
\begin{subfigure}{\textwidth}
\centering
\begin{turn}{0}
\LARGE$\xleftarrow[e^{-i\frac{\pi}{4}(\sigma_x \otimes \sigma_x )} \ \sim \ e^{i\frac{\pi}{4}(I \otimes \sigma_x )}]{-x_1 \land x_2\ \sim \ y_2 \land z_2}$%
\end{turn}
\end{subfigure}

\begin{subfigure}{.4\textwidth}
  \centering
\begin{minipage}[b][2cm][t]{0.4\textwidth}
\centering
\tdplotsetmaincoords{70}{110}
\begin{tikzpicture}[tdplot_main_coords, scale=0.75]
    \shade[ball color=white!50, opacity=0.6] (0,0,0) circle (1cm);
    
    \draw[thick,->, scale=1] (0,0,0) -- (1.2,0,0) node[anchor=north east, font=\small]{$x_1$};
    \draw[thick,->, scale=1] (0,0,0) -- (0,1.2,0) node[anchor=north, font=\small]{$y_1$};
    \draw[thick,->, scale=1] (0,0,0) -- (0,0,1.2) node[anchor=east, font=\small]{$z_1$};
    
    
    \tdplotdrawarc[thin, dashed, scale=1]{(0,0,0)}{1}{-4}{356}{}{}

    \coordinate (statevector) at (-1, 0, 0);
    
    \draw[orange, line width=1mm, -latex] (0,0,0) -- (statevector);

\end{tikzpicture}
\end{minipage}%
\begin{minipage}[b][2cm][t]{0.4\textwidth}
\centering
\tdplotsetmaincoords{70}{110}
\begin{tikzpicture}[tdplot_main_coords, scale=0.75]
    \shade[ball color=white!50, opacity=0.6] (0,0,0) circle (1cm);
    
    \draw[thick,->, scale=1] (0,0,0) -- (-1.2,0,0) node[anchor=south west, font=\small]{$y_2$};
    \draw[thick,->, scale=1] (0,0,0) -- (0,-1.2,0) node[anchor=south, font=\small]{$z_2$};
    \draw[thick,->, scale=1] (0,0,0) -- (0,0,1.2) node[anchor=west, font=\small]{$x_2$};
    
    
    \tdplotdrawarc[thin, dashed, scale=1]{(0,0,0)}{1}{-4}{356}{}{}

    \coordinate (statevector) at (-1, 0, 0);
    
    \draw[orange, line width=1mm, -latex] (0,0,0) -- (statevector);

\end{tikzpicture}
\end{minipage}

  \caption{$\ket{\uparrow - \downarrow} \otimes \ket{\uparrow + i \downarrow}$}
\end{subfigure}
\raisebox{2\height}{\LARGE$\xrightarrow[e^{i\frac{\pi}{4}(I \otimes \sigma_x)}]{y_2 \land z_2}$}%
\begin{subfigure}{.4\textwidth}
  \centering
\begin{minipage}[b][2cm][t]{0.4\textwidth}
\centering
\tdplotsetmaincoords{70}{110}
\begin{tikzpicture}[tdplot_main_coords, scale=0.75]
    \shade[ball color=white!50, opacity=0.6] (0,0,0) circle (1cm);
    
    \draw[thick,->, scale=1] (0,0,0) -- (1.2,0,0) node[anchor=north east, font=\small]{$x_1$};
    \draw[thick,->, scale=1] (0,0,0) -- (0,1.2,0) node[anchor=north, font=\small]{$y_1$};
    \draw[thick,->, scale=1] (0,0,0) -- (0,0,1.2) node[anchor=east, font=\small]{$z_1$};
    
    
    \tdplotdrawarc[thin, dashed, scale=1]{(0,0,0)}{1}{-4}{356}{}{}

    \coordinate (statevector) at (-1, 0, 0);
    
    \draw[orange, line width=1mm, -latex] (0,0,0) -- (statevector);

\end{tikzpicture}
\end{minipage}%
\begin{minipage}[b][2cm][t]{0.4\textwidth}
\centering
\tdplotsetmaincoords{70}{110}
\begin{tikzpicture}[tdplot_main_coords, scale=0.75]
    \shade[ball color=white!50, opacity=0.6] (0,0,0) circle (1cm);
    
    \draw[thick,->, scale=1] (0,0,0) -- (1.2,0,0) node[anchor=north east, font=\small]{$z_2$};
    \draw[thick,->, scale=1] (0,0,0) -- (0,-1.2,0) node[anchor=south, font=\small]{$y_2$};
    \draw[thick,->, scale=1] (0,0,0) -- (0,0,1.2) node[anchor=west, font=\small]{$x_2$};
    
    
    \tdplotdrawarc[thin, dashed, scale=1]{(0,0,0)}{1}{-4}{356}{}{}

    \coordinate (statevector) at (-1, 0, 0);
    
    \draw[orange, line width=1mm, -latex] (0,0,0) -- (statevector);

\end{tikzpicture}
\end{minipage}

  \caption{$\ket{\uparrow - \downarrow} \otimes \ket{\downarrow}$}
\end{subfigure}

\begin{subfigure}{\textwidth}
\centering
\begin{turn}{0}
\LARGE$\xleftarrow[e^{i\frac{\pi}{4}(\sigma_x \otimes I)}]{y_1 \land z_1}$%
\end{turn}
\end{subfigure}

\begin{subfigure}{.4\textwidth}
  \centering
\begin{minipage}[b][2cm][t]{0.4\textwidth}
\centering
\tdplotsetmaincoords{70}{110}
\begin{tikzpicture}[tdplot_main_coords, scale=0.75]
    \shade[ball color=white!50, opacity=0.6] (0,0,0) circle (1cm);
    
    \draw[thick,->, scale=1] (0,0,0) -- (1.2,0,0) node[anchor=north east, font=\small]{$x_1$};
    \draw[thick,->, scale=1] (0,0,0) -- (0,1.2,0) node[anchor=north, font=\small]{$y_1$};
    \draw[thick,->, scale=1] (0,0,0) -- (0,0,1.2) node[anchor=east, font=\small]{$z_1$};
    
    
    \tdplotdrawarc[thin, dashed, scale=1]{(0,0,0)}{1}{-4}{356}{}{}

    \coordinate (statevector) at (-1, 0, 0);
    
    \draw[orange, line width=1mm, -latex] (0,0,0) -- (statevector);

\end{tikzpicture}
\end{minipage}%
\begin{minipage}[b][2cm][t]{0.4\textwidth}
\centering
\tdplotsetmaincoords{70}{110}
\begin{tikzpicture}[tdplot_main_coords, scale=0.75]
    \shade[ball color=white!50, opacity=0.6] (0,0,0) circle (1cm);
    
    \draw[thick,->, scale=1] (0,0,0) -- (1.2,0,0) node[anchor=north east, font=\small]{$z_2$};
    \draw[thick,->, scale=1] (0,0,0) -- (0,-1.2,0) node[anchor=south, font=\small]{$y_2$};
    \draw[thick,->, scale=1] (0,0,0) -- (0,0,1.2) node[anchor=west, font=\small]{$x_2$};
    
    
    \tdplotdrawarc[thin, dashed, scale=1]{(0,0,0)}{1}{-4}{356}{}{}

    \coordinate (statevector) at (-1, 0, 0);
    
    \draw[orange, line width=1mm, -latex] (0,0,0) -- (statevector);

\end{tikzpicture}
\end{minipage}

  \caption{$\ket{\uparrow - \downarrow} \otimes \ket{\downarrow}$}
\end{subfigure}
\raisebox{2\height}{\LARGE$\xrightarrow[e^{i\frac{\pi}{4}(\sigma_y \otimes I)}]{z_1 \land x_1}$}%
\begin{subfigure}{.4\textwidth}
  \centering
\begin{minipage}[b][2cm][t]{0.4\textwidth}
\centering
\tdplotsetmaincoords{70}{110}
\begin{tikzpicture}[tdplot_main_coords, scale=0.75]
    \shade[ball color=white!50, opacity=0.6] (0,0,0) circle (1cm);
    
    \draw[thick,->, scale=1] (0,0,0) -- (1.2,0,0) node[anchor=north east, font=\small]{$x_1$};
    \draw[thick,->, scale=1] (0,0,0) -- (0,1.2,0) node[anchor=north, font=\small]{$y_1$};
    \draw[thick,->, scale=1] (0,0,0) -- (0,0,1.2) node[anchor=east, font=\small]{$z_1$};
    
    
    \tdplotdrawarc[thin, dashed, scale=1]{(0,0,0)}{1}{-4}{356}{}{}

    \coordinate (statevector) at (0, 0, -1);
    
    \draw[orange, line width=1mm, -latex] (0,0,0) -- (statevector);

\end{tikzpicture}
\end{minipage}%
\begin{minipage}[b][2cm][t]{0.4\textwidth}
\centering
\tdplotsetmaincoords{70}{110}
\begin{tikzpicture}[tdplot_main_coords, scale=0.75]
    \shade[ball color=white!50, opacity=0.6] (0,0,0) circle (1cm);
    
    \draw[thick,->, scale=1] (0,0,0) -- (-1.2,0,0) node[anchor=south west, font=\small]{$x_2$};
    \draw[thick,->, scale=1] (0,0,0) -- (0,-1.2,0) node[anchor=south, font=\small]{$y_2$};
    \draw[thick,->, scale=1] (0,0,0) -- (0,0,1.2) node[anchor=west, font=\small]{$z_2$};
    
    
    \tdplotdrawarc[thin, dashed, scale=1]{(0,0,0)}{1}{-4}{356}{}{}

    \coordinate (statevector) at (0, 0, -1);
    
    \draw[orange, line width=1mm, -latex] (0,0,0) -- (statevector);

\end{tikzpicture}
\end{minipage}

  \caption{$\ket{\downarrow\downarrow}$}
\end{subfigure}

\caption{Step-by-step representation of CNOT acting on the state $\ket{\downarrow\uparrow}$. (a) to (b): this rotation changes the state of the control qubit to $\ket{\leftarrow}$. (b) to (c) is a rotation by $e^{-i\frac{\pi}{4}(\sigma_x \otimes \sigma_x)}$ which in this case (see rules) is the same as rotation by $e^{i\frac{\pi}{4}(I \otimes \sigma_x)}$ or the plane $ y_2 \land z_2$ so only the second BS rotates. The absence of minus sign due to negative eigenvalue is an important difference from the previous case. State of the target qubit has changed. (c) to (d) $e^{i\frac{\pi}{4}(I \otimes \sigma_x)}$ or the plane $y_2 \land z_2$ which continues the rotation of the previous step so the state of the target qubit is flipped compared to beginning. (d) to (e)  is an eigenrotation. (e) to (f) rotates the state of the control qubit to its initial state.}
\label{fig:CNOTdu}
\end{figure}

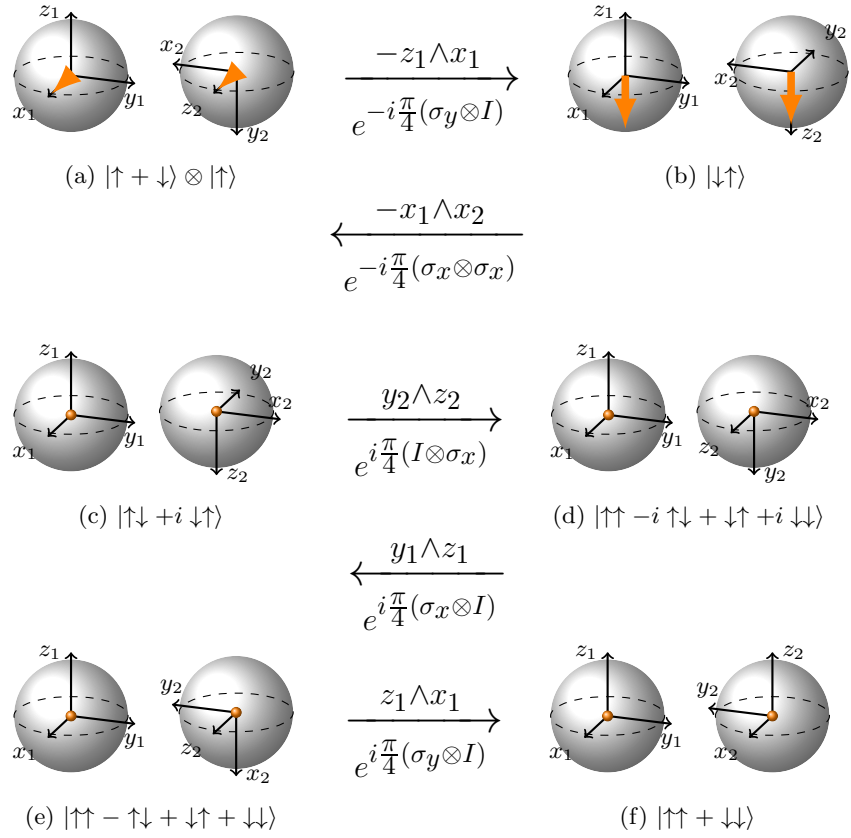
\begin{figure}
\begin{subfigure}{.4\textwidth}
  \centering
\begin{minipage}[b][2cm][t]{0.4\textwidth}
\centering
\tdplotsetmaincoords{70}{110}
\begin{tikzpicture}[tdplot_main_coords, scale=0.75]
    \shade[ball color=white!50, opacity=0.6] (0,0,0) circle (1cm);
    
    \draw[thick,->, scale=1] (0,0,0) -- (1.2,0,0) node[anchor=north east, font=\small]{$x_1$};
    \draw[thick,->, scale=1] (0,0,0) -- (0,1.2,0) node[anchor=north, font=\small]{$y_1$};
    \draw[thick,->, scale=1] (0,0,0) -- (0,0,1.2) node[anchor=east, font=\small]{$z_1$};
    
    
    \tdplotdrawarc[thin, dashed, scale=1]{(0,0,0)}{1}{-4}{356}{}{}

    \coordinate (statevector) at (1, 0, 0);
    
    \draw[orange, line width=1mm, -latex] (0,0,0) -- (statevector);

\end{tikzpicture}
\end{minipage}%
\begin{minipage}[b][1.75cm][t]{0.4\textwidth}
\centering
\tdplotsetmaincoords{70}{110}
\begin{tikzpicture}[tdplot_main_coords, scale=0.75]
    \shade[ball color=white!50, opacity=0.6] (0,0,0) circle (1cm);
    
    \draw[thick,->, scale=1] (0,0,0) -- (1.2,0,0) node[anchor=north east, font=\small]{$z_2$};
    \draw[thick,->, scale=1] (0,0,0) -- (0,-1.2,0) node[anchor=south, font=\small]{$x_2$};
    \draw[thick,->, scale=1] (0,0,0) -- (0,0,-1.2) node[anchor=west, font=\small]{$y_2$};
    
    
    \tdplotdrawarc[thin, dashed, scale=1]{(0,0,0)}{1}{-4}{356}{}{}

    \coordinate (statevector) at (1, 0, 0);
    
    \draw[orange, line width=1mm, -latex] (0,0,0) -- (statevector);

\end{tikzpicture}
\end{minipage}

  \caption{$\ket{\uparrow + \downarrow} \otimes \ket{\uparrow}$}
\end{subfigure}
\raisebox{2\height}{\LARGE$\xrightarrow[e^{-i\frac{\pi}{4}(\sigma_y \otimes I)}]{-z_1 \land x_1}$}%
\begin{subfigure}{.4\textwidth}
  \centering
\begin{minipage}[b][2cm][t]{0.4\textwidth}
\centering
\tdplotsetmaincoords{70}{110}
\begin{tikzpicture}[tdplot_main_coords, scale=0.75]
    \shade[ball color=white!50, opacity=0.6] (0,0,0) circle (1cm);
    
    \draw[thick,->, scale=1] (0,0,0) -- (1.2,0,0) node[anchor=north east, font=\small]{$x_1$};
    \draw[thick,->, scale=1] (0,0,0) -- (0,1.2,0) node[anchor=north, font=\small]{$y_1$};
    \draw[thick,->, scale=1] (0,0,0) -- (0,0,1.2) node[anchor=east, font=\small]{$z_1$};
    
    
    \tdplotdrawarc[thin, dashed, scale=1]{(0,0,0)}{1}{-4}{356}{}{}

    \coordinate (statevector) at (0, 0, -1);
    
    \draw[orange, line width=1mm, -latex] (0,0,0) -- (statevector);

\end{tikzpicture}
\end{minipage}%
\begin{minipage}[b][1.75cm][t]{0.4\textwidth}
\centering
\tdplotsetmaincoords{70}{110}
\begin{tikzpicture}[tdplot_main_coords, scale=0.75]
    \shade[ball color=white!50, opacity=0.6] (0,0,0) circle (1cm);
    
    \draw[thick,->, scale=1] (0,0,0) -- (-1.2,0,0) node[anchor=south west, font=\small]{$y_2$};
    \draw[thick,->, scale=1] (0,0,0) -- (0,-1.2,0) node[anchor=north, font=\small]{$x_2$};
    \draw[thick,->, scale=1] (0,0,0) -- (0,0,-1.2) node[anchor=west, font=\small]{$z_2$};
    
    
    \tdplotdrawarc[thin, dashed, scale=1]{(0,0,0)}{1}{-4}{356}{}{}

    \coordinate (statevector) at (0, 0, -1);
    
    \draw[orange, line width=1mm, -latex] (0,0,0) -- (statevector);

\end{tikzpicture}
\end{minipage}

  \caption{$\ket{\downarrow\uparrow}$}
\end{subfigure}
\vspace{0.5cm}
\begin{subfigure}{\textwidth}
\centering
\begin{turn}{0}
\LARGE$\xleftarrow[e^{-i\frac{\pi}{4}(\sigma_x \otimes \sigma_x )}]{-x_1 \land x_2}$%
\end{turn}
\end{subfigure}

\begin{subfigure}{.4\textwidth}
  \centering
\begin{minipage}[b][2cm][t]{0.4\textwidth}
\centering
\tdplotsetmaincoords{70}{110}
\begin{tikzpicture}[tdplot_main_coords, scale=0.75]
    \shade[ball color=white!50, opacity=0.6] (0,0,0) circle (1cm);
    
    \draw[thick,->, scale=1] (0,0,0) -- (1.2,0,0) node[anchor=north east, font=\small]{$x_1$};
    \draw[thick,->, scale=1] (0,0,0) -- (0,1.2,0) node[anchor=north, font=\small]{$y_1$};
    \draw[thick,->, scale=1] (0,0,0) -- (0,0,1.2) node[anchor=east, font=\small]{$z_1$};
    
    
    \tdplotdrawarc[thin, dashed, scale=1]{(0,0,0)}{1}{-4}{356}{}{}

    \shade[ball color=orange, opacity=1, scale=2] (0,0,0) circle (0.05cm);

\end{tikzpicture}
\end{minipage}%
\begin{minipage}[b][1.75cm][t]{0.4\textwidth}
\centering
\tdplotsetmaincoords{70}{110}
\begin{tikzpicture}[tdplot_main_coords, scale=0.75]
    \shade[ball color=white!50, opacity=0.6] (0,0,0) circle (1cm);
    
    \draw[thick,->, scale=1] (0,0,0) -- (-1.2,0,0) node[anchor=south west, font=\small]{$y_2$};
    \draw[thick,->, scale=1] (0,0,0) -- (0,1.2,0) node[anchor=south, font=\small]{$x_2$};
    \draw[thick,->, scale=1] (0,0,0) -- (0,0,-1.2) node[anchor=west, font=\small]{$z_2$};
    
    
    \tdplotdrawarc[thin, dashed, scale=1]{(0,0,0)}{1}{-4}{356}{}{}

    \shade[ball color=orange, opacity=1, scale=2] (0,0,0) circle (0.05cm);

\end{tikzpicture}
\end{minipage}

  \caption{$\ket{\uparrow\downarrow + i \downarrow\uparrow}$}
\end{subfigure}
\raisebox{2\height}{\LARGE$\xrightarrow[e^{i\frac{\pi}{4}(I \otimes \sigma_x)}]{y_2 \land z_2}$}%
\begin{subfigure}{.4\textwidth}
  \centering
\begin{minipage}[b][2cm][t]{0.4\textwidth}
\centering
\tdplotsetmaincoords{70}{110}
\begin{tikzpicture}[tdplot_main_coords, scale=0.75]
    \shade[ball color=white!50, opacity=0.6] (0,0,0) circle (1cm);
    
    \draw[thick,->, scale=1] (0,0,0) -- (1.2,0,0) node[anchor=north east, font=\small]{$x_1$};
    \draw[thick,->, scale=1] (0,0,0) -- (0,1.2,0) node[anchor=north, font=\small]{$y_1$};
    \draw[thick,->, scale=1] (0,0,0) -- (0,0,1.2) node[anchor=east, font=\small]{$z_1$};
    
    
    \tdplotdrawarc[thin, dashed, scale=1]{(0,0,0)}{1}{-4}{356}{}{}

    \shade[ball color=orange, opacity=1, scale=2] (0,0,0) circle (0.05cm);

\end{tikzpicture}
\end{minipage}%
\begin{minipage}[b][1.75cm][t]{0.4\textwidth}
\centering
\tdplotsetmaincoords{70}{110}
\begin{tikzpicture}[tdplot_main_coords, scale=0.75]
    \shade[ball color=white!50, opacity=0.6] (0,0,0) circle (1cm);
    
    \draw[thick,->, scale=1] (0,0,0) -- (1.2,0,0) node[anchor=north east, font=\small]{$z_2$};
    \draw[thick,->, scale=1] (0,0,0) -- (0,1.2,0) node[anchor=south, font=\small]{$x_2$};
    \draw[thick,->, scale=1] (0,0,0) -- (0,0,-1.2) node[anchor=west, font=\small]{$y_2$};
    
    
    \tdplotdrawarc[thin, dashed, scale=1]{(0,0,0)}{1}{-4}{356}{}{}

    \shade[ball color=orange, opacity=1, scale=2] (0,0,0) circle (0.05cm);

\end{tikzpicture}
\end{minipage}

  \caption{$\ket{\uparrow\uparrow - i\uparrow\downarrow + \downarrow\uparrow + i\downarrow\downarrow}$}
\end{subfigure}

\begin{subfigure}{\textwidth}
\centering
\begin{turn}{0}
\LARGE$\xleftarrow[e^{i\frac{\pi}{4}(\sigma_x \otimes I)}]{y_1 \land z_1}$%
\end{turn}
\end{subfigure}

\begin{subfigure}{.4\textwidth}
  \centering
\begin{minipage}[b][2cm][t]{0.4\textwidth}
\centering
\tdplotsetmaincoords{70}{110}
\begin{tikzpicture}[tdplot_main_coords, scale=0.75]
    \shade[ball color=white!50, opacity=0.6] (0,0,0) circle (1cm);
    
    \draw[thick,->, scale=1] (0,0,0) -- (1.2,0,0) node[anchor=north east, font=\small]{$x_1$};
    \draw[thick,->, scale=1] (0,0,0) -- (0,1.2,0) node[anchor=north, font=\small]{$y_1$};
    \draw[thick,->, scale=1] (0,0,0) -- (0,0,1.2) node[anchor=east, font=\small]{$z_1$};
    
    
    \tdplotdrawarc[thin, dashed, scale=1]{(0,0,0)}{1}{-4}{356}{}{}

    \shade[ball color=orange, opacity=1, scale=2] (0,0,0) circle (0.05cm);

\end{tikzpicture}
\end{minipage}%
\begin{minipage}[b][1.75cm][t]{0.4\textwidth}
\centering
\tdplotsetmaincoords{70}{110}
\begin{tikzpicture}[tdplot_main_coords, scale=0.75]
    \shade[ball color=white!50, opacity=0.6] (0,0,0) circle (1cm);
    
    \draw[thick,->, scale=1] (0,0,0) -- (1.2,0,0) node[anchor=north east, font=\small]{$z_2$};
    \draw[thick,->, scale=1] (0,0,0) -- (0,-1.2,0) node[anchor=south, font=\small]{$y_2$};
    \draw[thick,->, scale=1] (0,0,0) -- (0,0,-1.2) node[anchor=west, font=\small]{$x_2$};
    
    
    \tdplotdrawarc[thin, dashed, scale=1]{(0,0,0)}{1}{-4}{356}{}{}

    \shade[ball color=orange, opacity=1, scale=2] (0,0,0) circle (0.05cm);

\end{tikzpicture}
\end{minipage}

  \caption{$\ket{\uparrow\uparrow - \uparrow\downarrow + \downarrow\uparrow + \downarrow\downarrow}$}
\end{subfigure}
\raisebox{2\height}{\LARGE$\xrightarrow[e^{i\frac{\pi}{4}(\sigma_y \otimes I)}]{z_1 \land x_1}$}%
\begin{subfigure}{.4\textwidth}
  \centering
\begin{minipage}[b][2cm][t]{0.4\textwidth}
\centering
\tdplotsetmaincoords{70}{110}
\begin{tikzpicture}[tdplot_main_coords, scale=0.75]
    \shade[ball color=white!50, opacity=0.6] (0,0,0) circle (1cm);
    
    \draw[thick,->, scale=1] (0,0,0) -- (1.2,0,0) node[anchor=north east, font=\small]{$x_1$};
    \draw[thick,->, scale=1] (0,0,0) -- (0,1.2,0) node[anchor=north, font=\small]{$y_1$};
    \draw[thick,->, scale=1] (0,0,0) -- (0,0,1.2) node[anchor=east, font=\small]{$z_1$};
    
    
    \tdplotdrawarc[thin, dashed, scale=1]{(0,0,0)}{1}{-4}{356}{}{}

    \shade[ball color=orange, opacity=1, scale=2] (0,0,0) circle (0.05cm);

\end{tikzpicture}
\end{minipage}%
\begin{minipage}[b][2cm][t]{0.4\textwidth}
\centering
\tdplotsetmaincoords{70}{110}
\begin{tikzpicture}[tdplot_main_coords, scale=0.75]
    \shade[ball color=white!50, opacity=0.6] (0,0,0) circle (1cm);
    
    \draw[thick,->, scale=1] (0,0,0) -- (1.2,0,0) node[anchor=north east, font=\small]{$x_2$};
    \draw[thick,->, scale=1] (0,0,0) -- (0,-1.2,0) node[anchor=south, font=\small]{$y_2$};
    \draw[thick,->, scale=1] (0,0,0) -- (0,0,1.2) node[anchor=west, font=\small]{$z_2$};
    
    
    \tdplotdrawarc[thin, dashed, scale=1]{(0,0,0)}{1}{-4}{356}{}{}

    \shade[ball color=orange, opacity=1, scale=2] (0,0,0) circle (0.05cm);

\end{tikzpicture}
\end{minipage}

  \caption{$\ket{\uparrow\uparrow + \downarrow\downarrow}$}
\end{subfigure}

\caption{Step-by-step representation of CNOT acting on the state $\ket{\uparrow + \downarrow} \otimes \ket{\uparrow}$. (a) to (b): this rotation changes the state of the control qubit. (b) to (c) is entangling interaction (see rules): $x_2$ inverts and the statevector arrows are drawn into the center of BS; (c) through (f) rotations of local planes in entangled mode that finally arrive to the state $\ket{\uparrow\uparrow + \downarrow\downarrow}$}
\label{fig:CNOTru}
\end{figure}

We can see from the Figures \ref{fig:CNOTuu}, \ref{fig:CNOTdu}, \ref{fig:CNOTru}  that the key to such rich dynamics is the entangling interaction $e^{-i\tfrac{\pi}{4} \sigma_x \otimes  \sigma_x}$. In the three cases it is doing three different things: rotates along the plane $- y_2 \land z_2$, rotates in the opposite direction of that plane ($y_2 \land z_2$) and finally creates entanglement through rotation in the plane $x_1 \land x_2$ when conditions are right for it. It is interesting how in the first two cases (Figures \ref{fig:CNOTuu}, \ref{fig:CNOTdu}) the dependence on the state of the first qubit becomes apparent. The control qubit obtains its controlling function through becoming an eigenvector of the first sigma matrix in $e^{-i\tfrac{\pi}{4} \sigma_x \otimes  \sigma_x}$ rotation. In case the control qubit is $\ket{\uparrow}$ the eigenvalue is $1$ and if it is $\ket{\downarrow}$ the eigenvalue is $ -1$ which reverses the direction of rotation of the plane of the second qubit. Note how in case of non-entangling interactions the local rotation just after $e^{-i\tfrac{\pi}{4} \sigma_x \otimes  \sigma_x}$ undoes the rotation in case when control qubit is $\ket{\uparrow}$ and continues the rotation in case when control qubit is $\ket{\downarrow}$. Hence in case the control qubit is $\ket{\uparrow}$ the two rotations are $-\pi/2$ and $\pi/2$ rotations in the plane $y_2 \land z_2$ and in case the control qubit is $\ket{\downarrow}$ the rotations are $\pi/2$ and $\pi/2$ in the same plane. As a result in the first case the target qubit remains intact and in the second - it gets flipped. 

Speaking of the entangling case (Figure \ref{fig:CNOTru}), note how the first local rotation prepares the right configuration for the $e^{-i\tfrac{\pi}{4} \sigma_x \otimes  \sigma_x}$ rotation to work in entangling mode. How in the (b) subfigure of Figure \ref{fig:CNOTru} $-x_1$ and $x_2$ form an imaginary plane that is directed opposite to the statevector arrows and how in the (c) subfigure $x_2$ axis inverts and how the next three local rotations are gradually navigating the state towards $\ket{\uparrow\uparrow + \downarrow\downarrow}$.

There are many other interesting cases that may be analyzed graphically. For example one can see explicitly why for states lying along the x axes the control-target qubit relation reverses and the second qubit becomes control. The first rotation $-z_1 \land x_1$ keeps the second qubit aligned along the $x_2$ axis, therefore it will be an eigenstate of the second $\sigma_x$ matrix in $e^{-i\tfrac{\pi}{4} \sigma_x \otimes  \sigma_x}$, hence alignment or anti-alignment with the $x_2$ axis will determine the flip or no flip of the first qubit.

\FloatBarrier
\newpage
\section{Rotations by Arbitrary Angles, Partial Entanglement and Measures of Entanglement}

\begin{figure}[ht]
    \centering
    \begin{minipage}[b][2cm][t]{0.4\textwidth}
        \centering
        \tdplotsetmaincoords{70}{20}
        \begin{tikzpicture}[tdplot_main_coords, scale=2]
            \shade[ball color=white!50, opacity=0.6] (0,0,0) circle (1cm);
            \shade[ball color=cyan!50, opacity=0.6] (0,0,0) circle (0.707cm);
            
            \draw[thick,->, scale=1] (0,0,0) -- (1,0,0) node[anchor=north west, font=\small]{$x_1$};
            \draw[thick,->, scale=1] (0,0,0) -- (0,1,0) node[anchor=south west, font=\small]{$y_1$};
            \draw[thick,->, scale=1] (0,0,0) -- (0,0,1) node[anchor=east, font=\small]{$z_1$};
            
            \tdplotdrawarc[thin, dashed, scale=1]{(0,0,0)}{0.707}{-3}{357}{}{}
            
            \coordinate (statevector) at (0, 0, 0.707);
            
            \draw[orange, line width=1mm, -latex] (0,0,0) -- (statevector);
        \end{tikzpicture}
    \end{minipage}%
    \begin{minipage}[b][2cm][t]{0.4\textwidth}
        \centering
        \tdplotsetmaincoords{70}{20}
        \begin{tikzpicture}[tdplot_main_coords, scale=2]
            \shade[ball color=white!50, opacity=0.6] (0,0,0) circle (1cm);
            \shade[ball color=cyan!50, opacity=0.6] (0,0,0) circle (0.707cm);
            
            \draw[thick,-, scale=1] (0,0,0) -- (0.707,0,0) node[anchor=north west, font=\small]{};
            \draw[thick,->, scale=1] (-0.707,0,0) -- (-1,0,0) node[anchor=north east, font=\small]{$x_2$};
            \draw[thick,->, scale=1] (0,0,0) -- (0,1,0) node[anchor=south west, font=\small]{$y_2$};
            \draw[thick,->, scale=1] (0,0,0) -- (0,0,1) node[anchor=west, font=\small]{$z_2$};
            
            \tdplotdrawarc[thin, dashed, scale=1]{(0,0,0)}{0.707}{-3}{357}{}{}
            
            \coordinate (statevector) at (0, 0, 0.707);
            
            \draw[orange, line width=1mm, -latex] (0,0,0) -- (statevector);
        \end{tikzpicture}
    \end{minipage}
    \vspace{2cm}
    \caption{Example of a state of partial entanglement. It has been generated by a $\pi/4$ rotation of state $\ket{\uparrow\uparrow}$ in the plane $y_1 \land x_2$ or in matrix formalism by a rotation $e^{i\frac{\pi}{8}(\sigma_y \otimes \sigma_x)}$. Note how $x_2$ axis is partially inverted. How the inner sphere represents the separable part of the state and the outer shell - entangled part.}
    \label{fig:partialent}
\end{figure}
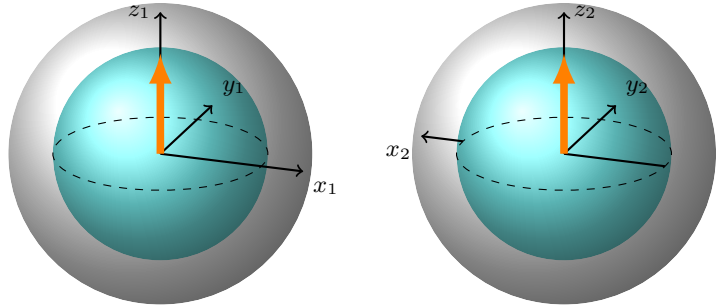

\begin{figure}
\begin{subfigure}{.45\textwidth}
  \centering
\begin{minipage}[b][2cm][t]{0.4\textwidth}
\centering
\tdplotsetmaincoords{70}{20}
\begin{tikzpicture}[tdplot_main_coords, scale=1]
    \shade[ball color=cyan!50, opacity=0.6] (0,0,0) circle (1cm);
    
    \draw[thick,->, scale=1] (0,0,0) -- (1,0,0) node[anchor=north west, font=\small]{$x_1$};
    \draw[thick,->, scale=1] (0,0,0) -- (0,1,0) node[anchor=south west, font=\small]{$y_1$};
    \draw[thick,->, scale=1] (0,0,0) -- (0,0,1) node[anchor=east, font=\small]{$z_1$};

    \tdplotdrawarc[thin, dashed, scale=1]{(0,0,0)}{1}{-3}{357}{}{}

    \coordinate (statevector) at (0, 0, 1);
    
    \draw[orange, line width=1mm, -latex] (0,0,0) -- (statevector);

\end{tikzpicture}
\end{minipage}%
\hspace{0.3cm}
\begin{minipage}[b][2cm][t]{0.4\textwidth}
\centering
\tdplotsetmaincoords{70}{20}
\begin{tikzpicture}[tdplot_main_coords, scale=1]
    \shade[ball color=cyan!50, opacity=0.6] (0,0,0) circle (1cm);
    
    \draw[thick,->, scale=1] (0,0,0) -- (1,0,0) node[anchor=north west, font=\small]{$x_2$};
    \draw[thick,->, scale=1] (0,0,0) -- (0,1,0) node[anchor=south west, font=\small]{$y_2$};
    \draw[thick,->, scale=1] (0,0,0) -- (0,0,1) node[anchor=west, font=\small]{$z_2$};

    \tdplotdrawarc[thin, dashed, scale=1]{(0,0,0)}{1}{-3}{357}{}{}

    \coordinate (statevector) at (0, 0, 1);
    
    \draw[orange, line width=1mm, -latex] (0,0,0) -- (statevector);

\end{tikzpicture}
\end{minipage}
\end{subfigure}
\begin{subfigure}{.45\textwidth}
  \centering
\begin{minipage}[b][2cm][t]{0.4\textwidth}
\centering
\tdplotsetmaincoords{70}{20}
\begin{tikzpicture}[tdplot_main_coords, scale=1]
    \shade[ball color=cyan!50, opacity=0.6] (0,0,0) circle (1cm);
    
    \draw[thick,->, scale=1] (0,0,0) -- (1,0,0) node[anchor=north west, font=\small]{$x_1$};
    \draw[thick,->, scale=1] (0,0,0) -- (0,1,0) node[anchor=south west, font=\small]{$y_1$};
    \draw[thick,->, scale=1] (0,0,0) -- (0,0,1) node[anchor=east, font=\small]{$z_1$};

    \tdplotdrawarc[thin, dashed, scale=1]{(0,0,0)}{1}{-3}{357}{}{}

    \coordinate (statevector) at (0, 0, -1);
    
    \draw[orange, line width=1mm, -latex] (0,0,0) -- (statevector);

\end{tikzpicture}
\end{minipage}%
\hspace{0.3cm}
\begin{minipage}[b][2cm][t]{0.4\textwidth}
\centering
\tdplotsetmaincoords{70}{20}
\begin{tikzpicture}[tdplot_main_coords, scale=1]
    \shade[ball color=cyan!50, opacity=0.6] (0,0,0) circle (1cm);
    
    \draw[thick,->, scale=1] (0,0,0) -- (1,0,0) node[anchor=north west, font=\small]{$x_2$};
    \draw[thick,->, scale=1] (0,0,0) -- (0,1,0) node[anchor=south west, font=\small]{$y_2$};
    \draw[thick,->, scale=1] (0,0,0) -- (0,0,1) node[anchor=west, font=\small]{$z_2$};

    \tdplotdrawarc[thin, dashed, scale=1]{(0,0,0)}{1}{-3}{357}{}{}

    \coordinate (statevector) at (0, 0, -1);
    
    \draw[orange, line width=1mm, -latex] (0,0,0) -- (statevector);

\end{tikzpicture}
\end{minipage}
\end{subfigure}

\vspace{1cm}

\begin{subfigure}{.45\textwidth}
  \centering
\begin{minipage}[b][2cm][t]{0.4\textwidth}
\centering
\tdplotsetmaincoords{70}{20}
\begin{tikzpicture}[tdplot_main_coords, scale=1]
    \shade[ball color=white!50, opacity=0.6] (0,0,0) circle (1cm);
    \shade[ball color=cyan!50, opacity=0.6] (0,0,0) circle (0.707cm);
    
    \draw[thick,->, scale=1] (0,0,0) -- (1,0,0) node[anchor=north west, font=\small]{$x_1$};
    \draw[thick,->, scale=1] (0,0,0) -- (0,1,0) node[anchor=south west, font=\small]{$y_1$};
    \draw[thick,->, scale=1] (0,0,0) -- (0,0,1) node[anchor=east, font=\small]{$z_1$};
    
    \tdplotdrawarc[thin, dashed, scale=1]{(0,0,0)}{0.707}{-3}{357}{}{}

    \coordinate (statevector) at (0, 0, 0.707);
    
    \draw[orange, line width=1mm, -latex] (0,0,0) -- (statevector);

\end{tikzpicture}
\end{minipage}%
\begin{minipage}[b][2cm][t]{0.4\textwidth}
\centering
\tdplotsetmaincoords{70}{20}
\begin{tikzpicture}[tdplot_main_coords, scale=1]
    \shade[ball color=white!50, opacity=0.6] (0,0,0) circle (1cm);
    \shade[ball color=cyan!50, opacity=0.6] (0,0,0) circle (0.707cm);
    
    \draw[thick,-, scale=1] (0,0,0) -- (0.707,0,0) node[anchor=north west, font=\small]{};
    \draw[thick,->, scale=1] (-0.707,0,0) -- (-1,0,0) node[anchor=south east, font=\small]{$x_2$};
    \draw[thick,->, scale=1] (0,0,0) -- (0,1,0) node[anchor=south west, font=\small]{$y_2$};
    \draw[thick,->, scale=1] (0,0,0) -- (0,0,1) node[anchor=west, font=\small]{$z_2$};
    
    \tdplotdrawarc[thin, dashed, scale=1]{(0,0,0)}{0.707}{-3}{357}{}{}

    \coordinate (statevector) at (0, 0, 0.707);
    
    \draw[orange, line width=1mm, -latex] (0,0,0) -- (statevector);

\end{tikzpicture}
\end{minipage}

\end{subfigure}
\begin{subfigure}{.45\textwidth}
  \centering
\begin{minipage}[b][2cm][t]{0.4\textwidth}
\centering
\tdplotsetmaincoords{70}{20}
\begin{tikzpicture}[tdplot_main_coords, scale=1]
    \shade[ball color=white!50, opacity=0.6] (0,0,0) circle (1cm);
    \shade[ball color=cyan!50, opacity=0.6] (0,0,0) circle (0.707cm);
    
    \draw[thick,->, scale=1] (0,0,0) -- (1,0,0) node[anchor=north west, font=\small]{$x_1$};
    \draw[thick,->, scale=1] (0,0,0) -- (0,1,0) node[anchor=south west, font=\small]{$y_1$};
    \draw[thick,->, scale=1] (0,0,0) -- (0,0,1) node[anchor=east, font=\small]{$z_1$};
    
    \tdplotdrawarc[thin, dashed, scale=1]{(0,0,0)}{0.707}{-3}{357}{}{}

    \coordinate (statevector) at (0, 0, -0.707);
    
    \draw[orange, line width=1mm, -latex] (0,0,0) -- (statevector);

\end{tikzpicture}
\end{minipage}%
\hspace{0.3cm}
\begin{minipage}[b][2cm][t]{0.4\textwidth}
\centering
\tdplotsetmaincoords{70}{20}
\begin{tikzpicture}[tdplot_main_coords, scale=1]
    \shade[ball color=white!50, opacity=0.6] (0,0,0) circle (1cm);
    \shade[ball color=cyan!50, opacity=0.6] (0,0,0) circle (0.707cm);
    
    \draw[thick,->, scale=1] (0,-0.707,0) -- (0,-1,0) node[anchor= east, font=\small]{$y_2$};
    \draw[thick,->, scale=1] (0,0,0) -- (1,0,0) node[anchor=north west, font=\small]{$x_2$};
    \draw[thick,-, scale=1] (0,0,0) -- (0,0.707,0) node[anchor=south west, font=\small]{};
    \draw[thick,->, scale=1] (0,0,0) -- (0,0,1) node[anchor=west, font=\small]{$z_2$};
    
    \tdplotdrawarc[thin, dashed, scale=1]{(0,0,0)}{0.707}{-3}{357}{}{}

    \coordinate (statevector) at (0, 0, -0.707);
    
    \draw[orange, line width=1mm, -latex] (0,0,0) -- (statevector);

\end{tikzpicture}
\end{minipage}

\end{subfigure}

\vspace{1cm}

\begin{subfigure}{.45\textwidth}
  \centering
\begin{minipage}[b][2cm][t]{0.4\textwidth}
\centering
\tdplotsetmaincoords{70}{20}
\begin{tikzpicture}[tdplot_main_coords, scale=1]
    \shade[ball color=white!50, opacity=0.6] (0,0,0) circle (1cm);
    \shade[ball color=cyan!50, opacity=0.6] (0,0,0) circle (0.5cm);
    
    \draw[thick,->, scale=1] (0,0,0) -- (1,0,0) node[anchor=north west, font=\small]{$x_1$};
    \draw[thick,->, scale=1] (0,0,0) -- (0,1,0) node[anchor=south west, font=\small]{$y_1$};
    \draw[thick,->, scale=1] (0,0,0) -- (0,0,1) node[anchor=east, font=\small]{$z_1$};
    
    \tdplotdrawarc[thin, dashed, scale=1]{(0,0,0)}{0.5}{-3}{357}{}{}

    \coordinate (statevector) at (0, 0, 0.5);
    
    \draw[orange, line width=1mm, -latex] (0,0,0) -- (statevector);

\end{tikzpicture}
\end{minipage}%
\begin{minipage}[b][2cm][t]{0.4\textwidth}
\centering
\tdplotsetmaincoords{70}{20}
\begin{tikzpicture}[tdplot_main_coords, scale=1]
    \shade[ball color=white!50, opacity=0.6] (0,0,0) circle (1cm);
    \shade[ball color=cyan!50, opacity=0.6] (0,0,0) circle (0.5cm);
    
    \draw[thick,-, scale=1] (0,0,0) -- (0.5,0,0) node[anchor=north west, font=\small]{};
    \draw[thick,->, scale=1] (-0.5,0,0) -- (-1,0,0) node[anchor=south east, font=\small]{$x_2$};
    \draw[thick,->, scale=1] (0,0,0) -- (0,1,0) node[anchor=south west, font=\small]{$y_2$};
    \draw[thick,->, scale=1] (0,0,0) -- (0,0,1) node[anchor=west, font=\small]{$z_2$};
    
    \tdplotdrawarc[thin, dashed, scale=1]{(0,0,0)}{0.5}{-3}{357}{}{}

    \coordinate (statevector) at (0, 0, 0.5);
    
    \draw[orange, line width=1mm, -latex] (0,0,0) -- (statevector);

\end{tikzpicture}
\end{minipage}
\end{subfigure}
\begin{subfigure}{.45\textwidth}
  \centering
\begin{minipage}[b][2cm][t]{0.4\textwidth}
\centering
\tdplotsetmaincoords{70}{20}
\begin{tikzpicture}[tdplot_main_coords, scale=1]
    \shade[ball color=white!50, opacity=0.6] (0,0,0) circle (1cm);
    \shade[ball color=cyan!50, opacity=0.6] (0,0,0) circle (0.5cm);
    
    \draw[thick,->, scale=1] (0,0,0) -- (1,0,0) node[anchor=north west, font=\small]{$x_1$};
    \draw[thick,->, scale=1] (0,0,0) -- (0,1,0) node[anchor=south west, font=\small]{$y_1$};
    \draw[thick,->, scale=1] (0,0,0) -- (0,0,1) node[anchor=east, font=\small]{$z_1$};
    
    \tdplotdrawarc[thin, dashed, scale=1]{(0,0,0)}{0.5}{-3}{357}{}{}

    \coordinate (statevector) at (0, 0, -0.5);
    
    \draw[orange, line width=1mm, -latex] (0,0,0) -- (statevector);

\end{tikzpicture}
\end{minipage}%
\hspace{0.3cm}
\begin{minipage}[b][2cm][t]{0.4\textwidth}
\centering
\tdplotsetmaincoords{70}{20}
\begin{tikzpicture}[tdplot_main_coords, scale=1]
    \shade[ball color=white!50, opacity=0.6] (0,0,0) circle (1cm);
    \shade[ball color=cyan!50, opacity=0.6] (0,0,0) circle (0.5cm);
    
    \draw[thick,->, scale=1] (0,-0.5,0) -- (0,-1,0) node[anchor= east, font=\small]{$y_2$};
    \draw[thick,->, scale=1] (0,0,0) -- (1,0,0) node[anchor=north west, font=\small]{$x_2$};
    \draw[thick,-, scale=1] (0,0,0) -- (0,0.5,0) node[anchor=south west, font=\small]{};
    \draw[thick,->, scale=1] (0,0,0) -- (0,0,1) node[anchor=west, font=\small]{$z_2$};
    
    \tdplotdrawarc[thin, dashed, scale=1]{(0,0,0)}{0.5}{-3}{357}{}{}

    \coordinate (statevector) at (0, 0, -0.5);
    
    \draw[orange, line width=1mm, -latex] (0,0,0) -- (statevector);

\end{tikzpicture}
\end{minipage}
\end{subfigure}

\vspace{1cm}

\begin{subfigure}{.45\textwidth}
  \centering
\begin{minipage}[b][2cm][t]{0.4\textwidth}
\centering
\tdplotsetmaincoords{70}{20}
\begin{tikzpicture}[tdplot_main_coords, scale=1]
    \shade[ball color=white!50, opacity=0.6] (0,0,0) circle (1cm);

    \draw[thick,->, scale=1] (0,0,0) -- (1,0,0) node[anchor=north west, font=\small]{$x_1$};
    \draw[thick,->, scale=1] (0,0,0) -- (0,1,0) node[anchor=south west, font=\small]{$y_1$};
    \draw[thick,->, scale=1] (0,0,0) -- (0,0,1) node[anchor=east, font=\small]{$z_1$};

    \tdplotdrawarc[thin, dashed, scale=1]{(0,0,0)}{1}{-3}{357}{}{}
    \shade[ball color=orange, opacity=1, scale=2] (0,0,0) circle (0.05cm);

\end{tikzpicture}
\end{minipage}%
\begin{minipage}[b][2cm][t]{0.4\textwidth}
\centering
\tdplotsetmaincoords{70}{20}
\begin{tikzpicture}[tdplot_main_coords, scale=1]
    \shade[ball color=white!50, opacity=0.6] (0,0,0) circle (1cm);

    \draw[thick,->, scale=1] (0,0,0) -- (-1,0,0) node[anchor=south east, font=\small]{$x_2$};
    \draw[thick,->, scale=1] (0,0,0) -- (0,1,0) node[anchor=south west, font=\small]{$y_2$};
    \draw[thick,->, scale=1] (0,0,0) -- (0,0,1) node[anchor=west, font=\small]{$z_2$};

    \tdplotdrawarc[thin, dashed, scale=1]{(0,0,0)}{1}{-3}{357}{}{}
    \shade[ball color=orange, opacity=1, scale=2] (0,0,0) circle (0.05cm);

\end{tikzpicture}
\end{minipage}

\end{subfigure}
\begin{subfigure}{.45\textwidth}
  \centering
\begin{minipage}[b][2cm][t]{0.4\textwidth}
\centering
\tdplotsetmaincoords{70}{20}
\begin{tikzpicture}[tdplot_main_coords, scale=1]
    \shade[ball color=white!50, opacity=0.6] (0,0,0) circle (1cm);

    \draw[thick,->, scale=1] (0,0,0) -- (1,0,0) node[anchor=north west, font=\small]{$x_1$};
    \draw[thick,->, scale=1] (0,0,0) -- (0,1,0) node[anchor=south west, font=\small]{$y_1$};
    \draw[thick,->, scale=1] (0,0,0) -- (0,0,1) node[anchor=east, font=\small]{$z_1$};

    \tdplotdrawarc[thin, dashed, scale=1]{(0,0,0)}{1}{-3}{357}{}{}
    \shade[ball color=orange, opacity=1, scale=2] (0,0,0) circle (0.05cm);

\end{tikzpicture}
\end{minipage}%
\hspace{0.3cm}
\begin{minipage}[b][2cm][t]{0.4\textwidth}
\centering
\tdplotsetmaincoords{70}{20}
\begin{tikzpicture}[tdplot_main_coords, scale=1]
    \shade[ball color=white!50, opacity=0.6] (0,0,0) circle (1cm);

    \draw[thick,->, scale=1] (0,0,0) -- (1,0,0) node[anchor=north west, font=\small]{$x_2$};
    \draw[thick,->, scale=1] (0,0,0) -- (0,-1,0) node[anchor=east, font=\small]{$y_2$};
    \draw[thick,->, scale=1] (0,0,0) -- (0,0,1) node[anchor=west, font=\small]{$z_2$};

    \tdplotdrawarc[thin, dashed, scale=1]{(0,0,0)}{1}{-3}{357}{}{}
    \shade[ball color=orange, opacity=1, scale=2] (0,0,0) circle (0.05cm);

\end{tikzpicture}
\end{minipage}

\end{subfigure}
\vspace{1cm}

\begin{subfigure}{.45\textwidth}
  \centering
\begin{minipage}[b][2cm][t]{0.4\textwidth}
\centering
\tdplotsetmaincoords{70}{20}
\begin{tikzpicture}[tdplot_main_coords, scale=1]
    \shade[ball color=white!50, opacity=0.6] (0,0,0) circle (1cm);
    \shade[ball color=cyan!50, opacity=0.6] (0,0,0) circle (0.5cm);
    
    \draw[thick,->, scale=1] (0,0,0) -- (1,0,0) node[anchor=north west, font=\small]{$x_1$};
    \draw[thick,->, scale=1] (0,0,0) -- (0,1,0) node[anchor=south west, font=\small]{$y_1$};
    \draw[thick,->, scale=1] (0,0,0) -- (0,0,1) node[anchor=east, font=\small]{$z_1$};
    
    \tdplotdrawarc[thin, dashed, scale=1]{(0,0,0)}{0.5}{-3}{357}{}{}

    \coordinate (statevector) at (0, 0, -0.5);
    
    \draw[orange, line width=1mm, -latex] (0,0,0) -- (statevector);

\end{tikzpicture}
\end{minipage}%
\begin{minipage}[b][2cm][t]{0.4\textwidth}
\centering
\tdplotsetmaincoords{70}{20}
\begin{tikzpicture}[tdplot_main_coords, scale=1]
    \shade[ball color=white!50, opacity=0.6] (0,0,0) circle (1cm);
    \shade[ball color=cyan!50, opacity=0.6] (0,0,0) circle (0.5cm);
    
    \draw[thick,-, scale=1] (0,0,0) -- (0.5,0,0) node[anchor=north west, font=\small]{};
    \draw[thick,->, scale=1] (-0.5,0,0) -- (-1,0,0) node[anchor=south east, font=\small]{$x_2$};
    \draw[thick,->, scale=1] (0,0,0) -- (0,1,0) node[anchor=south west, font=\small]{$y_2$};
    \draw[thick,->, scale=1] (0,0,0) -- (0,0,1) node[anchor=west, font=\small]{$z_2$};
    
    \tdplotdrawarc[thin, dashed, scale=1]{(0,0,0)}{0.5}{-3}{357}{}{}

    \coordinate (statevector) at (0, 0, -0.5);
    
    \draw[orange, line width=1mm, -latex] (0,0,0) -- (statevector);

\end{tikzpicture}
\end{minipage}
\end{subfigure}
\begin{subfigure}{.45\textwidth}
  \centering
\begin{minipage}[b][2cm][t]{0.4\textwidth}
\centering
\tdplotsetmaincoords{70}{20}
\begin{tikzpicture}[tdplot_main_coords, scale=1]
    \shade[ball color=white!50, opacity=0.6] (0,0,0) circle (1cm);
    \shade[ball color=cyan!50, opacity=0.6] (0,0,0) circle (0.5cm);
    
    \draw[thick,->, scale=1] (0,0,0) -- (1,0,0) node[anchor=north west, font=\small]{$x_1$};
    \draw[thick,->, scale=1] (0,0,0) -- (0,1,0) node[anchor=south west, font=\small]{$y_1$};
    \draw[thick,->, scale=1] (0,0,0) -- (0,0,1) node[anchor=east, font=\small]{$z_1$};
    
    \tdplotdrawarc[thin, dashed, scale=1]{(0,0,0)}{0.5}{-3}{357}{}{}

    \coordinate (statevector) at (0, 0, 0.5);
    
    \draw[orange, line width=1mm, -latex] (0,0,0) -- (statevector);

\end{tikzpicture}
\end{minipage}%
\hspace{0.3cm}
\begin{minipage}[b][2cm][t]{0.4\textwidth}
\centering
\tdplotsetmaincoords{70}{20}
\begin{tikzpicture}[tdplot_main_coords, scale=1]
    \shade[ball color=white!50, opacity=0.6] (0,0,0) circle (1cm);
    \shade[ball color=cyan!50, opacity=0.6] (0,0,0) circle (0.5cm);
    
    \draw[thick,->, scale=1] (0,-0.5,0) -- (0,-1,0) node[anchor= east, font=\small]{$y_2$};
    \draw[thick,->, scale=1] (0,0,0) -- (1,0,0) node[anchor=north west, font=\small]{$x_2$};
    \draw[thick,-, scale=1] (0,0,0) -- (0,0.5,0) node[anchor=south west, font=\small]{};
    \draw[thick,->, scale=1] (0,0,0) -- (0,0,1) node[anchor=west, font=\small]{$z_2$};
    
    \tdplotdrawarc[thin, dashed, scale=1]{(0,0,0)}{0.5}{-3}{357}{}{}

    \coordinate (statevector) at (0, 0, 0.5);
    
    \draw[orange, line width=1mm, -latex] (0,0,0) -- (statevector);

\end{tikzpicture}
\end{minipage}
\end{subfigure}

\vspace{1cm}
\begin{subfigure}{.45\textwidth}
  \centering
\begin{minipage}[b][2cm][t]{0.4\textwidth}
\centering
\tdplotsetmaincoords{70}{20}
\begin{tikzpicture}[tdplot_main_coords, scale=1]
    \shade[ball color=white!50, opacity=0.6] (0,0,0) circle (1cm);
    \shade[ball color=cyan!50, opacity=0.6] (0,0,0) circle (0.707cm);
    
    \draw[thick,->, scale=1] (0,0,0) -- (1,0,0) node[anchor=north west, font=\small]{$x_1$};
    \draw[thick,->, scale=1] (0,0,0) -- (0,1,0) node[anchor=south west, font=\small]{$y_1$};
    \draw[thick,->, scale=1] (0,0,0) -- (0,0,1) node[anchor=east, font=\small]{$z_1$};
    
    \tdplotdrawarc[thin, dashed, scale=1]{(0,0,0)}{0.707}{-3}{357}{}{}

    \coordinate (statevector) at (0, 0, -0.707);
    
    \draw[orange, line width=1mm, -latex] (0,0,0) -- (statevector);

\end{tikzpicture}
\end{minipage}%
\begin{minipage}[b][2cm][t]{0.4\textwidth}
\centering
\tdplotsetmaincoords{70}{20}
\begin{tikzpicture}[tdplot_main_coords, scale=1]
    \shade[ball color=white!50, opacity=0.6] (0,0,0) circle (1cm);
    \shade[ball color=cyan!50, opacity=0.6] (0,0,0) circle (0.707cm);
    
    \draw[thick,-, scale=1] (0,0,0) -- (0.707,0,0) node[anchor=north west, font=\small]{};
    \draw[thick,->, scale=1] (-0.707,0,0) -- (-1,0,0) node[anchor=south east, font=\small]{$x_2$};
    \draw[thick,->, scale=1] (0,0,0) -- (0,1,0) node[anchor=south west, font=\small]{$y_2$};
    \draw[thick,->, scale=1] (0,0,0) -- (0,0,1) node[anchor=west, font=\small]{$z_2$};
    
    \tdplotdrawarc[thin, dashed, scale=1]{(0,0,0)}{0.707}{-3}{357}{}{}

    \coordinate (statevector) at (0, 0, -0.707);
    
    \draw[orange, line width=1mm, -latex] (0,0,0) -- (statevector);

\end{tikzpicture}
\end{minipage}
\end{subfigure}
\begin{subfigure}{.45\textwidth}
  \centering
\begin{minipage}[b][2cm][t]{0.4\textwidth}
\centering
\tdplotsetmaincoords{70}{20}
\begin{tikzpicture}[tdplot_main_coords, scale=1]
    \shade[ball color=white!50, opacity=0.6] (0,0,0) circle (1cm);
    \shade[ball color=cyan!50, opacity=0.6] (0,0,0) circle (0.707cm);
    
    \draw[thick,->, scale=1] (0,0,0) -- (1,0,0) node[anchor=north west, font=\small]{$x_1$};
    \draw[thick,->, scale=1] (0,0,0) -- (0,1,0) node[anchor=south west, font=\small]{$y_1$};
    \draw[thick,->, scale=1] (0,0,0) -- (0,0,1) node[anchor=east, font=\small]{$z_1$};
    
    \tdplotdrawarc[thin, dashed, scale=1]{(0,0,0)}{0.707}{-3}{357}{}{}

    \coordinate (statevector) at (0, 0, 0.707);
    
    \draw[orange, line width=1mm, -latex] (0,0,0) -- (statevector);

\end{tikzpicture}
\end{minipage}%
\hspace{0.3cm}
\begin{minipage}[b][2cm][t]{0.4\textwidth}
\centering
\tdplotsetmaincoords{70}{20}
\begin{tikzpicture}[tdplot_main_coords, scale=1]
    \shade[ball color=white!50, opacity=0.6] (0,0,0) circle (1cm);
    \shade[ball color=cyan!50, opacity=0.6] (0,0,0) circle (0.707cm);
    
    \draw[thick,->, scale=1] (0,-0.707,0) -- (0,-1,0) node[anchor= east, font=\small]{$y_2$};
    \draw[thick,->, scale=1] (0,0,0) -- (1,0,0) node[anchor=north west, font=\small]{$x_2$};
    \draw[thick,-, scale=1] (0,0,0) -- (0,0.707,0) node[anchor=south west, font=\small]{};
    \draw[thick,->, scale=1] (0,0,0) -- (0,0,1) node[anchor=west, font=\small]{$z_2$};
    
    \tdplotdrawarc[thin, dashed, scale=1]{(0,0,0)}{0.707}{-3}{357}{}{}

    \coordinate (statevector) at (0, 0, 0.707);
    
    \draw[orange, line width=1mm, -latex] (0,0,0) -- (statevector);

\end{tikzpicture}
\end{minipage}
\end{subfigure}

\vspace{1cm}
\caption{Full rotation of state $\ket{\uparrow\uparrow}$ in the plane $y_1 \land x_2$, in matrix formalism rotation generator is $\sigma_y \otimes \sigma_x$. The figure should be read column by column, from top to bottom. We gradually move through states $\ket{\uparrow\uparrow}$, $\ket{\uparrow\uparrow - \downarrow\downarrow}$, $\ket{\downarrow\downarrow}$, $\ket{\uparrow\uparrow + \downarrow\downarrow}$ back to $\ket{\uparrow\uparrow}$}
\label{fig:ent-full-rot}
\end{figure}
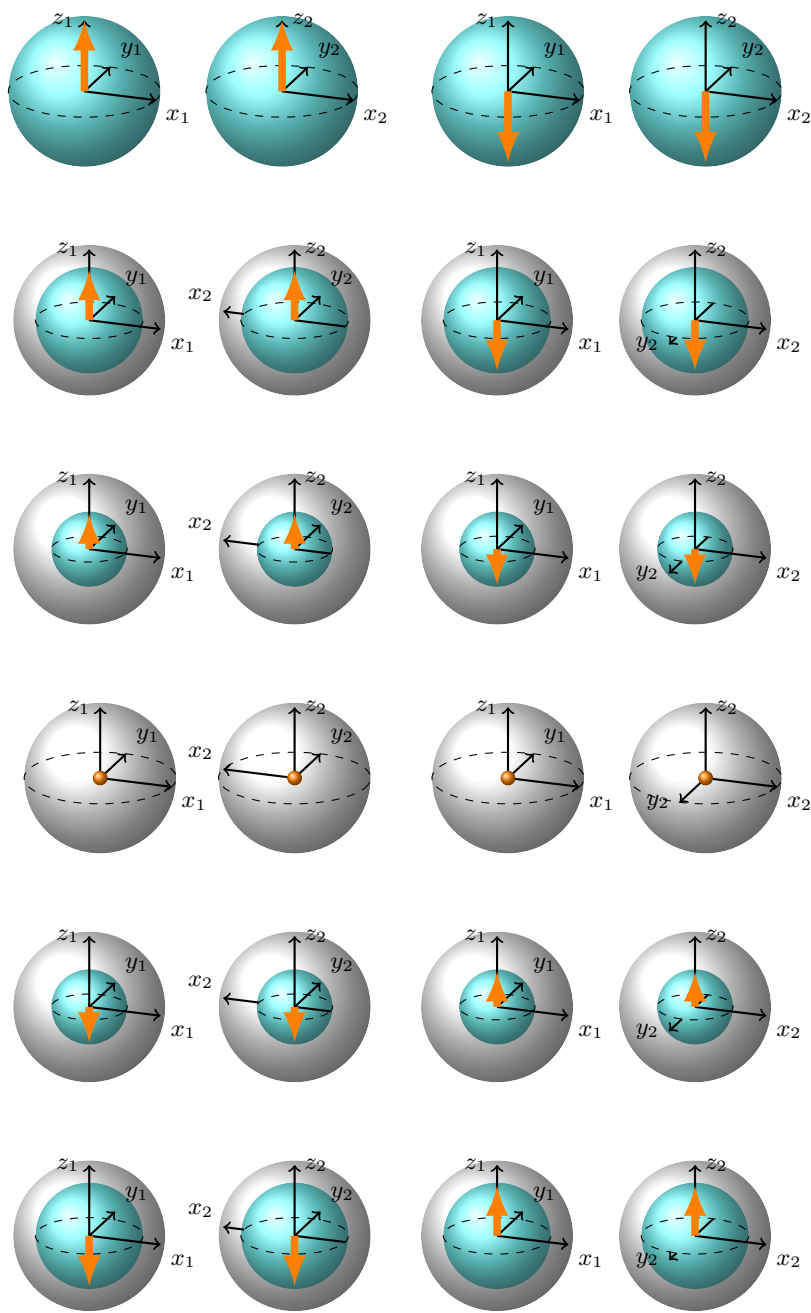

So far we have depicted $\pi/2$ rotations in all of the planes. What about arbitrary angle rotations? Being able to generalize the representation to rotations by arbitrary angles we would cover all the two-qubit Hilbert space and, hence, complete the Two Bloch Sphere Representation of Pure Two Qubit States. In this section we develop a method to represent Partially Entangled States (PES) shown in Figure \ref{fig:partialent}. Furthermore, we discuss measures of entanglement and relate them to the graphical representation. Finally we show dynamics of a full rotation in a double-Pauli plane $y_1 \land x_2$ that goes through Separable, Partially Entangled and Maximally entangled states in Figure \ref{fig:ent-full-rot}.

Generalization to arbitrary angles is rather straightforward for local rotations. We just rotate the relevant coordinate axes of the wedge product by the relevant angle. This generalization works for local rotations of both separable and entangled states. Let us pause here and discuss alternative representations of a single qubit that is half way between $\ket{\uparrow}$ and $\ket{\leftarrow}$ that is $\ket{k} = \frac{\sqrt{2+\sqrt{2}}}{2}\ket{\uparrow} - \frac{\sqrt{2-\sqrt{2}}}{2}\ket{\downarrow}$. 

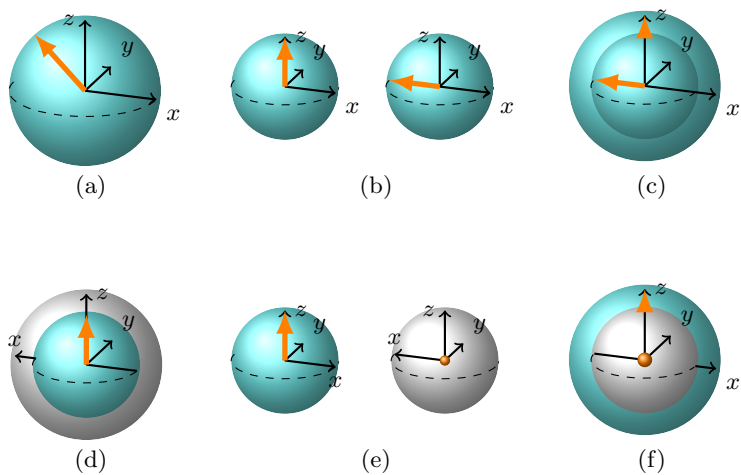
\begin{figure}
\centering
\begin{subfigure}{.2\textwidth}
\begin{minipage}[b][2cm][t]{0.2\textwidth}
\centering
\tdplotsetmaincoords{70}{20}
\begin{tikzpicture}[tdplot_main_coords, scale=1]
    \tdplotdrawarc[thin, dashed, scale=1]{(0,0,0)}{1}{-3}{177}{}{}
    \shade[ball color=cyan!50, opacity=0.6] (0,0,0) circle (1cm);
    
    \draw[thick,->, scale=1] (0,0,0) -- (1,0,0) node[anchor=north west, font=\small]{$x$};
    \draw[thick,->, scale=1] (0,0,0) -- (0,1,0) node[anchor=south west, font=\small]{$y$};
    \draw[thick,->, scale=1] (0,0,0) -- (0,0,1) node[anchor=east, font=\small]{$z$};

    \tdplotdrawarc[thin, dashed, scale=1]{(0,0,0)}{1}{-180}{0}{}{}

    \coordinate (statevector) at (-0.707, 0, 0.707);
    
    \draw[orange, line width=0.7mm, -latex] (0,0,0) -- (statevector);

\end{tikzpicture}
\end{minipage}%
    \caption{}
\end{subfigure}
\hspace{0.3cm}
\begin{subfigure}{0.33\textwidth}
\begin{minipage}[b][2cm][c]{0.2\textwidth}
\centering
\tdplotsetmaincoords{70}{20}
\begin{tikzpicture}[tdplot_main_coords, scale=0.707]
    \tdplotdrawarc[thin, dashed, scale=1]{(0,0,0)}{1}{-3}{177}{}{}
    \shade[ball color=cyan!50, opacity=0.6] (0,0,0) circle (1cm);
    
    \draw[thick,->, scale=1] (0,0,0) -- (1,0,0) node[anchor=north west, font=\small]{$x$};
    \draw[thick,->, scale=1] (0,0,0) -- (0,1,0) node[anchor=south west, font=\small]{$y$};
    \draw[thick,->, scale=1] (0,0,0) -- (0,0,1) node[anchor=west, font=\small]{$z$};

    \tdplotdrawarc[thin, dashed, scale=1]{(0,0,0)}{1}{-180}{0}{}{}

    \coordinate (statevector) at (0, 0, 1);
    
    \draw[orange, line width=0.7mm, -latex] (0,0,0) -- (statevector);

\end{tikzpicture}
\end{minipage}
\hspace{1cm}
\begin{minipage}[b][2cm][c]{.2\textwidth}
\centering
\tdplotsetmaincoords{70}{20}
\begin{tikzpicture}[tdplot_main_coords, scale=0.707]
    \tdplotdrawarc[thin, dashed, scale=1]{(0,0,0)}{1}{-3}{177}{}{}
    \shade[ball color=cyan!50, opacity=0.6] (0,0,0) circle (1cm);
    
    \draw[thick,->, scale=1] (0,0,0) -- (1,0,0) node[anchor=north west, font=\small]{$x$};
    \draw[thick,->, scale=1] (0,0,0) -- (0,1,0) node[anchor=south west, font=\small]{$y$};
    \draw[thick,->, scale=1] (0,0,0) -- (0,0,1) node[anchor=east, font=\small]{$z$};

    \tdplotdrawarc[thin, dashed, scale=1]{(0,0,0)}{1}{-180}{0}{}{}

    \coordinate (statevector) at (-1, 0, 0);
    
    \draw[orange, line width=0.7mm, -latex] (0,0,0) -- (statevector);

\end{tikzpicture}
\end{minipage}%
    \caption{}
\end{subfigure}
\hspace{0.3cm}
\begin{subfigure}{0.18\textwidth}
\begin{minipage}[b][2cm][c]{0.2\textwidth}
\tdplotsetmaincoords{70}{20}
\begin{tikzpicture}[tdplot_main_coords, scale=1]
    \tdplotdrawarc[thin, dashed, scale=0.707]{(0,0,0)}{1}{-3}{177}{}{}

    \shade[ball color=cyan!50, opacity=0.4] (0,0,0) circle (1cm);
    \shade[ball color=cyan!50, opacity=0.4] (0,0,0) circle (0.707cm);
    
    \draw[thick,->, scale=1] (0,0,0) -- (1,0,0) node[anchor=north west, font=\small]{$x$};
    \draw[thick,->, scale=1] (0,0,0) -- (0,1,0) node[anchor=south west, font=\small]{$y$};
    \draw[thick,->, scale=1] (0,0,0) -- (0,0,1) node[anchor=west, font=\small]{$z$};
    \draw[orange, line width=0.7mm, -latex, scale=1] (0,0,0) -- (-0.707,0,0);
    \draw[orange, line width=0.7mm, -latex, scale=1] (0,0,0.707) -- (0,0,1);

    \tdplotdrawarc[thin, dashed, scale=0.707]{(0,0,0)}{1}{-177}{-3}{}{}

\end{tikzpicture}
\end{minipage}
    \caption{}
\end{subfigure}

\vspace{1cm}
\begin{subfigure}{.2\textwidth}
\begin{minipage}[b][2cm][t]{0.2\textwidth}
\centering
\tdplotsetmaincoords{70}{20}
\begin{tikzpicture}[tdplot_main_coords, scale=1]
    \tdplotdrawarc[thin, dashed, scale=0.707]{(0,0,0)}{1}{-3}{177}{}{}

    \shade[ball color=white!50, opacity=0.4] (0,0,0) circle (1cm);
    \shade[ball color=cyan!50, opacity=0.4] (0,0,0) circle (0.707cm);
    
    \draw[thick,-, scale=1]  (0,0,0) -- (0.707,0,0);     
    \draw[thick,->, scale=1] (-0.707,0,0) -- (-1,0,0) node[anchor=south, font=\small]{$x$};
    \draw[thick,->, scale=1] (0,0,0) -- (0,1,0) node[anchor=south west, font=\small]{$y$};
    \draw[thick,->, scale=1] (0,0,0) -- (0,0,1) node[anchor=west, font=\small]{$z$};
    \draw[orange, line width=0.7mm, -latex, scale=1] (0,0,0) -- (0,0,0.707);

    \tdplotdrawarc[thin, dashed, scale=0.707]{(0,0,0)}{1}{-180}{0}{}{}
\end{tikzpicture}
\end{minipage}%
    \caption{}
\end{subfigure}
\hspace{0.3cm}
\begin{subfigure}{0.33\textwidth}
\begin{minipage}[b][2cm][c]{0.2\textwidth}
\centering
\tdplotsetmaincoords{70}{20}
\begin{tikzpicture}[tdplot_main_coords, scale=0.707]
    \tdplotdrawarc[thin, dashed, scale=1]{(0,0,0)}{1}{-3}{177}{}{}
    \shade[ball color=cyan!50, opacity=0.6] (0,0,0) circle (1cm);
    
    \draw[thick,->, scale=1] (0,0,0) -- (1,0,0) node[anchor=north, font=\small]{$x$};
    \draw[thick,->, scale=1] (0,0,0) -- (0,1,0) node[anchor=south west, font=\small]{$y$};
    \draw[thick,->, scale=1] (0,0,0) -- (0,0,1) node[anchor=west, font=\small]{$z$};
    
    \tdplotdrawarc[thin, dashed, scale=1]{(0,0,0)}{1}{-180}{0}{}{}

    \coordinate (statevector) at (0, 0, 1);
    
    \draw[orange, line width=0.7mm, -latex] (0,0,0) -- (statevector);

\end{tikzpicture}
\end{minipage}
\hspace{1cm}
\begin{minipage}[b][2cm][c]{.2\textwidth}
\centering
\tdplotsetmaincoords{70}{20}
\begin{tikzpicture}[tdplot_main_coords, scale=0.707]
    \tdplotdrawarc[thin, dashed, scale=1]{(0,0,0)}{1}{-3}{177}{}{}
    \shade[ball color=white!50, opacity=0.6] (0,0,0) circle (1cm);
    
    \draw[thick,->, scale=1] (0,0,0) -- (-1,0,0) node[anchor=south, font=\small]{$x$};
    \draw[thick,->, scale=1] (0,0,0) -- (0,1,0) node[anchor=south west, font=\small]{$y$};
    \draw[thick,->, scale=1] (0,0,0) -- (0,0,1) node[anchor=east, font=\small]{$z$};
    
    \tdplotdrawarc[thin, dashed, scale=1]{(0,0,0)}{1}{-180}{0}{}{}
    \shade[ball color=orange, opacity=1, scale=2] (0,0,0) circle (0.05cm);

\end{tikzpicture}
\end{minipage}%
    \caption{}
\end{subfigure}
\hspace{0.3cm}
\begin{subfigure}{0.18\textwidth}
\begin{minipage}[b][2cm][c]{0.2\textwidth}
\tdplotsetmaincoords{70}{20}
\begin{tikzpicture}[tdplot_main_coords, scale=1]
    \tdplotdrawarc[thin, dashed, scale=0.707]{(0,0,0)}{1}{-3}{177}{}{}

    \shade[ball color=cyan!50, opacity=0.4] (0,0,0) circle (1cm);
    \shade[ball color=white!50, opacity=0.4] (0,0,0) circle (0.707cm);
    
    \draw[thick,-, scale=1]  (0,0,0) -- (-0.707,0,0);     
    \draw[thick,->, scale=1] (0.707,0,0) -- (1,0,0) node[anchor=north west, font=\small]{$x$};
    \draw[thick,->, scale=1] (0,0,0) -- (0,1,0) node[anchor=south west, font=\small]{$y$};
    \draw[thick,->, scale=1] (0,0,0) -- (0,0,1) node[anchor=west, font=\small]{$z$};
    \draw[orange, line width=0.7mm, -latex, scale=1] (0,0,0.707) -- (0,0,1);

    \tdplotdrawarc[thin, dashed, scale=0.707]{(0,0,0)}{1}{-177}{-3}{}{}

    \shade[ball color=orange, opacity=1, scale=2] (0,0,0) circle (0.05cm);

\end{tikzpicture}
\end{minipage}
    \caption{}
\end{subfigure}
    \caption{Alternative representations of a single qubit from the two qubit Stabilizer state rotated by $\pi/4$ angle in local plane - (a) through (c); double Pauli plane - (d) through (f).}
    \label{fig:analogy}
\end{figure}

The three possible representations are shown in Figure \ref{fig:analogy} (a) through (c). In (a) $\ket{k}$ is shown in the usual way: as a statevector arrow tilted by $\pi/4$ angle. In (b) - as a combination of two BS with statevectors pointing along $z$ and $-x$ directions. The lengths of the statevectors are $\sqrt{2}/2$. Note that the combined surface area of these BS is equal to the surface area of the original BS. (c) is a way to merge the two BS from (b) in such a way that the size of the resulting BS is the same as in (a) and the statevectors are kept on the coordinate axes. 

In this case the length of the outer statevector arrow aligned with $z$ axis in (c) doesn't seem to match its length inside a smaller BS in (b). Nevertheless, if we define surface area enclosing the vector as surface area at the tip of the vector minus surface area at the origin of the vector, this quantity remains equal to that in (b). In other words, in all three representations (a), (b) and (c) the surface area enclosing the statevector is the same. One can see that this works for less trivial case when there are projections of the statevector on all three coordinate axes. We will see usefulness of such attention to the surface area when we speak about entanglement measures later in this chapter.

Similar approach could be taken for double-Pauli rotations by arbitrary angle. Partially entangled states, for example, could be represented as a separable state rotated in double-Pauli plane by a less than $\pi/2$ angle. Hence we can take two representations - one for separable state, another for entangled state - and combine them using the angle of rotation for scaling. 

This is done in Figure \ref{fig:analogy} (d) - (f) where three alternative representations of a single qubit from a partially entangled pair are shown. We take two BS with squared radii summing to one (e), or equivalently, with surface areas adding up to that of a standard BS. Then, we combine them analogously to (c). There are multiple methods to combine them, but the two most straightforward approaches are depicted in (d) and (f). They differ in which of the two BS is situated inside the compound BS and which touches the surface. Note, that the size of the Spheres in (e) need not be the same and will generally vary depending on how far the state is from being separable.

Now let us use this technique to represent the partially entangled state $\ket{P} = (0.924 \ket{\uparrow\uparrow} - 0.383 \ket{\downarrow\downarrow}) = \frac{\sqrt{2+\sqrt{2}}}{2}\ket{\uparrow\uparrow} - \frac{\sqrt{2-\sqrt{2}}}{2}\ket{\downarrow\downarrow}$. It is generated by applying $\pi/4$ rotation in the plane $y_1 \land x_2$ to the state $\ket{\uparrow\uparrow}$. Given that same rotation by twice the angle ($\pi/2$) results in the state $\ket{\uparrow\uparrow} - \ket{\downarrow\downarrow}$ we will say that $\ket{P}$ is half-way between entanglement and separability. 

Figure \ref{fig:partialent} provides a representation of $\ket{P}$. The separable part is represented by the inner sphere enclosing the statevector arrow and the entangled part is represented by the outer shell lacking the statevector arrow. Note that although it might not be obvious from the picture, the ends of coordinate axes reach the surface of the outer shell. To represent entanglement we need the outer shell of the second BS to have inverted coordinates. In this particular case $x_2$ should be inverted. Therefore in the Figure \ref{fig:partialent} $x_2$ axis reaches the edge of the inner shell in usual right-handed way and then becomes inverted and reaches the edge of the outer shell in the left-handed way. In the picture we don't draw in negative parts of the axes, but logically negative part of the $x_2$ axis also inverts in the same way.

This representation maintains a connection with standard purity measures, such as $Tr(\rho^2)$, serving as a separability measure in our context. $\rho$ in this case becomes the reduced density matrix $\rho_A$ of one of the particles of the pair; $Tr(\rho_A^2) = Tr(\rho_B^2)$.  Additionally, the representation hints at a relation between the separability/entanglement measure and the surface area of the Bloch Sphere section enfolding separable/entangled state. The conventional purity measure, the Trace of the Square of the density matrix, is related to the square of the statevector arrow $r^2$:

\begin{equation}
\centering
    Tr(\rho^2) = \frac{1 + r^2}{2}
\end{equation}

$r^2$ is proportional to the surface area of the sub-sphere containing the statevector arrow. Consequently, $Tr(\rho_A^2)$ as a separability measure is directly related to the surface area of the sub-sphere representing separable part of the PES. A measure of the entanglement in a given state therefore may be obtained by subtracting the surface area of the inner sub-sphere from the surface area of the whole BS. Note that this is the ``surface area enclosing the state" we've defined discussing Figure \ref{fig:analogy}. Hence we can say that entanglement in a PES is proportional to the surface area of the sub-sphere enclosing the entangled part of the PES. Or equivalently, proportional to the surface area of the BS without a statevector in Figure \ref{fig:analogy} (e) which represents the entangled part of the PES.

Other measure of entanglement, Concurrence \cite{Hill1997, Rungta2001, Bhaskara2017} may be related to the graphical representation too. Being proportional to square root of separability measure $Tr(\rho_A^2)$ for pure states

\begin{equation}
\centering
    C = \sqrt{2(1-Tr(\rho^2)} = \sqrt{2(1-\frac{1+r^2}{2})} = \sqrt{1-r^2} = \sqrt{\Tilde{r}^2} = \Tilde{r}
\end{equation}

Concurrence may be associated with $\Tilde{r}$ which is the radius of the Bloch sphere that represents entangled part of the partially entangled state, i.e. BS without a statevector in Figure \ref{fig:analogy} (e). Hence Concurrence for a Partially entangled state $\ket{P}$ represented in Figure \ref{fig:partialent} that we called being exactly half way between separable and entangled state is $\sqrt{2}/2$, not $1/2$. But given graphical interpretation we understand why. In this case $r=\Tilde{r}$ nevertheless, meaning that the Bloch Spheres representing separable and entangled part are of equal size. And $\sqrt{2}/2$ is an artifact of the fact that when we rotate a vector by $\pi/4$, its projection doesn't halve. Moreover, we understand that one can relate Concurrence and Separability through length of radius and surface area. Concurrence being length of radius of the BS representing entangled part of the PES and Separability being proportional the surface area of the BS representing the separable part of the PES.

In fact, being able to represent arbitrary separable state, arbitrary MES and one PES is sufficient to be able to represent arbitrary PES. We can make Schmidt decomposition of the PES arriving at $\ket{P} = \alpha\ket{a_0}\ket{b_0} + \beta\ket{a_1}\ket{b_1}$. This makes obvious between which separable and MES states the PES lies: $\ket{a_0}\ket{b_0}$ and $\ket{a_0}\ket{b_0} + \ket{a_1}\ket{b_1}$. We just take the representations of those two states, which we know, scale and combine them accordingly. This observation implies that rules and representations shown in this work are sufficient to create representations of arbitrary pure two-qubit state. Having representations of arbitrary states we can simulate the action of a unitary on a given state by creating representations of all the intermediate states on the trajectory of the action of the unitary. Moreover, having representation of a single separable state, single MES and being able to redefine the coordinate axes is enough to create representation of an arbitrary pure state too. 

As a side note, having the rules, being able to redefine the coordinates and perform Cartan decomposition should be enough to perform arbitrary rotations on arbitrary state. Redefinition of coordinates can always give us stabilizer state. Even for a PES, which may be disassembled into a separable and a ME parts and redefinition of coordinates for each of those parts gives the stabilizers. In the newly redefined basis we can always use Cartan decomposition to reduce any unitary to a combination of rotations in some of the 15 planes. Then we can apply the rules.

Now let us examine the Figure \ref{fig:ent-full-rot} where a full rotation of the state $\ket{\uparrow\uparrow}$ in the plane $y_1 \land x_2$ is represented. The Figure should be read column-by-column up to down. The rotation transitions through the states $\ket{\uparrow\uparrow}$, $\ket{\uparrow\uparrow} - \ket{\downarrow\downarrow}$, $\ket{\downarrow\downarrow}$, $\ket{\uparrow\uparrow} + \ket{\downarrow\downarrow}$, back to $\ket{\uparrow\uparrow}$. 

During this double-Pauli rotation, the statevector arrows are drawn into the center of the Bloch Spheres before flipping to the opposite side. Additionally, the coordinate axes undergo inversion: first, the $x_2$ axis flips, followed by a reversal, and then the $y_2$ axis undergoes a similar inversion. These flips align with the previously discussed rules, where we utilized the flexibility in representing separable state coordinate axes to enhance the visual dynamics. Note how double-Pauli rotations are composed of inversions rather than rotations. And how such representation utilizes the whole volume of Bloch Spheres to represent states and rotations.

\FloatBarrier
\section{Discussion}

The model presented here has a lot of structure which we have only started to explore. We see this paper as an invitation for the members of community to join the exploration. We've decided to organize this section in the form of Q\&A to highlight some questions that might arise regarding the model. 

\textit{Although in the previous section it is shown how a representation of an arbitrary pure two-qubit state and, in some sense, arbitrary unitary may be obtained if one is able to redefine the coordinates, exploring manual rule application alone in different contexts may also be valuable.}

\textit{Q1. Does the model work only for 15 rotations or for any arbitrary rotation?}

As we mentioned at the end of Chapter 2 any arbitrary two-qubit unitary may be expressed as a product of some of the 15 rotations using Cartan decomposition \cite{HavelDoran2004, Khaneja2000, Khaneja2001, Zhang2003}. In fact, not all of the 15 planes are necessary and lesser number suffices. 

\textit{Q2. How does one rotate when the initial set is not from a stabilizer set, i.e. if the statevector arrows don't align with coordinate axes?}

Decompose the state into pairs of BS of smaller size similar to how it is done in \ref{fig:analogy} so that the statevector arrows align with the coordinate axes, perform rotations using rules, recompose the Spheres. We have tried it on some simple examples and it works. Second option: redefine coordinates so that the statevector arrows are aligned with the coordinate axes, transform the rotation planes to conform to new description, use the rules as usual.

\textit{Q3. What about PES states - do rules still hold?}

PES, just like non-stabilizer states in the question above may be disassembled into pairs of smaller BS which can be rotated using the rules and then reassembled. Again, we have tried a couple of simpler cases and they work. Even for the case when a double-Pauli rotation applied to a PES doesn't lead to MES or separable state, but moves towards another PES. That happens because application of the same rule to MES part of the PES reduces its entanglement, but when applied on the separable part of the PES - increases its entanglement. As a result we move from PES to PES.

\textit{Q4. Can we do some kind of visual calculations using the representations and rules?}

Some visual calculations might be done using the representations. We can estimate the amount of entanglement in a PES; guess the eigenrotations and stabilizers of the state, tell what planes of rotation/Pauli matrices are needed to entangle/disentangle the state. It would be great if someone found a way to connect representation in a direct and straightforward way to its numeric representation in the vector or density matrix form. 

\textit{Q5. How can we be sure that these rules work? They are just stated as fact.}

We have developed the rules for moving between separable and ME stabilizer states after we have obtained the representations of stabilizer MES. In principle one does not need the rules to create representation of an arbitrary state - representation of single separable and single ME state combined with all the possible basis changes are sufficient to create representations. But these rules seem to obey the structure of the states in question. Every rotation from separable to ME state involves coordinate handedness change in one of the BS - rules reflect that; they also reflect the dynamics of statevector arrows being pulled into and outside the center of BS. Those are two main features that any set of rules for given representation must incorporate and rules listed in Chapter 4 do so. But we do agree that tying those rules to some fundamental considerations would be important, especially for the possible extension of the model to more qubits where the number of state combinations grows beyond what is possible to check manually. 

\textit{Q6. Are these representations and rules unique?}

No. There are alternative equivalent representations. We have chosen a convention in which the coordinate axes of the first BS are always fixed for the ease of manual manipulation of the representations. But MES representations are equivalent up to simultaneous rotation of both BS that doesn't change the relative axis orientation. Separable states are defined up to an arbitrary rotation of each BS axes around the statevector arrow. In fact, if we were to represent separable states as faithfully as possible we shouldn't draw the two BS axes perpendicular to the statevector at all. Instead we should only depict the plane defined by those axes. This ambiguity is actually very important for the functioning of the rules.

As a result when we combine the separable and MES BS representations to create a PES representation there is a lot of freedom in the way they get combined. In Figure \ref{fig:ent-full-rot} we have chosen the representation so that the inner and outer BS seem to have a common set of coordinate axes, but it need not be the case. Generally the Spheres may be merged in ways that the description of dynamics would change - different axes would invert for example. So the rules and dynamics of the process depend on the convention one chooses to represent the states. What does not change in the description of the process of entanglement are the BS dynamics of handedness and statevector arrows mentioned in previous question. But even here there is some flexibility and one could create a model where during a full double-Pauli rotation the handedness of both BS coordinates would change from Right-Right to RL, to LL, to LR back to RR.

Considering alternative representations we think it would be an interesting challenge to create a representation that would combine the different rules of action of double-Pauli planes on separable states into one.

\section{Conclusion}

We have shown a way to graphically represent two-qubit pure states and unitaries using two Bloch Spheres. This approach applies the concept of Plane of Rotation from Geometric Algebra to the context of two Bloch Spheres. Separable states are represented as a pair of Bloch Spheres with statevector arrows; coordinate axes handedness is the same; the pair of coordinates of the plane perpendicular to the statevector are defined up to an arbitrary rotation of the pair.  Maximally entangled states are represented as a pair of Bloch Spheres with a dot in the middle instead of statevector arrows; the handedness of the coordinate axes of the Bloch Spheres is opposite; relative directions of coordinate axes of two BS determine the exact maximally entangled state the BS pair represents. 

We have provided graphical representations of separable and maximally entangled states. We used a method from a paper by Havel and Doran \cite{HavelDoran2004} to translate 15 matrix generators of rotation into 15 planes of rotation. With this, we derived rules for performing graphical rotations in each of these planes by a $\pi/2$ angle. Using these rules we can navigate the set of two-qubit stabilizer states. Next we have shown through an example of CNOT gate how quantum gates may be graphically implemented on two Bloch Spheres. Furthermore, we have generalized the representations to include arbitrary angle rotations in both local and double-Pauli planes and represent any pure two-qubit state. This is done by interpolation between representations of two relevant stabilizer states by combining and properly scaling them.

While the model technically permits the representation of arbitrary pure two-qubit states and unitaries, the level of effort and computation varies depending on the specific representation required. Rotations of any stabilizer state in any of the 15 planes by any angle can be visually performed without the need to compute Pauli Matrices for rotation, offering a straightforward approach. However, for now, rotations of arbitrary states in arbitrary planes by arbitrary angles typically demand a higher level of sophistication. Alongside rule application, tasks may include coordinate redefinition, Cartan decomposition, Schmidt decomposition, and decomposition of BS into multiple smaller BS.

When working on this model we were guided by the symmetries of the uncovered representations. Some ambiguities in representation eventually turned out to be important for the functioning of the model. Some ambiguities were resolved by choosing a suitable and consistent convention. Our main criterion for selecting the convention was its ease of manual manipulation of the representations, without relying on simulation. Those developing a computational application of this model might choose different conventions to emphasize different aspects of two-qubit dynamics. Indeed we would be very happy to see a simulation developed for this model.

Representation provided in this work allows for more intuitive manipulation of quantum states - one that does not even require computation of rotation matrices. Furthermore, it allows to see the properties of the state, like eigenmatrices of its rotation or even degree of entanglement. This model utilizes the entire volume of the Bloch Sphere for pure two-qubit state representation. Simultaneously, if one examines the representation of a single qubit from an entangled pair, it appears as a maximally mixed single qubit state.

Three important takeaways for the foundations of QM in our opinion are: a) the idea that double-Pauli entangling unitaries are rotations in the plane formed by axes of different Bloch Spheres like $z_1 \land x_2$ and that such a rotation leads to a change of handedness in coordiante axes of one of the BS, hence b) entanglement is linked to an act of inverting BS inside-out and c) to relations of handedness of particles' internal degrees of freedom or particles' coordinate frames.

\newpage

\section{Acknowledgements}

We would like to acknowledge the help of Quantum Computing Stack Exchange users Adam Zalcman in understanding more exotic maximally entangled states; Rammus and Norbert Schuch for pointing out that Schmidt Decomposition of a PES allows one to find separable and MES state between which the PES lies; 3blue1brown visualizations of 4D rotations; Vaxjo 2023 conference where this work has been first presented and especially Andrei Khrennikov and John Small; Navin Khaneja; Francesco Buscemi and members of his group at the University of Nagoya; anonymous Reviewers.



\newpage



\appendix


\begin{figure}
\begin{subfigure}{.5\textwidth}
  \centering
\begin{minipage}[b][2cm][t]{0.4\textwidth}
\centering
\tdplotsetmaincoords{70}{110}
\begin{tikzpicture}[tdplot_main_coords, scale=0.75]
    \shade[ball color=white!50, opacity=0.6] (0,0,0) circle (1cm);
    
    \draw[thick,->, scale=1] (0,0,0) -- (1.2,0,0) node[anchor=north east, font=\small]{$x_1$};
    \draw[thick,->, scale=1] (0,0,0) -- (0,1.2,0) node[anchor=north, font=\small]{$y_1$};
    \draw[thick,->, scale=1] (0,0,0) -- (0,0,1.2) node[anchor=east, font=\small]{$z_1$};
    
    
    \tdplotdrawarc[thin, dashed, scale=1]{(0,0,0)}{1}{-4}{356}{}{}
    
    \shade[ball color=orange, opacity=1, scale=2] (0,0,0) circle (0.05cm);
\end{tikzpicture}
\end{minipage}%
\begin{minipage}[b][1.75cm][t]{0.4\textwidth}
\centering
\tdplotsetmaincoords{70}{110}
\begin{tikzpicture}[tdplot_main_coords, scale=0.75]
    \shade[ball color=white!50, opacity=0.6] (0,0,0) circle (1cm);
    
    \draw[thick,->, scale=1] (0,0,0) -- (-1.2,0,0) node[anchor=south west, font=\small]{$x_2$};
    \draw[thick,->, scale=1] (0,0,0) -- (0,-1.2,0) node[anchor=south, font=\small]{$y_2$};
    \draw[thick,->, scale=1] (0,0,0) -- (0,0,-1.2) node[anchor=west, font=\small]{$z_2$};
    
    
    \tdplotdrawarc[thin, dashed, scale=1]{(0,0,0)}{1}{-4}{356}{}{}
    
    \shade[ball color=orange, opacity=1, scale=2] (0,0,0) circle (0.05cm);
\end{tikzpicture}
\end{minipage}

  \caption{$\Psi^- = \ket{\uparrow\downarrow - \downarrow\uparrow}$}
  \label{figap1:sub-a}
\end{subfigure}
\begin{subfigure}{.5\textwidth}
  \centering
\begin{minipage}[b][2cm][t]{0.4\textwidth}
\centering
\tdplotsetmaincoords{70}{110}
\begin{tikzpicture}[tdplot_main_coords, scale=0.75]
    \shade[ball color=white!50, opacity=0.6] (0,0,0) circle (1cm);
    
    \draw[thick,->, scale=1] (0,0,0) -- (1.2,0,0) node[anchor=north east, font=\small]{$x_1$};
    \draw[thick,->, scale=1] (0,0,0) -- (0,1.2,0) node[anchor=north, font=\small]{$y_1$};
    \draw[thick,->, scale=1] (0,0,0) -- (0,0,1.2) node[anchor=east, font=\small]{$z_1$};
    
    
    \tdplotdrawarc[thin, dashed, scale=1]{(0,0,0)}{1}{-4}{356}{}{}
    
    \shade[ball color=orange, opacity=1, scale=2] (0,0,0) circle (0.05cm);
\end{tikzpicture}
\end{minipage}
\begin{minipage}[b][1.75cm][t]{0.4\textwidth}
\centering
\tdplotsetmaincoords{70}{110}
\begin{tikzpicture}[tdplot_main_coords, scale=0.75]
    \shade[ball color=white!50, opacity=0.6] (0,0,0) circle (1cm);
    
    \draw[thick,->, scale=1] (0,0,0) -- (1.2,0,0) node[anchor=north east, font=\small]{$x_2$};
    \draw[thick,->, scale=1] (0,0,0) -- (0,1.2,0) node[anchor=north, font=\small]{$y_2$};
    \draw[thick,->, scale=1] (0,0,0) -- (0,0,-1.2) node[anchor=west, font=\small]{$z_2$};
    
    
    \tdplotdrawarc[thin, dashed, scale=1]{(0,0,0)}{1}{-4}{356}{}{}
    
    \shade[ball color=orange, opacity=1, scale=2] (0,0,0) circle (0.05cm);
\end{tikzpicture}
\end{minipage}

  \caption{$\Psi^+ = \ket{\uparrow\downarrow + \downarrow\uparrow}$}
  \label{figap1:sub-b}
\end{subfigure}

\vspace{0.5cm}

\begin{subfigure}{.5\textwidth}
  \centering
\begin{minipage}[b][2cm][t]{0.4\textwidth}
\centering
\tdplotsetmaincoords{70}{110}
\begin{tikzpicture}[tdplot_main_coords, scale=0.75]
    \shade[ball color=white!50, opacity=0.6] (0,0,0) circle (1cm);
    
    \draw[thick,->, scale=1] (0,0,0) -- (1.2,0,0) node[anchor=north east, font=\small]{$x_1$};
    \draw[thick,->, scale=1] (0,0,0) -- (0,1.2,0) node[anchor=north, font=\small]{$y_1$};
    \draw[thick,->, scale=1] (0,0,0) -- (0,0,1.2) node[anchor=east, font=\small]{$z_1$};
    
    
    \tdplotdrawarc[thin, dashed, scale=1]{(0,0,0)}{1}{-4}{356}{}{}
    
    \shade[ball color=orange, opacity=1, scale=2] (0,0,0) circle (0.05cm);
\end{tikzpicture}
\end{minipage}%
\begin{minipage}[b][1.75cm][t]{0.4\textwidth}
\centering
\tdplotsetmaincoords{70}{110}
\begin{tikzpicture}[tdplot_main_coords, scale=0.75]
    \shade[ball color=white!50, opacity=0.6] (0,0,0) circle (1cm);
    
    \draw[thick,->, scale=1] (0,0,0) -- (-1.2,0,0) node[anchor=south west, font=\small]{$y_2$};
    \draw[thick,->, scale=1] (0,0,0) -- (0,-1.2,0) node[anchor=south, font=\small]{$z_2$};
    \draw[thick,->, scale=1] (0,0,0) -- (0,0,-1.2) node[anchor=west, font=\small]{$x_2$};
    
    
    \tdplotdrawarc[thin, dashed, scale=1]{(0,0,0)}{1}{-4}{356}{}{}
    
    \shade[ball color=orange, opacity=1, scale=2] (0,0,0) circle (0.05cm);
\end{tikzpicture}
\end{minipage}

\caption{$\ket{\uparrow\uparrow - \uparrow\downarrow - i\downarrow\uparrow - i\downarrow\downarrow}$}
  \label{figap1:sub-c}
\end{subfigure}
\begin{subfigure}{.5\textwidth}
  \centering
\begin{minipage}[b][2cm][t]{0.4\textwidth}
\centering
\tdplotsetmaincoords{70}{110}
\begin{tikzpicture}[tdplot_main_coords, scale=0.75]
    \shade[ball color=white!50, opacity=0.6] (0,0,0) circle (1cm);
    
    \draw[thick,->, scale=1] (0,0,0) -- (1.2,0,0) node[anchor=north east, font=\small]{$x_1$};
    \draw[thick,->, scale=1] (0,0,0) -- (0,1.2,0) node[anchor=north, font=\small]{$y_1$};
    \draw[thick,->, scale=1] (0,0,0) -- (0,0,1.2) node[anchor=east, font=\small]{$z_1$};
    
    
    \tdplotdrawarc[thin, dashed, scale=1]{(0,0,0)}{1}{-4}{356}{}{}
    
    \shade[ball color=orange, opacity=1, scale=2] (0,0,0) circle (0.05cm);
\end{tikzpicture}
\end{minipage}
\begin{minipage}[b][1.75cm][t]{0.4\textwidth}
\centering
\tdplotsetmaincoords{70}{110}
\begin{tikzpicture}[tdplot_main_coords, scale=0.75]
    \shade[ball color=white!50, opacity=0.6] (0,0,0) circle (1cm);
    
    \draw[thick,->, scale=1] (0,0,0) -- (1.2,0,0) node[anchor=north east, font=\small]{$y_2$};
    \draw[thick,->, scale=1] (0,0,0) -- (0,1.2,0) node[anchor=north, font=\small]{$z_2$};
    \draw[thick,->, scale=1] (0,0,0) -- (0,0,-1.2) node[anchor=west, font=\small]{$x_2$};
    
    
    \tdplotdrawarc[thin, dashed, scale=1]{(0,0,0)}{1}{-4}{356}{}{}
    
    \shade[ball color=orange, opacity=1, scale=2] (0,0,0) circle (0.05cm);
\end{tikzpicture}
\end{minipage}
\caption{$\ket{\uparrow\uparrow - \uparrow\downarrow + i\downarrow\uparrow + i\downarrow\downarrow}$}
  \label{figap1:sub-d}
\end{subfigure}

\vspace{0.5cm}

\begin{subfigure}{.5\textwidth}
  \centering
\begin{minipage}[b][2cm][t]{0.4\textwidth}
\centering
\tdplotsetmaincoords{70}{110}
\begin{tikzpicture}[tdplot_main_coords, scale=0.75]
    \shade[ball color=white!50, opacity=0.6] (0,0,0) circle (1cm);
    
    \draw[thick,->, scale=1] (0,0,0) -- (1.2,0,0) node[anchor=north east, font=\small]{$x_1$};
    \draw[thick,->, scale=1] (0,0,0) -- (0,1.2,0) node[anchor=north, font=\small]{$y_1$};
    \draw[thick,->, scale=1] (0,0,0) -- (0,0,1.2) node[anchor=east, font=\small]{$z_1$};
    
    
    \tdplotdrawarc[thin, dashed, scale=1]{(0,0,0)}{1}{-4}{356}{}{}
    
    \shade[ball color=orange, opacity=1, scale=2] (0,0,0) circle (0.05cm);
\end{tikzpicture}
\end{minipage}%
\begin{minipage}[b][1.75cm][t]{0.4\textwidth}
\centering
\tdplotsetmaincoords{70}{110}
\begin{tikzpicture}[tdplot_main_coords, scale=0.75]
    \shade[ball color=white!50, opacity=0.6] (0,0,0) circle (1cm);
    
    \draw[thick,->, scale=1] (0,0,0) -- (-1.2,0,0) node[anchor=south west, font=\small]{$z_2$};
    \draw[thick,->, scale=1] (0,0,0) -- (0,-1.2,0) node[anchor=south, font=\small]{$x_2$};
    \draw[thick,->, scale=1] (0,0,0) -- (0,0,-1.2) node[anchor=west, font=\small]{$y_2$};
    
    
    \tdplotdrawarc[thin, dashed, scale=1]{(0,0,0)}{1}{-4}{356}{}{}
    
    \shade[ball color=orange, opacity=1, scale=2] (0,0,0) circle (0.05cm);
\end{tikzpicture}
\end{minipage}

\caption{$\ket{\uparrow\uparrow - i\uparrow\downarrow - \downarrow\uparrow - i\downarrow\downarrow}$}
  \label{figap1:sub-e}
\end{subfigure}
\begin{subfigure}{.5\textwidth}
  \centering
\begin{minipage}[b][2cm][t]{0.4\textwidth}
\centering
\tdplotsetmaincoords{70}{110}
\begin{tikzpicture}[tdplot_main_coords, scale=0.75]
    \shade[ball color=white!50, opacity=0.6] (0,0,0) circle (1cm);
    
    \draw[thick,->, scale=1] (0,0,0) -- (1.2,0,0) node[anchor=north east, font=\small]{$x_1$};
    \draw[thick,->, scale=1] (0,0,0) -- (0,1.2,0) node[anchor=north, font=\small]{$y_1$};
    \draw[thick,->, scale=1] (0,0,0) -- (0,0,1.2) node[anchor=east, font=\small]{$z_1$};
    
    
    \tdplotdrawarc[thin, dashed, scale=1]{(0,0,0)}{1}{-4}{356}{}{}
    
    \shade[ball color=orange, opacity=1, scale=2] (0,0,0) circle (0.05cm);
\end{tikzpicture}
\end{minipage}
\begin{minipage}[b][1.75cm][t]{0.4\textwidth}
\centering
\tdplotsetmaincoords{70}{110}
\begin{tikzpicture}[tdplot_main_coords, scale=0.75]
    \shade[ball color=white!50, opacity=0.6] (0,0,0) circle (1cm);
    
    \draw[thick,->, scale=1] (0,0,0) -- (1.2,0,0) node[anchor=north east, font=\small]{$z_2$};
    \draw[thick,->, scale=1] (0,0,0) -- (0,1.2,0) node[anchor=north, font=\small]{$x_2$};
    \draw[thick,->, scale=1] (0,0,0) -- (0,0,-1.2) node[anchor=west, font=\small]{$y_2$};
    
    
    \tdplotdrawarc[thin, dashed, scale=1]{(0,0,0)}{1}{-4}{356}{}{}
    
    \shade[ball color=orange, opacity=1, scale=2] (0,0,0) circle (0.05cm);
\end{tikzpicture}
\end{minipage}

\caption{$\ket{\uparrow\uparrow - i\uparrow\downarrow + \downarrow\uparrow + i\downarrow\downarrow}$}
  \label{figap1:sub-f}
\end{subfigure}

\vspace{0.5cm}

\begin{subfigure}{.5\textwidth}
  \centering
\begin{minipage}[b][2cm][t]{0.4\textwidth}
\centering
\tdplotsetmaincoords{70}{110}
\begin{tikzpicture}[tdplot_main_coords, scale=0.75]
    \shade[ball color=white!50, opacity=0.6] (0,0,0) circle (1cm);
    
    \draw[thick,->, scale=1] (0,0,0) -- (1.2,0,0) node[anchor=north east, font=\small]{$x_1$};
    \draw[thick,->, scale=1] (0,0,0) -- (0,1.2,0) node[anchor=north, font=\small]{$y_1$};
    \draw[thick,->, scale=1] (0,0,0) -- (0,0,1.2) node[anchor=east, font=\small]{$z_1$};
    
    
    \tdplotdrawarc[thin, dashed, scale=1]{(0,0,0)}{1}{-4}{356}{}{}
    
    \shade[ball color=orange, opacity=1, scale=2] (0,0,0) circle (0.05cm);
\end{tikzpicture}
\end{minipage}%
\begin{minipage}[b][2cm][t]{0.4\textwidth}
\centering
\tdplotsetmaincoords{70}{110}
\begin{tikzpicture}[tdplot_main_coords, scale=0.75]
    \shade[ball color=white!50, opacity=0.6] (0,0,0) circle (1cm);
    
    \draw[thick,->, scale=1] (0,0,0) -- (-1.2,0,0) node[anchor=south west, font=\small]{$x_2$};
    \draw[thick,->, scale=1] (0,0,0) -- (0,1.2,0) node[anchor=north, font=\small]{$y_2$};
    \draw[thick,->, scale=1] (0,0,0) -- (0,0,1.2) node[anchor=east, font=\small]{$z_2$};
    
    
    \tdplotdrawarc[thin, dashed, scale=1]{(0,0,0)}{1}{-4}{356}{}{}
    
    \shade[ball color=orange, opacity=1, scale=2] (0,0,0) circle (0.05cm);
\end{tikzpicture}
\end{minipage}
  \caption{$\Phi^- = \ket{\uparrow\uparrow - \downarrow\downarrow}$}
  \label{figap1:sub-g}
\end{subfigure}
\begin{subfigure}{.5\textwidth}
  \centering
\begin{minipage}[b][2cm][t]{0.4\textwidth}
\centering
\tdplotsetmaincoords{70}{110}
\begin{tikzpicture}[tdplot_main_coords, scale=0.75]
    \shade[ball color=white!50, opacity=0.6] (0,0,0) circle (1cm);
    
    \draw[thick,->, scale=1] (0,0,0) -- (1.2,0,0) node[anchor=north east, font=\small]{$x_1$};
    \draw[thick,->, scale=1] (0,0,0) -- (0,1.2,0) node[anchor=north, font=\small]{$y_1$};
    \draw[thick,->, scale=1] (0,0,0) -- (0,0,1.2) node[anchor=east, font=\small]{$z_1$};
    
    
    \tdplotdrawarc[thin, dashed, scale=1]{(0,0,0)}{1}{-4}{356}{}{}
    
    \shade[ball color=orange, opacity=1, scale=2] (0,0,0) circle (0.05cm);
\end{tikzpicture}
\end{minipage}%
\begin{minipage}[b][2cm][t]{0.4\textwidth}
\centering
\tdplotsetmaincoords{70}{110}
\begin{tikzpicture}[tdplot_main_coords, scale=0.75]
    \shade[ball color=white!50, opacity=0.6] (0,0,0) circle (1cm);
    
    \draw[thick,->, scale=1] (0,0,0) -- (1.2,0,0) node[anchor=north east, font=\small]{$x_2$};
    \draw[thick,->, scale=1] (0,0,0) -- (0,-1.2,0) node[anchor=south, font=\small]{$y_2$};
    \draw[thick,->, scale=1] (0,0,0) -- (0,0,1.2) node[anchor=east, font=\small]{$z_2$};
    
    
    \tdplotdrawarc[thin, dashed, scale=1]{(0,0,0)}{1}{-4}{356}{}{}
    
    \shade[ball color=orange, opacity=1, scale=2] (0,0,0) circle (0.05cm);
\end{tikzpicture}
\end{minipage}
  \caption{$\Phi^+ = \ket{\uparrow\uparrow + \downarrow\downarrow}$}
  \label{figap1:sub-h}
\end{subfigure}

\vspace{0.5cm}

\begin{subfigure}{.5\textwidth}
  \centering
\begin{minipage}[b][2cm][t]{0.4\textwidth}
\centering
\tdplotsetmaincoords{70}{110}
\begin{tikzpicture}[tdplot_main_coords, scale=0.75]
    \shade[ball color=white!50, opacity=0.6] (0,0,0) circle (1cm);
    
    \draw[thick,->, scale=1] (0,0,0) -- (1.2,0,0) node[anchor=north east, font=\small]{$x_1$};
    \draw[thick,->, scale=1] (0,0,0) -- (0,1.2,0) node[anchor=north, font=\small]{$y_1$};
    \draw[thick,->, scale=1] (0,0,0) -- (0,0,1.2) node[anchor=east, font=\small]{$z_1$};
    
    
    \tdplotdrawarc[thin, dashed, scale=1]{(0,0,0)}{1}{-4}{356}{}{}
    
    \shade[ball color=orange, opacity=1, scale=2] (0,0,0) circle (0.05cm);
\end{tikzpicture}
\end{minipage}%
\begin{minipage}[b][2cm][t]{0.4\textwidth}
\centering
\tdplotsetmaincoords{70}{110}
\begin{tikzpicture}[tdplot_main_coords, scale=0.75]
    \shade[ball color=white!50, opacity=0.6] (0,0,0) circle (1cm);
    
    \draw[thick,->, scale=1] (0,0,0) -- (-1.2,0,0) node[anchor=south west, font=\small]{$y_2$};
    \draw[thick,->, scale=1] (0,0,0) -- (0,1.2,0) node[anchor=north, font=\small]{$z_2$};
    \draw[thick,->, scale=1] (0,0,0) -- (0,0,1.2) node[anchor=east, font=\small]{$x_2$};
    
    
    \tdplotdrawarc[thin, dashed, scale=1]{(0,0,0)}{1}{-4}{356}{}{}
    
    \shade[ball color=orange, opacity=1, scale=2] (0,0,0) circle (0.05cm);
\end{tikzpicture}
\end{minipage}
\caption{$\ket{\uparrow\uparrow + \uparrow\downarrow + i\downarrow\uparrow - i\downarrow\downarrow}$}
  \label{figap1:sub-i}
\end{subfigure}
\begin{subfigure}{.5\textwidth}
  \centering
\begin{minipage}[b][2cm][t]{0.4\textwidth}
\centering
\tdplotsetmaincoords{70}{110}
\begin{tikzpicture}[tdplot_main_coords, scale=0.75]
    \shade[ball color=white!50, opacity=0.6] (0,0,0) circle (1cm);
    
    \draw[thick,->, scale=1] (0,0,0) -- (1.2,0,0) node[anchor=north east, font=\small]{$x_1$};
    \draw[thick,->, scale=1] (0,0,0) -- (0,1.2,0) node[anchor=north, font=\small]{$y_1$};
    \draw[thick,->, scale=1] (0,0,0) -- (0,0,1.2) node[anchor=east, font=\small]{$z_1$};
    
    
    \tdplotdrawarc[thin, dashed, scale=1]{(0,0,0)}{1}{-4}{356}{}{}
    
    \shade[ball color=orange, opacity=1, scale=2] (0,0,0) circle (0.05cm);
\end{tikzpicture}
\end{minipage}%
\begin{minipage}[b][2cm][t]{0.4\textwidth}
\centering
\tdplotsetmaincoords{70}{110}
\begin{tikzpicture}[tdplot_main_coords, scale=0.75]
    \shade[ball color=white!50, opacity=0.6] (0,0,0) circle (1cm);
    
    \draw[thick,->, scale=1] (0,0,0) -- (1.2,0,0) node[anchor=north east, font=\small]{$y_2$};
    \draw[thick,->, scale=1] (0,0,0) -- (0,-1.2,0) node[anchor=south, font=\small]{$z_2$};
    \draw[thick,->, scale=1] (0,0,0) -- (0,0,1.2) node[anchor=east, font=\small]{$x_2$};
    
    
    \tdplotdrawarc[thin, dashed, scale=1]{(0,0,0)}{1}{-4}{356}{}{}
    
    \shade[ball color=orange, opacity=1, scale=2] (0,0,0) circle (0.05cm);
\end{tikzpicture}
\end{minipage}
\caption{$\ket{\uparrow\uparrow + \uparrow\downarrow - i\downarrow\uparrow + i\downarrow\downarrow}$}
  \label{figap1:sub-j}
\end{subfigure}

\vspace{0.5cm}

\begin{subfigure}{.5\textwidth}
  \centering
\begin{minipage}[b][2cm][t]{0.4\textwidth}
\centering
\tdplotsetmaincoords{70}{110}
\begin{tikzpicture}[tdplot_main_coords, scale=0.75]
    \shade[ball color=white!50, opacity=0.6] (0,0,0) circle (1cm);
    
    \draw[thick,->, scale=1] (0,0,0) -- (1.2,0,0) node[anchor=north east, font=\small]{$x_1$};
    \draw[thick,->, scale=1] (0,0,0) -- (0,1.2,0) node[anchor=north, font=\small]{$y_1$};
    \draw[thick,->, scale=1] (0,0,0) -- (0,0,1.2) node[anchor=east, font=\small]{$z_1$};
    
    
    \tdplotdrawarc[thin, dashed, scale=1]{(0,0,0)}{1}{-4}{356}{}{}
    
    \shade[ball color=orange, opacity=1, scale=2] (0,0,0) circle (0.05cm);
\end{tikzpicture}
\end{minipage}%
\begin{minipage}[b][2cm][t]{0.4\textwidth}
\centering
\tdplotsetmaincoords{70}{110}
\begin{tikzpicture}[tdplot_main_coords, scale=0.75]
    \shade[ball color=white!50, opacity=0.6] (0,0,0) circle (1cm);
    
    \draw[thick,->, scale=1] (0,0,0) -- (-1.2,0,0) node[anchor=south west, font=\small]{$z_2$};
    \draw[thick,->, scale=1] (0,0,0) -- (0,1.2,0) node[anchor=north, font=\small]{$x_2$};
    \draw[thick,->, scale=1] (0,0,0) -- (0,0,1.2) node[anchor=east, font=\small]{$y_2$};
    
    
    \tdplotdrawarc[thin, dashed, scale=1]{(0,0,0)}{1}{-4}{356}{}{}
    
    \shade[ball color=orange, opacity=1, scale=2] (0,0,0) circle (0.05cm);
\end{tikzpicture}
\end{minipage}
\caption{$\ket{\uparrow\uparrow + i\uparrow\downarrow - \downarrow\uparrow + i\downarrow\downarrow}$}
  \label{figap1:sub-k}
\end{subfigure}
\begin{subfigure}{.5\textwidth}
  \centering
\begin{minipage}[b][2cm][t]{0.4\textwidth}
\centering
\tdplotsetmaincoords{70}{110}
\begin{tikzpicture}[tdplot_main_coords, scale=0.75]
    \shade[ball color=white!50, opacity=0.6] (0,0,0) circle (1cm);
    
    \draw[thick,->, scale=1] (0,0,0) -- (1.2,0,0) node[anchor=north east, font=\small]{$x_1$};
    \draw[thick,->, scale=1] (0,0,0) -- (0,1.2,0) node[anchor=north, font=\small]{$y_1$};
    \draw[thick,->, scale=1] (0,0,0) -- (0,0,1.2) node[anchor=east, font=\small]{$z_1$};
    
    
    \tdplotdrawarc[thin, dashed, scale=1]{(0,0,0)}{1}{-4}{356}{}{}
    
    \shade[ball color=orange, opacity=1, scale=2] (0,0,0) circle (0.05cm);
\end{tikzpicture} 
\end{minipage}%
\begin{minipage}[b][2cm][t]{0.4\textwidth}
\centering
\tdplotsetmaincoords{70}{110}
\begin{tikzpicture}[tdplot_main_coords, scale=0.75]
    \shade[ball color=white!50, opacity=0.6] (0,0,0) circle (1cm);
    
    \draw[thick,->, scale=1] (0,0,0) -- (1.2,0,0) node[anchor=north east, font=\small]{$z_2$};
    \draw[thick,->, scale=1] (0,0,0) -- (0,-1.2,0) node[anchor=south, font=\small]{$x_2$};
    \draw[thick,->, scale=1] (0,0,0) -- (0,0,1.2) node[anchor=east, font=\small]{$y_2$};
    
    
    \tdplotdrawarc[thin, dashed, scale=1]{(0,0,0)}{1}{-4}{356}{}{}
    
    \shade[ball color=orange, opacity=1, scale=2] (0,0,0) circle (0.05cm);
\end{tikzpicture}
\end{minipage}
\caption{$\ket{\uparrow\uparrow + i\uparrow\downarrow + \downarrow\uparrow - i\downarrow\downarrow}$}
  \label{figap1:sub-l}
\end{subfigure}

\caption{Representation of the maximally entangled states from two qubit Stabilizer set. States are divided into four groups three each to indicate more symmetries shared within each group.}
\label{figap1:fig}
\end{figure}

\begin{figure}
\begin{subfigure}{.5\textwidth}
  \centering
\begin{minipage}[b][2cm][t]{0.4\textwidth}
\centering
\tdplotsetmaincoords{70}{110}
\begin{tikzpicture}[tdplot_main_coords, scale=0.75]
    \shade[ball color=white!50, opacity=0.6] (0,0,0) circle (1cm);
    
    \draw[thick,->, scale=1] (0,0,0) -- (1.2,0,0) node[anchor=north east, font=\small]{$x_1$};
    \draw[thick,->, scale=1] (0,0,0) -- (0,1.2,0) node[anchor=north, font=\small]{$y_1$};
    \draw[thick,->, scale=1] (0,0,0) -- (0,0,1.2) node[anchor=east, font=\small]{$z_1$};
    
    
    \tdplotdrawarc[thin, dashed, scale=1]{(0,0,0)}{1}{-4}{356}{}{}
    
    \shade[ball color=orange, opacity=1, scale=2] (0,0,0) circle (0.05cm);
\end{tikzpicture}
\end{minipage}%
\begin{minipage}[b][1.75cm][t]{0.4\textwidth}
\centering
\tdplotsetmaincoords{70}{110}
\begin{tikzpicture}[tdplot_main_coords, scale=0.75]
    \shade[ball color=white!50, opacity=0.6] (0,0,0) circle (1cm);
    
    \draw[thick,->, scale=1] (0,0,0) -- (0,-1.2,0) node[anchor=south, font=\small]{$x_2$};
    \draw[thick,->, scale=1] (0,0,0) -- (1.2,0,0) node[anchor=north, font=\small]{$y_2$};
    \draw[thick,->, scale=1] (0,0,0) -- (0,0,-1.2) node[anchor=west, font=\small]{$z_2$};
    
    
    \tdplotdrawarc[thin, dashed, scale=1]{(0,0,0)}{1}{-4}{356}{}{}
    
    \shade[ball color=orange, opacity=1, scale=2] (0,0,0) circle (0.05cm);
\end{tikzpicture}
\end{minipage}

  \caption{$\ket{\uparrow\downarrow - i\downarrow\uparrow}$}
  \label{figap2:sub-a}
\end{subfigure}
\begin{subfigure}{.5\textwidth}
  \centering
\begin{minipage}[b][2cm][t]{0.4\textwidth}
\centering
\tdplotsetmaincoords{70}{110}
\begin{tikzpicture}[tdplot_main_coords, scale=0.75]
    \shade[ball color=white!50, opacity=0.6] (0,0,0) circle (1cm);
    
    \draw[thick,->, scale=1] (0,0,0) -- (1.2,0,0) node[anchor=north east, font=\small]{$x_1$};
    \draw[thick,->, scale=1] (0,0,0) -- (0,1.2,0) node[anchor=north, font=\small]{$y_1$};
    \draw[thick,->, scale=1] (0,0,0) -- (0,0,1.2) node[anchor=east, font=\small]{$z_1$};
    
    
    \tdplotdrawarc[thin, dashed, scale=1]{(0,0,0)}{1}{-4}{356}{}{}
    
    \shade[ball color=orange, opacity=1, scale=2] (0,0,0) circle (0.05cm);
\end{tikzpicture}
\end{minipage}
\begin{minipage}[b][1.75cm][t]{0.4\textwidth}
\centering
\tdplotsetmaincoords{70}{110}
\begin{tikzpicture}[tdplot_main_coords, scale=0.75]
    \shade[ball color=white!50, opacity=0.6] (0,0,0) circle (1cm);
    
    \draw[thick,->, scale=1] (0,0,0) -- (0,1.2,0) node[anchor=north, font=\small]{$x_2$};
    \draw[thick,->, scale=1] (0,0,0) -- (-1.2,0,0) node[anchor=south east, font=\small]{$y_2$};
    \draw[thick,->, scale=1] (0,0,0) -- (0,0,-1.2) node[anchor=west, font=\small]{$z_2$};
    
    
    \tdplotdrawarc[thin, dashed, scale=1]{(0,0,0)}{1}{-4}{356}{}{}
    
    \shade[ball color=orange, opacity=1, scale=2] (0,0,0) circle (0.05cm);
\end{tikzpicture}
\end{minipage}

  \caption{$\ket{\uparrow\downarrow + i\downarrow\uparrow}$}
  \label{figap2:sub-b}
\end{subfigure}

\vspace{0.5cm}

\begin{subfigure}{.5\textwidth}
  \centering
\begin{minipage}[b][2cm][t]{0.4\textwidth}
\centering
\tdplotsetmaincoords{70}{110}
\begin{tikzpicture}[tdplot_main_coords, scale=0.75]
    \shade[ball color=white!50, opacity=0.6] (0,0,0) circle (1cm);
    
    \draw[thick,->, scale=1] (0,0,0) -- (1.2,0,0) node[anchor=north east, font=\small]{$x_1$};
    \draw[thick,->, scale=1] (0,0,0) -- (0,1.2,0) node[anchor=north, font=\small]{$y_1$};
    \draw[thick,->, scale=1] (0,0,0) -- (0,0,1.2) node[anchor=east, font=\small]{$z_1$};
    
    
    \tdplotdrawarc[thin, dashed, scale=1]{(0,0,0)}{1}{-4}{356}{}{}
    
    \shade[ball color=orange, opacity=1, scale=2] (0,0,0) circle (0.05cm);
\end{tikzpicture}
\end{minipage}%
\begin{minipage}[b][1.75cm][t]{0.4\textwidth}
\centering
\tdplotsetmaincoords{70}{110}
\begin{tikzpicture}[tdplot_main_coords, scale=0.75]
    \shade[ball color=white!50, opacity=0.6] (0,0,0) circle (1cm);
    
    \draw[thick,->, scale=1] (0,0,0) -- (0,-1.2,0) node[anchor=south, font=\small]{$y_2$};
    \draw[thick,->, scale=1] (0,0,0) -- (1.2,0,0) node[anchor=north, font=\small]{$z_2$};
    \draw[thick,->, scale=1] (0,0,0) -- (0,0,-1.2) node[anchor=west, font=\small]{$x_2$};
    
    
    \tdplotdrawarc[thin, dashed, scale=1]{(0,0,0)}{1}{-4}{356}{}{}
    
    \shade[ball color=orange, opacity=1, scale=2] (0,0,0) circle (0.05cm);
\end{tikzpicture}
\end{minipage}

\caption{$\ket{\uparrow\uparrow - \uparrow\downarrow + \downarrow\uparrow + \downarrow\downarrow}$}
  \label{figap2:sub-c}
\end{subfigure}
\begin{subfigure}{.5\textwidth}
  \centering
\begin{minipage}[b][2cm][t]{0.4\textwidth}
\centering
\tdplotsetmaincoords{70}{110}
\begin{tikzpicture}[tdplot_main_coords, scale=0.75]
    \shade[ball color=white!50, opacity=0.6] (0,0,0) circle (1cm);
    
    \draw[thick,->, scale=1] (0,0,0) -- (1.2,0,0) node[anchor=north east, font=\small]{$x_1$};
    \draw[thick,->, scale=1] (0,0,0) -- (0,1.2,0) node[anchor=north, font=\small]{$y_1$};
    \draw[thick,->, scale=1] (0,0,0) -- (0,0,1.2) node[anchor=east, font=\small]{$z_1$};
    
    
    \tdplotdrawarc[thin, dashed, scale=1]{(0,0,0)}{1}{-4}{356}{}{}
    
    \shade[ball color=orange, opacity=1, scale=2] (0,0,0) circle (0.05cm);
\end{tikzpicture}
\end{minipage}
\begin{minipage}[b][1.75cm][t]{0.4\textwidth}
\centering
\tdplotsetmaincoords{70}{110}
\begin{tikzpicture}[tdplot_main_coords, scale=0.75]
    \shade[ball color=white!50, opacity=0.6] (0,0,0) circle (1cm);
    
    \draw[thick,->, scale=1] (0,0,0) -- (0,1.2,0) node[anchor=north, font=\small]{$y_2$};
    \draw[thick,->, scale=1] (0,0,0) -- (-1.2,0,0) node[anchor=south east, font=\small]{$z_2$};
    \draw[thick,->, scale=1] (0,0,0) -- (0,0,-1.2) node[anchor=west, font=\small]{$x_2$};
    
    
    \tdplotdrawarc[thin, dashed, scale=1]{(0,0,0)}{1}{-4}{356}{}{}
    
    \shade[ball color=orange, opacity=1, scale=2] (0,0,0) circle (0.05cm);
\end{tikzpicture}
\end{minipage}

\caption{$\ket{\uparrow\uparrow - \uparrow\downarrow - \downarrow\uparrow - \downarrow\downarrow}$}
  \label{figap2:sub-d}
\end{subfigure}

\vspace{0.5cm}

\begin{subfigure}{.5\textwidth}
  \centering
\begin{minipage}[b][2cm][t]{0.4\textwidth}
\centering
\tdplotsetmaincoords{70}{110}
\begin{tikzpicture}[tdplot_main_coords, scale=0.75]
    \shade[ball color=white!50, opacity=0.6] (0,0,0) circle (1cm);
    
    \draw[thick,->, scale=1] (0,0,0) -- (1.2,0,0) node[anchor=north east, font=\small]{$x_1$};
    \draw[thick,->, scale=1] (0,0,0) -- (0,1.2,0) node[anchor=north, font=\small]{$y_1$};
    \draw[thick,->, scale=1] (0,0,0) -- (0,0,1.2) node[anchor=east, font=\small]{$z_1$};
    
    
    \tdplotdrawarc[thin, dashed, scale=1]{(0,0,0)}{1}{-4}{356}{}{}
    
    \shade[ball color=orange, opacity=1, scale=2] (0,0,0) circle (0.05cm);
\end{tikzpicture}
\end{minipage}%
\begin{minipage}[b][1.75cm][t]{0.4\textwidth}
\centering
\tdplotsetmaincoords{70}{110}
\begin{tikzpicture}[tdplot_main_coords, scale=0.75]
    \shade[ball color=white!50, opacity=0.6] (0,0,0) circle (1cm);
    
    \draw[thick,->, scale=1] (0,0,0) -- (0,-1.2,0) node[anchor=south, font=\small]{$z_2$};
    \draw[thick,->, scale=1] (0,0,0) -- (1.2,0,0) node[anchor=north, font=\small]{$x_2$};
    \draw[thick,->, scale=1] (0,0,0) -- (0,0,-1.2) node[anchor=west, font=\small]{$y_2$};
    
    
    \tdplotdrawarc[thin, dashed, scale=1]{(0,0,0)}{1}{-4}{356}{}{}
    
    \shade[ball color=orange, opacity=1, scale=2] (0,0,0) circle (0.05cm);
\end{tikzpicture}
\end{minipage}

\caption{$\ket{\uparrow\uparrow - i\uparrow\downarrow - i\downarrow\uparrow + \downarrow\downarrow}$}
  \label{figap2:sub-e}
\end{subfigure}
\begin{subfigure}{.5\textwidth}
  \centering
\begin{minipage}[b][2cm][t]{0.4\textwidth}
\centering
\tdplotsetmaincoords{70}{110}
\begin{tikzpicture}[tdplot_main_coords, scale=0.75]
    \shade[ball color=white!50, opacity=0.6] (0,0,0) circle (1cm);
    
    \draw[thick,->, scale=1] (0,0,0) -- (1.2,0,0) node[anchor=north east, font=\small]{$x_1$};
    \draw[thick,->, scale=1] (0,0,0) -- (0,1.2,0) node[anchor=north, font=\small]{$y_1$};
    \draw[thick,->, scale=1] (0,0,0) -- (0,0,1.2) node[anchor=east, font=\small]{$z_1$};
    
    
    \tdplotdrawarc[thin, dashed, scale=1]{(0,0,0)}{1}{-4}{356}{}{}
    
    \shade[ball color=orange, opacity=1, scale=2] (0,0,0) circle (0.05cm);
\end{tikzpicture}
\end{minipage}
\begin{minipage}[b][1.75cm][t]{0.4\textwidth}
\centering
\tdplotsetmaincoords{70}{110}
\begin{tikzpicture}[tdplot_main_coords, scale=0.75]
    \shade[ball color=white!50, opacity=0.6] (0,0,0) circle (1cm);
    
    \draw[thick,->, scale=1] (0,0,0) -- (0,1.2,0) node[anchor=north, font=\small]{$z_2$};
    \draw[thick,->, scale=1] (0,0,0) -- (-1.2,0,0) node[anchor=south east, font=\small]{$x_2$};
    \draw[thick,->, scale=1] (0,0,0) -- (0,0,-1.2) node[anchor=west, font=\small]{$y_2$};
    
    
    \tdplotdrawarc[thin, dashed, scale=1]{(0,0,0)}{1}{-4}{356}{}{}
    
    \shade[ball color=orange, opacity=1, scale=2] (0,0,0) circle (0.05cm);
\end{tikzpicture}
\end{minipage}

\caption{$\ket{\uparrow\uparrow - i\uparrow\downarrow + i\downarrow\uparrow - \downarrow\downarrow}$}
  \label{figap2:sub-f}
\end{subfigure}

\vspace{0.5cm}

\begin{subfigure}{.5\textwidth}
  \centering
\begin{minipage}[b][2cm][t]{0.4\textwidth}
\centering
\tdplotsetmaincoords{70}{110}
\begin{tikzpicture}[tdplot_main_coords, scale=0.75]
    \shade[ball color=white!50, opacity=0.6] (0,0,0) circle (1cm);
    
    \draw[thick,->, scale=1] (0,0,0) -- (1.2,0,0) node[anchor=north east, font=\small]{$x_1$};
    \draw[thick,->, scale=1] (0,0,0) -- (0,1.2,0) node[anchor=north, font=\small]{$y_1$};
    \draw[thick,->, scale=1] (0,0,0) -- (0,0,1.2) node[anchor=east, font=\small]{$z_1$};
    
    
    \tdplotdrawarc[thin, dashed, scale=1]{(0,0,0)}{1}{-4}{356}{}{}
    
    \shade[ball color=orange, opacity=1, scale=2] (0,0,0) circle (0.05cm);
\end{tikzpicture}
\end{minipage}%
\begin{minipage}[b][2cm][t]{0.4\textwidth}
\centering
\tdplotsetmaincoords{70}{110}
\begin{tikzpicture}[tdplot_main_coords, scale=0.75]
    \shade[ball color=white!50, opacity=0.6] (0,0,0) circle (1cm);
    
    \draw[thick,->, scale=1] (0,0,0) -- (0,-1.2,0) node[anchor=south, font=\small]{$x_2$};
    \draw[thick,->, scale=1] (0,0,0) -- (-1.2,0,0) node[anchor=south, font=\small]{$y_2$};
    \draw[thick,->, scale=1] (0,0,0) -- (0,0,1.2) node[anchor=east, font=\small]{$z_2$};
    
    
    \tdplotdrawarc[thin, dashed, scale=1]{(0,0,0)}{1}{-4}{356}{}{}
    
    \shade[ball color=orange, opacity=1, scale=2] (0,0,0) circle (0.05cm);
\end{tikzpicture}
\end{minipage}

\caption{$\ket{\uparrow\uparrow - i\downarrow\downarrow}$}
  \label{figap2:sub-g}
\end{subfigure}
\begin{subfigure}{.5\textwidth}
  \centering
\begin{minipage}[b][2cm][t]{0.4\textwidth}
\centering
\tdplotsetmaincoords{70}{110}
\begin{tikzpicture}[tdplot_main_coords, scale=0.75]
    \shade[ball color=white!50, opacity=0.6] (0,0,0) circle (1cm);
    
    \draw[thick,->, scale=1] (0,0,0) -- (1.2,0,0) node[anchor=north east, font=\small]{$x_1$};
    \draw[thick,->, scale=1] (0,0,0) -- (0,1.2,0) node[anchor=north, font=\small]{$y_1$};
    \draw[thick,->, scale=1] (0,0,0) -- (0,0,1.2) node[anchor=east, font=\small]{$z_1$};
    
    
    \tdplotdrawarc[thin, dashed, scale=1]{(0,0,0)}{1}{-4}{356}{}{}
    
    \shade[ball color=orange, opacity=1, scale=2] (0,0,0) circle (0.05cm);
\end{tikzpicture}
\end{minipage}
\begin{minipage}[b][2cm][t]{0.4\textwidth}
\centering
\tdplotsetmaincoords{70}{110}
\begin{tikzpicture}[tdplot_main_coords, scale=0.75]
    \shade[ball color=white!50, opacity=0.6] (0,0,0) circle (1cm);
    
    \draw[thick,->, scale=1] (0,0,0) -- (0,1.2,0) node[anchor=north, font=\small]{$x_2$};
    \draw[thick,->, scale=1] (0,0,0) -- (1.2,0,0) node[anchor=north, font=\small]{$y_2$};
    \draw[thick,->, scale=1] (0,0,0) -- (0,0,1.2) node[anchor=west, font=\small]{$z_2$};
    
    
    \tdplotdrawarc[thin, dashed, scale=1]{(0,0,0)}{1}{-4}{356}{}{}
    
    \shade[ball color=orange, opacity=1, scale=2] (0,0,0) circle (0.05cm);
\end{tikzpicture}
\end{minipage}
\caption{$\ket{\uparrow\uparrow + i\downarrow\downarrow}$}
  \label{figap2:sub-h}
\end{subfigure}

\vspace{0.5cm}

\begin{subfigure}{.5\textwidth}
  \centering
\begin{minipage}[b][2cm][t]{0.4\textwidth}
\centering
\tdplotsetmaincoords{70}{110}
\begin{tikzpicture}[tdplot_main_coords, scale=0.75]
    \shade[ball color=white!50, opacity=0.6] (0,0,0) circle (1cm);
    
    \draw[thick,->, scale=1] (0,0,0) -- (1.2,0,0) node[anchor=north east, font=\small]{$x_1$};
    \draw[thick,->, scale=1] (0,0,0) -- (0,1.2,0) node[anchor=north, font=\small]{$y_1$};
    \draw[thick,->, scale=1] (0,0,0) -- (0,0,1.2) node[anchor=east, font=\small]{$z_1$};
    
    
    \tdplotdrawarc[thin, dashed, scale=1]{(0,0,0)}{1}{-4}{356}{}{}
    
    \shade[ball color=orange, opacity=1, scale=2] (0,0,0) circle (0.05cm);
\end{tikzpicture}
\end{minipage}%
\begin{minipage}[b][2cm][t]{0.4\textwidth}
\centering
\tdplotsetmaincoords{70}{110}
\begin{tikzpicture}[tdplot_main_coords, scale=0.75]
    \shade[ball color=white!50, opacity=0.6] (0,0,0) circle (1cm);
    
    \draw[thick,->, scale=1] (0,0,0) -- (0,-1.2,0) node[anchor=south, font=\small]{$y_2$};
    \draw[thick,->, scale=1] (0,0,0) -- (-1.2,0,0) node[anchor=south, font=\small]{$z_2$};
    \draw[thick,->, scale=1] (0,0,0) -- (0,0,1.2) node[anchor=east, font=\small]{$x_2$};
    
    
    \tdplotdrawarc[thin, dashed, scale=1]{(0,0,0)}{1}{-4}{356}{}{}
    
    \shade[ball color=orange, opacity=1, scale=2] (0,0,0) circle (0.05cm);
\end{tikzpicture}
\end{minipage}

\caption{$\ket{\uparrow\uparrow + \uparrow\downarrow - \downarrow\uparrow + \downarrow\downarrow}$}
  \label{figap2:sub-i}
\end{subfigure}
\begin{subfigure}{.5\textwidth}
  \centering
\begin{minipage}[b][2cm][t]{0.4\textwidth}
\centering
\tdplotsetmaincoords{70}{110}
\begin{tikzpicture}[tdplot_main_coords, scale=0.75]
    \shade[ball color=white!50, opacity=0.6] (0,0,0) circle (1cm);
    
    \draw[thick,->, scale=1] (0,0,0) -- (1.2,0,0) node[anchor=north east, font=\small]{$x_1$};
    \draw[thick,->, scale=1] (0,0,0) -- (0,1.2,0) node[anchor=north, font=\small]{$y_1$};
    \draw[thick,->, scale=1] (0,0,0) -- (0,0,1.2) node[anchor=east, font=\small]{$z_1$};
    
    
    \tdplotdrawarc[thin, dashed, scale=1]{(0,0,0)}{1}{-4}{356}{}{}
    
    \shade[ball color=orange, opacity=1, scale=2] (0,0,0) circle (0.05cm);
\end{tikzpicture}
\end{minipage}
\begin{minipage}[b][2cm][t]{0.4\textwidth}
\centering
\tdplotsetmaincoords{70}{110}
\begin{tikzpicture}[tdplot_main_coords, scale=0.75]
    \shade[ball color=white!50, opacity=0.6] (0,0,0) circle (1cm);
    
    \draw[thick,->, scale=1] (0,0,0) -- (0,1.2,0) node[anchor=north, font=\small]{$y_2$};
    \draw[thick,->, scale=1] (0,0,0) -- (1.2,0,0) node[anchor=north, font=\small]{$z_2$};
    \draw[thick,->, scale=1] (0,0,0) -- (0,0,1.2) node[anchor=west, font=\small]{$x_2$};
    
    
    \tdplotdrawarc[thin, dashed, scale=1]{(0,0,0)}{1}{-4}{356}{}{}
    
    \shade[ball color=orange, opacity=1, scale=2] (0,0,0) circle (0.05cm);
\end{tikzpicture}
\end{minipage}
\caption{$\ket{\uparrow\uparrow + \uparrow\downarrow + \downarrow\uparrow - \downarrow\downarrow}$}
  \label{figap2:sub-j}
\end{subfigure}

\vspace{0.5cm}

\begin{subfigure}{.5\textwidth}
  \centering
\begin{minipage}[b][2cm][t]{0.4\textwidth}
\centering
\tdplotsetmaincoords{70}{110}
\begin{tikzpicture}[tdplot_main_coords, scale=0.75]
    \shade[ball color=white!50, opacity=0.6] (0,0,0) circle (1cm);
    
    \draw[thick,->, scale=1] (0,0,0) -- (1.2,0,0) node[anchor=north east, font=\small]{$x_1$};
    \draw[thick,->, scale=1] (0,0,0) -- (0,1.2,0) node[anchor=north, font=\small]{$y_1$};
    \draw[thick,->, scale=1] (0,0,0) -- (0,0,1.2) node[anchor=east, font=\small]{$z_1$};
    
    
    \tdplotdrawarc[thin, dashed, scale=1]{(0,0,0)}{1}{-4}{356}{}{}
    
    \shade[ball color=orange, opacity=1, scale=2] (0,0,0) circle (0.05cm);
\end{tikzpicture}
\end{minipage}%
\begin{minipage}[b][2cm][t]{0.4\textwidth}
\centering
\tdplotsetmaincoords{70}{110}
\begin{tikzpicture}[tdplot_main_coords, scale=0.75]
    \shade[ball color=white!50, opacity=0.6] (0,0,0) circle (1cm);
    
    \draw[thick,->, scale=1] (0,0,0) -- (0,-1.2,0) node[anchor=south, font=\small]{$z_2$};
    \draw[thick,->, scale=1] (0,0,0) -- (-1.2,0,0) node[anchor=south, font=\small]{$x_2$};
    \draw[thick,->, scale=1] (0,0,0) -- (0,0,1.2) node[anchor=east, font=\small]{$y_2$};
    
    
    \tdplotdrawarc[thin, dashed, scale=1]{(0,0,0)}{1}{-4}{356}{}{}
    
    \shade[ball color=orange, opacity=1, scale=2] (0,0,0) circle (0.05cm);
\end{tikzpicture}
\end{minipage}

\caption{$\ket{\uparrow\uparrow + i\uparrow\downarrow - i\downarrow\uparrow - \downarrow\downarrow}$}
  \label{figap2:sub-k}
\end{subfigure}
\begin{subfigure}{.5\textwidth}
  \centering
\begin{minipage}[b][2cm][t]{0.4\textwidth}
\centering
\tdplotsetmaincoords{70}{110}
\begin{tikzpicture}[tdplot_main_coords, scale=0.75]
    \shade[ball color=white!50, opacity=0.6] (0,0,0) circle (1cm);
    
    \draw[thick,->, scale=1] (0,0,0) -- (1.2,0,0) node[anchor=north east, font=\small]{$x_1$};
    \draw[thick,->, scale=1] (0,0,0) -- (0,1.2,0) node[anchor=north, font=\small]{$y_1$};
    \draw[thick,->, scale=1] (0,0,0) -- (0,0,1.2) node[anchor=east, font=\small]{$z_1$};
    
    
    \tdplotdrawarc[thin, dashed, scale=1]{(0,0,0)}{1}{-4}{356}{}{}
    
    \shade[ball color=orange, opacity=1, scale=2] (0,0,0) circle (0.05cm);
\end{tikzpicture}
\end{minipage}
\begin{minipage}[b][2cm][t]{0.4\textwidth}
\centering
\tdplotsetmaincoords{70}{110}
\begin{tikzpicture}[tdplot_main_coords, scale=0.75]
    \shade[ball color=white!50, opacity=0.6] (0,0,0) circle (1cm);
    
    \draw[thick,->, scale=1] (0,0,0) -- (0,1.2,0) node[anchor=north, font=\small]{$z_2$};
    \draw[thick,->, scale=1] (0,0,0) -- (1.2,0,0) node[anchor=north, font=\small]{$x_2$};
    \draw[thick,->, scale=1] (0,0,0) -- (0,0,1.2) node[anchor=west, font=\small]{$y_2$};
    
    
    \tdplotdrawarc[thin, dashed, scale=1]{(0,0,0)}{1}{-4}{356}{}{}
    
    \shade[ball color=orange, opacity=1, scale=2] (0,0,0) circle (0.05cm);
\end{tikzpicture}
\end{minipage}
\caption{$\ket{\uparrow\uparrow + i\uparrow\downarrow + i\downarrow\uparrow + \downarrow\downarrow}$}
  \label{figap2:sub-l}
\end{subfigure}

\caption{(continued) Representation of the maximally entangled states from two qubit Stabilizer set. States are divided into four groups three each to indicate more symmetries shared within each group.}
\label{figap2:fig}
\end{figure}

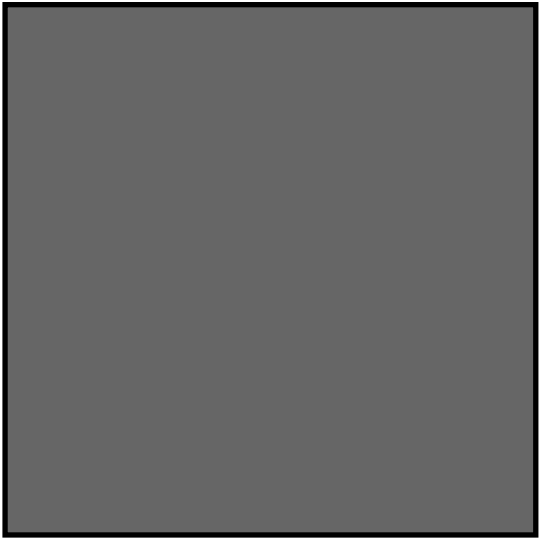
\end{document}